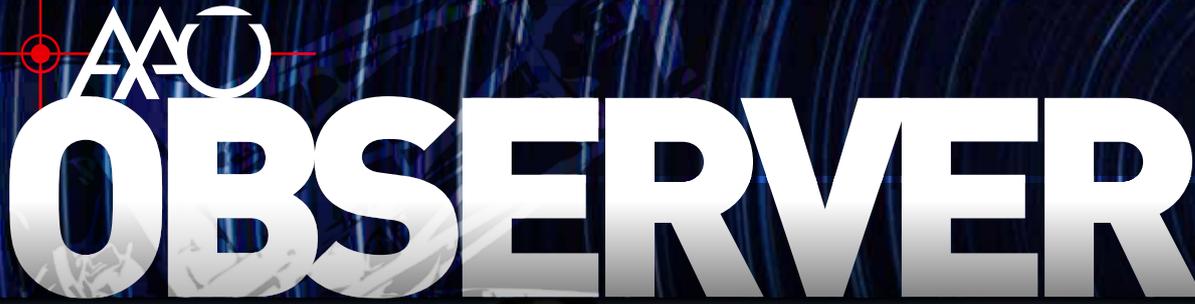

# AAO OBSERVER

THE AUSTRALIAN ASTRONOMICAL OBSERVATORY NEWSLETTER

NUMBER **122** | AUGUST **2012**

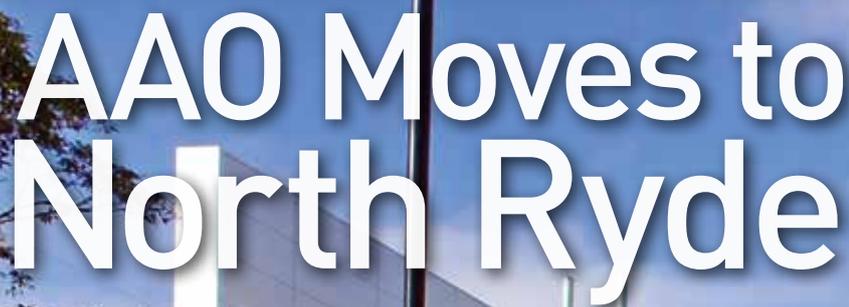

## AAO Moves to North Ryde

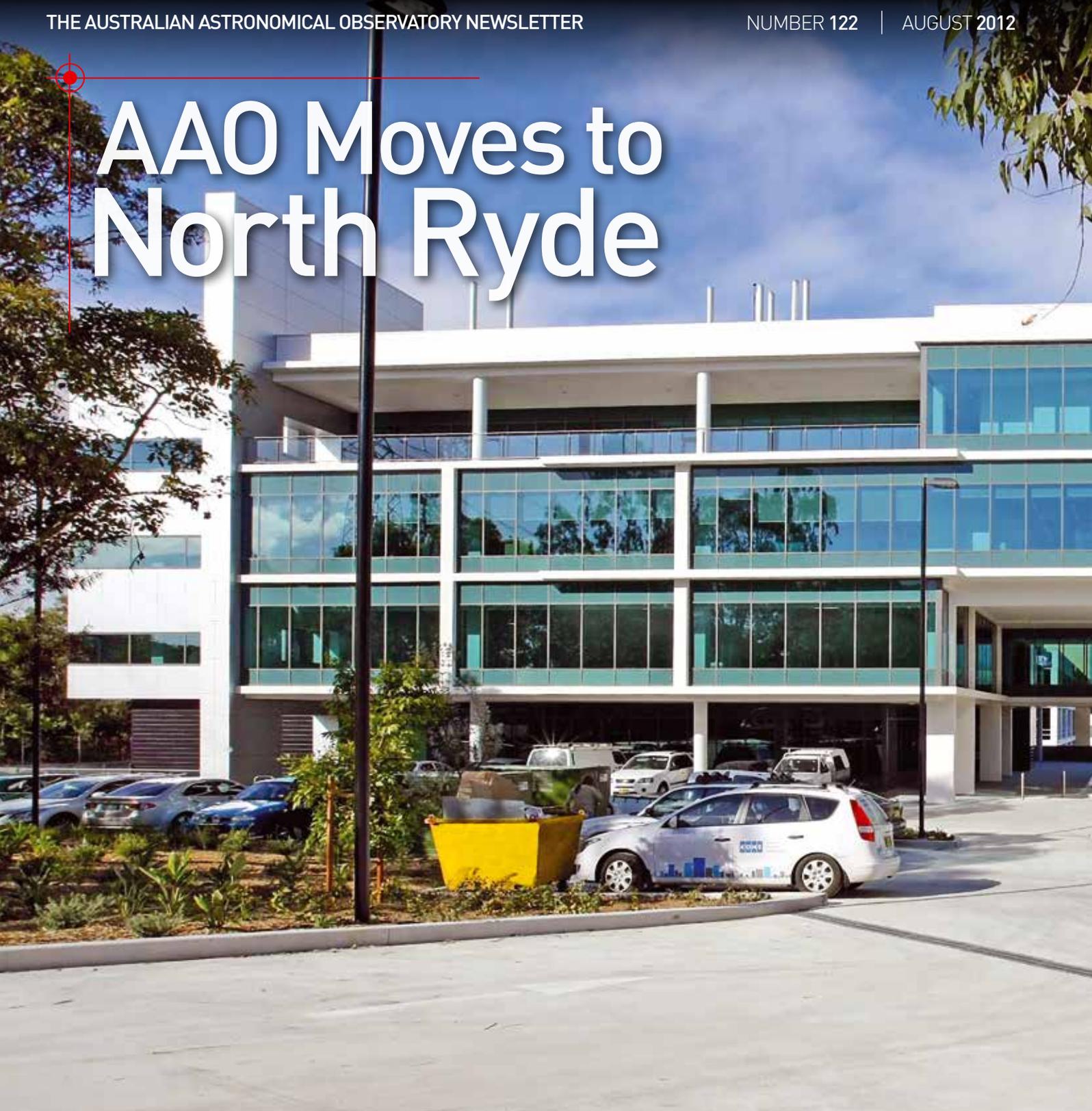

**SPIE Extravaganza** | **Cosmology with 6dFGS** | **HERMES comes together**

# DIRECTOR'S MESSAGE

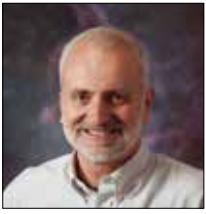

# Director's message
Matthew Colless

This will be my final Director's Message for the AAO Newsletter. In December, after nine years in the job, I will be leaving the AAO to become Director of the Research School of Astronomy & Astrophysics at the Australian National University. While I will be sad to leave the AAO – an institution and a group of people that I love – I'm happy to be able to say that I leave it in excellent health, and with a bright future ahead.

Since I became Director in January 2004 the AAO has gone through some major changes. Just as I was starting, the UK decided to withdraw from the Anglo-Australian Observatory after an outstandingly successful partnership spanning 35 years. Rather than seek to reverse this decision, I chose to set about convincing the Australian government to transform the AAO into Australia's national optical observatory. This challenge was made more difficult by the rapid ramping down of the UK's nominal 50% share in the AAO's funding over the five years prior to the formal ending of the AAT Agreement in June 2010. With wise advice from the AAT Board, creative energy from AAO staff and much-appreciated support and funding from what was then the Department of Innovation, Industry, Science and Research, the AAO came through this difficult period and the AAO was reincarnated as a wholly Australian organisation.

After examining a variety of governance options, the decision was eventually taken to make the new Australian Astronomical Observatory part of the public service, as a division of the Department. Although there was some trepidation inside and outside the AAO about this new arrangement, in the event it has proved highly beneficial overall. The AAO now has a strong legislative basis (the Australian Astronomical Observatory Act 2010) that mandates its existence and lays out the functions it performs as Australia's national optical observatory. It also has an appropriate 'home' alongside other national research organisations within what is now the Department of Industry, Innovation, Science, Research and Tertiary Education. This arrangement provides ongoing support from government and a stable basis for long-term funding.

During this period, the AAO has continued to enable outstanding science by Australian, British and other astronomers. Big surveys are one of the AAT's great strengths, and highlights from the past nine years have included the 2dF Galaxy Redshift Survey's precise measurements of the density of dark matter, baryons and neutrinos; an independent detection of the accelerating expansion of the universe by the WiggleZ survey; the richly detailed multi-wavelength view of galaxies and their environments from the GAMA survey (see page 13); and, in a very different style of survey, the discovery of tens of exoplanets by the Anglo-Australian Planet Search. The UKST has also been devoted to surveys: first the 6dF Galaxy Survey, which has mapped the redshifts and peculiar motions of galaxies in the local universe (see page 10), and then the Radial Velocity (RAVE) survey of the stellar populations and kinematics of the Milky Way.

# CONTENTS

Besides these large surveys, however, there has also been high-impact science emerging from the many smaller observing programs on the AAT. Just a few of the most memorable for me were the discovery of ultra-compact dwarf (UCD) galaxies, the first demonstration of the viability of chemical tagging for Galactic archaeology studies, the remarkable interior views of stars provided by asteroseismology, and the discovery of analogues to high-redshift galaxies in the local universe.

At the same time, the AAO has continued its tradition of innovation in astronomical instrumentation, providing new capabilities for the AAT and other telescopes and developing exciting new technologies. For the AAT, the major developments in my tenure were the completion of the AAOmega fibre-fed spectrograph and the conception, design and construction of the HERMES high-resolution multi-object spectrograph, which is expected to dramatically improve our understanding of the formation of the Milky Way when it comes on-line in 2013. Aside from these headline instruments, the AAT instrument suite has also had a series of significant upgrades, including the CYCLOPS image-slicer for UCLES and the KOALA fibre integral field that will shortly replace SPIRAL. The AAT itself has undergone a thorough refurbishment of all critical components, from the telescope drives to the new dome shutter brake, and is now in excellent condition for another decade of service.

The AAO's instrumentation program builds on its investment in innovative technologies. In the last few years two potentially revolutionary technologies have been tested on-sky at the AAT: fibre Bragg gratings for suppressing the OH emission that dominates the sky background at near-infrared wavelengths, and a prototype integrated photonic spectrograph that is the forerunner of a future generation of solid-state astronomical instrumentation. The AAO is currently exploring two new instrument concepts for the AAT: an ultra-precise high-resolution spectrograph (Veloce) and a multi-object integral field spectrograph (Hector); a prototype for the latter, called SAMI, is already deployed on the AAT.

The AAO's instrumentation program has also provided instruments and designs for other telescopes around the world. Noteworthy examples include the Echidna fibre positioner constructed for the FMOS spectrograph on Subaru, the WFMOS multi-object spectrograph designed for Gemini and Subaru, the PILOT concept for a 2.5-metre Antarctic telescopes, the MANIFEST fibre-feed system for the first-generation instruments on the Giant Magellan Telescope, and the GHOST concept for the Gemini high-resolution optical spectrograph.

I have reviewed in some depth the current state of the AAO and the challenges it faces over the next few years in the recently released AAO Forward Look to 2015 (see http://www.aao.gov.au/AAOForwardLook.html). The Forward Look sets out 30 specific recommendations for improving the effectiveness and efficiency of the AAO in serving the Australian astronomical community and other users of its facilities and instruments around the world. These include: providing remote observing capabilities for the AAT; improving the telescope's infrastructure and instrumentation in order to reduce the burden of repairs and maintenance; building major new instruments for the AAT and UKST; revamping the services provided by the Australian Gemini Office; supporting Australian efforts to join the European Southern Observatory; further improving the research environment for AAO staff, students and users of our facilities; strengthening the AAO's network of technology linkages and enhancing interaction with Australian industry; and, of course, continuing to maintain a highly transparent and consultative relationship with all our stakeholders.

My last major task before leaving the AAO has been to ensure a smooth transition for the AAO's headquarters from our current aging buildings on CSIRO's Marsfield campus to our brand new premises at North Ryde, which we share with the National Measurement Institute. This big move occurred over 11-12 August. I hope many of you will come and visit our new premises now we are settled in. An article in this newsletter (page 34) explains how these new premises, with their modern office facilities, purpose-built labs and large instrument assembly areas, will open up exciting new opportunities for the AAO.

With a clear mandate as Australia's national optical observatory, strong support from the Department, stable long-term funding, ambitious new science projects and an innovative instrumentation program, the new Director of the AAO has a strong foundation on which to build the organisation's future. But the greatest strength of the AAO is, and will remain, the remarkable people who work here. It has been my great pleasure and privilege to work with them, and I thank them all for everything that we have accomplished together.

### Thanks and best wishes to Matthew from the staff of the AAO

Over the nine years that Matthew has been Director, the AAO has seen many exciting successes and many changes. These include key scientific results, such as the announcement by the 2dF team of the detection of the BAO signal in the galaxy distribution to the more recent WiggleZ announcement confirming Dark Energy at higher redshift. They include major instrumentation successes, from the commissioning of AAOmega for the AAT, including the SPIRAL IFU feed as well as its multi-object mode, through building the FMOS Echidna fibre positioner for Subaru, to the decision to build HERMES as the next major AAT instrument. The first on-sky demonstration of OH-suppression fibres was a very exciting moment, and the AAO now aspires to build MANIFEST for the GMT, a sophisticated fibre positioning system using the new Starbugs technology.

Through his leadership of the AAO, Matthew has successfully navigated the organisation through a variety of challenges. These include most notably the staged withdrawal of the UK from the bilateral Anglo-Australian days and our transition to a Division within a Federal Government Department, to, most recently, the physical relocation of the AAO Sydney headquarters, another new "first" for the AAO. The past nine years have been a very productive and fruitful time for the AAO, which is saying a lot for an organisation with a history of major innovations and scientific successes. The staff at the AAO have benefited substantially from Matthew's time as Director, and we thank him for all his efforts. We particularly appreciate the genuine care and concern that Matthew displays for all his staff. We hope that he will find a warm welcome at Stromlo, and will have as successful and productive a time there as he has in his time at the AAO. Best wishes Matthew!

*Andrew Hopkins, on behalf of the AAO staff.*

# SPIE Extravaganza

Jon Lawrence, James Gilbert and Andy Green (AAO)

The SPIE Astronomical Telescopes and Instrumentation conference is the premier international meeting for this field that is held every 2 years. The first week of July 2012 saw this meeting convene in Amsterdam, the Netherlands. With over 2300 attendees, and more than 2000 papers, this six day conference is large. This year, the AAO was well represented, with more than a dozen members of the Instrument Science and Instrumentation groups in attendance. AAO staff lead 25 contributions, and assisted on 41.

Our contributions ranged from well-developed projects such as HERMES through exciting new technologies like Star Bugs to far out concepts like photonic spectrographs. Papers surrounding several instruments have already been demonstrated on the telescope. GNOSIS suppresses night sky lines from infrared light using fibre Bragg gratings imprinted in a fibre feed for IRIS2 demonstrated late last year. The Sydney-AAO Multi-object Integral Field Unit (SAMI) has been awarded competitive time on the AAT to begin a large, spatially resolved spectroscopic galaxy survey. And both CYCLOPS2 and the corresponding support unit CURE fed first light into UCLES in late July. Many of these technologies will benefit Australian and international astronomers in the near future.

Of particular interest at the workshop was the AAO's Starbugs concept. James Gilbert presented Starbugs to a very full room at the Modern Technologies section, and was happy to say it was extremely well received. There were no grumbles, one or two surprised mumbles and even some excitable chatter when the demo video was shown. Such a positive reaction is testament to the compelling results achieved with Starbugs over the last two years, as well as the fact that they are, to quote one SPIE attendee, "just so cool". Of course there's always a danger of falling short when calling a paper 'Starbugs: all-singing, all-dancing fibre positioning robots', but the level of progress we had to report fully justified the title. As the conference drew to a close, the talk was one of five nominated for the 'best oral presentation' award, although in the end the trophy was clinched by another presenter. C'est la vie. By all accounts though, SPIE 2012 was a definite success for AAO fibre positioners, with a paper on MANIFEST by Michael Goodwin and with Will Saunders cheerleading MOHAWK, a small-pitch-high-fibre-count Echidna style positioner for the Blanco telescope.

More information on the conference and the book of abstracts are available from the SPIE website: spie.org/x89022.xml

**List of AAO contributions, with selected abstracts:**

- Robert Content, Ray Sharples, Mathew J. Page, Richard Cole, David M. Walton, Berend Winter, Kristian Pedersen, Jens Hjorth, Michael Andersen, Allan Hornstrup, Jan-Willem A. den Herder, Luigi Piro, "Optical telescope BIRT in ORIGIN for high-Z gamma ray burst observing", Proc SPIE 8442-164, 2012.

- Michael C. B. Ashley, Yael Augarten, Colin S. Bonner, Michael G. Burton, Luke Bycroft, Jon S. Lawrence, Daniel M. Luong-Van, Scott McDaid, Campbell McLaren, Geoff Sims, John W. V. Storey, "PLATO-R: a new concept for Antarctic science", Proc SPIE 8444-63, 2012.

- Will Saunders, "A simple wide-field telescope design for Antarctica", Proc SPIE 8444-205, 2012.

- Geoff Sims, Craig Kulesa, Michael C. B. Ashley, Jon S. Lawrence, Will Saunders, John W. V. Storey, "Where is Ridge A?", Proc SPIE 8444-209, 2012.

    *First identified in 2009 as the site with the lowest precipitable water and best terahertz transmission on earth, Ridge A is located approximately 150 km south of Dome A, Antarctica . To further refine this optimum location prior to deployment in 2012 of a robotic THz observatory, we have modelled the atmospheric transmission as a function of location over a 1000 km square grid using three years of data from the Microwave Humidity Sounder on the NOAA satellite . The modelling identifies a broad area of exceptionally low water vapour close to the 4,000 metre elevation contour, reaching below 25 microns for extended periods of time.*

- Barnaby R. Norris, Peter G. Tuthill, Michael J. Ireland, Sylvestre Lacour, "Probing dusty circumstellar environments with polarimetric aperture masking interferometry (THESIS)", Proc SPIE 8445-02, 2012.

- Nemanja Jovanovic, Peter G. Tuthill, Barnaby Norris, Simon Gross, Paul Stewart, Ned Charles, Sylvestre Lacour, Jon Lawrence, Gordon Robertson, Alexander Fuerbach and Michael J. Withford, "Progress and challenges with the Dragonfly instrument: an integrated-photonic pupil-remapping interferometer", Proc SPIE 8445-04, 2012.

- Michael J. Ireland, "Detecting extrasolar planets with sparse aperture masking", Proc SPIE 8445-05, 2012.

- Vicente Maestro, Michael J. Ireland, Peter G. Tuthill, Daniel Huber, Gail Schaefer, "Imaging rapid rotators with the PAVO beam combiner at CHARA", Proc SPIE 8445-15, 2012.

- J. Gordon Robertson, Michael J. Ireland, William J. Tango, Peter G. Tuthill, Benjamin A. Warrington, Yitping Kok, Andrew P. Jacob, "Science and technology progress at the Sydney University stellar interferometer", Proc SPIE 8445-21, 2012.

- Yitping Kok, Michael J. Ireland, Peter G. Tuthill, James G. Robertson, Benjamin A. Warrington, William J. Tango, "Self-phase-referencing interferometry with SUSI", Proc SPIE 8445-72, 2012.

- Naoyuki Tamura, Naruhisa Takato, Fumihide Iwamuro, Masayuki Akiyama, Masahiko Kimura, Philip Tait, Gavin B. Dalton, Graham J. Murray, Scott Smedley, Toshinori Maihara, Koji Ohta, Yuuki Moritani,

Kiyoto Yabe, Masanao Sumiyoshi, Hajime Sugai, Hiroshi Karoji, Shiang-Yu Wang, Youichi Ohyama, "Subaru FMOS now and future", Proc SPIE 8446-20, 2012.

- S.C. Ellis, M. Ireland, J.S. Lawrence, J. Tims, N. Staszak, J. Bland-Hawthorn, J. Brzeski, S. Case, M. Colless, S. Croom, W. Couch, O. De Marco, K. Glazebrook, Q.A Parker, R. Sharp, W. Saunders, R. Webster, D. Zucker, "KOALA: a wide-field 1000 element integral field unit for the Anglo-Australian Telescope" Proc SPIE 8446-29, 2012.

- Jurek K. Brzeski, Nicholas Staszak, Luke Gers, "Hermes: the engineering challenges", Proc SPIE 8446-30, 2012.

- Julia J. Bryant, Joss Bland-Hawthorn, Jon Lawrence, Scott Croom, Lisa Fogarty, Michael Goodwin, Samuel Richards, Tony Farrell, Stan Miziarski, Ron Heald, Heath Jones, Steve Lee, Matthew Colless, Michael Birchall, Andrew M. Hopkins, Sarah Brough, Amanda E. Bauer, "SAMI: a new multi-object IFU for the Anglo-Australian Telescope", Proc SPIE 8446-31, 2012.

  *Featured in issue 120, page 4.*

- George H. Jacoby, Antonin H. Bouchez, Matthew Colless, Darren L. DePoy, Daniel G. Fabricant, Philip M. Hinz, Daniel T. Jaffe, Matt Johns, Patrick J. McCarthy, Peter J. McGregor, Stephen A. Shectman, Andrew H. Szentgyorgyi, "The instrument development and selection process for the Giant Magellan Telescope", Proc SPIE 8446-50, 2012.

- Michael J. Irelanda, Stuart Barnes, David Cochrane, Matthew Colless, Peter Connor, Anthony Horton, Steve Gibson, Jon Lawrence, Sarah Martell, Peter McGregor, Tom Nicolle, Kathryn Nield, David Orr, J. Gordon Robertson, Stuart Ryder, Andrew Sheinis, Greg Smith, Nick Staszak, Julia Tims, Pascal Xavier, Peter Young and Jessica Zheng, "The AAOs Gemini high resolution optical spectrograph (GHOST) concept",Proc SPIE 8446-80, 2012.

  *Featured in this issue, page 7.*

- Samuel Richards, William Martin, David Campbell, Hugh Jones, Joss Bland-Hawthorn, Jon Lawrence, Elias Brinks, Julia J. Bryant, Lisa Fogarty, Mark Gallaway, Michael Goodwin, Sergio Leon-

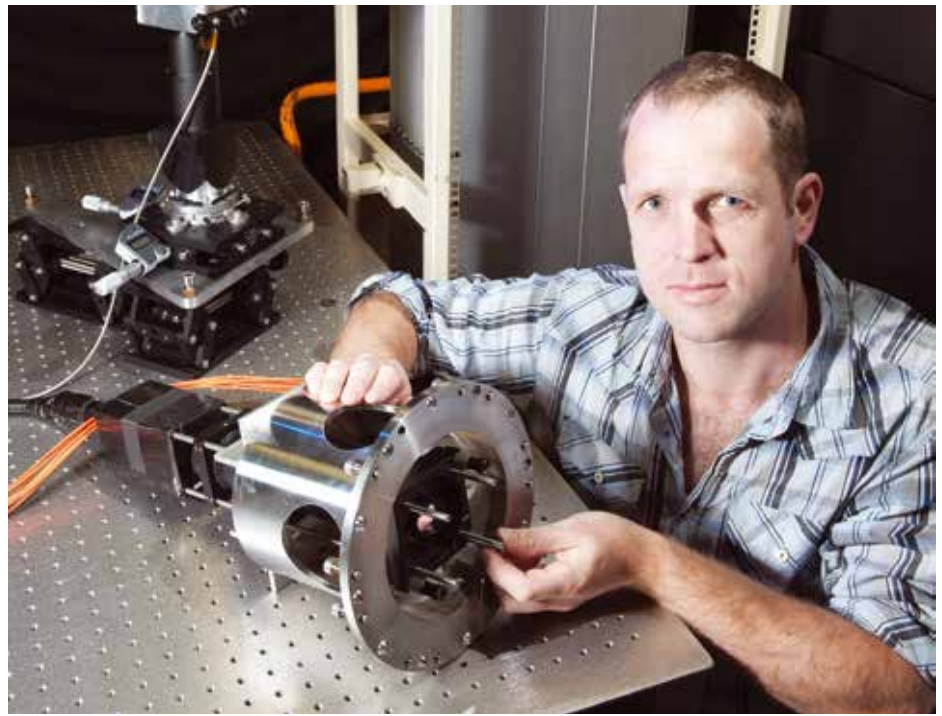

Simon Ellis with the GNOSIS OH-suppressing fibre feed for IRIS2, which was demonstrated on the AAT late last year.

Saval, Marc Sarzi, Daniel Smith, "BASIS: Bayfordbury single-object integral field spectrograph" ,Proc SPIE 8446-82, 2012.

- Michael Goodwin, Tony J. Farrell, Scott Smedley, Jeroen Heijmans, Ron Heald, Gayandhi De Silva, "The AAO fiber instrument data simulator", Proc SPIE 8446-84, 2012.

*The fiber instrument data simulator is an in-house software tool that simulates detector images of fiber-fed spectrographs developed by the Australian Astronomical Observatory (AAO) . In addition to helping validate the instrument designs, the resulting simulated images are used to develop the required data reduction software. Example applications that have benefited from the tool usage are the HERMES and SAMI instrumental projects for the Anglo-Australian Telescope (AAT). Given the sophistication of these projects an end-to-end data simulator that accurately models the predicted detector images is required. The data simulator encompasses all aspects of the transmission and optical aberrations of the light path: from the science object, through the atmosphere, telescope, fibres, spectrograph and finally the camera detectors. The simulator runs under a Linux environment that uses pre-calculated information derived from ZEMAX models and processed data from MATLAB. In this paper, we discuss the aspects of the model, software, example simulations and verification.*

- Corey M. Wood, Arthur D. Eigenbrot, Matthew A. Bershady, Scott Buckley, Michael P. Smith, Marsha J. Wolf, John S. Gallagher III, Eric J. Hooper, Andrew I. Sheinis, "HexPak and GradPak: variable-pitch dual-head IFUs for the WIYN 3.5m telescope bench spectrograph", Proc SPIE 8446-106, 2012.

- Anthony Horton, C.G. Tinney, Scott Case, Tony Farrell, Luke Gers, Damien Jones, Jon Lawrence, Stan Miziarski, Nick Staszak, David Orr, Minh Vuong, Lew Waller and Ross Zhelem, "CYCLOPS2: the fibre image slicer upgrade for the UCLES high resolution spectrograph", Proc SPIE 8446-121, 2012.

*CYCLOPS2 is an upgrade for the UCLES high resolution spectrograph on the Anglo-Australian Telescope, scheduled for commissioning in semester 2012A. By replacing the 5 mirror Coudé train with a Cassegrain mounted fibre based image slicer CYCLOPS2 simultaneously provides improved throughput, reduced aperture losses and increased spectral resolution. Sixteen optical fibres collect light from a 5 .0 arcsecond^2 area of sky and reformat it into the equivalent of a 0 .6 arcsecond wide slit, delivering a spectral resolution*

of R=70000 and up to twice as much flux as the standard 1 arcsecond slit of the Coudé train. CYCLOPS2 also adds support for simultaneous ThAr wavelength calibration via a dedicated fibre. [abridged]

- Nick Cvetojevic, Nemanja Jovanovic, Jon S. Lawrence, Michael J. Withford, and Joss Bland-Hawthorn, "Redesign of the Integrated Photonic Spectrograph for Improved Astronomical Performance", Proc SPIE 8446-130, 2012.

- C. Q. Trinh, S. C. Ellis, J. S. Lawrence, A. J. Horton, J. Bland-Hawthorn, S. G. Leon-Saval, J. Bryant, S. Case, M. Colless, W. Couch, K. Freeman, L. Gers, K. Glazebrook, R. Haynes, S. Lee, H.-G. Lohmannsroben, S. Miziarski, J. O'Byrne, W. Rambold, M. M. Roth, B. Schmidt, K. Shortridge, S. Smedley, C. G. Tinney, P. Xavier, J. Zheng, "GNOSIS: a new near-infrared OH suppression unit at the AAT", Proc SPIE 8446-131, 2012.

- Will Saunders, Greg A. Smith, Rolf Muller, Jurek K. Brzeski, Lewis G. Waller, Stan Miziarski, Tony J. Farrell, James Gilbert, "Mohawk : a 4000-fiber positioner for DESpec", Proc SPIE 8446-188, 2012.

- Jon Lawrence, Joss Bland-Hawthorn, Julia Bryant, Jurek Brzeski, Matthew Colless, Scott Croom, Luke Gers, James Gilbert, Peter Gillingham, Michael Goodwin, Jeroen Heijmans, Anthony Horton, Mike Ireland, Stan Miziarski, Will Saunders, Greg Smith, "Hector: a high-multiplex survey instrument for spatially resolved galaxy spectroscopy", Proc SPIE 8446-195, 2012.

- Will Saunders, "A fast new cadioptric design for fiber-fed spectrographs", Proc SPIE 8446-196, 2012.

- Jennifer L. Marshall, Stephen M. Kent, H. Thomas Diehl, Brenna L. Flaugher, Joshua Frieman, Darren L. DePoy, Will Saunders, Matthew Colless, Ofer Lahav, Filipe Abdalla, Stephanie Jouvel, Donnacha Kirk, Huan Lin, James Annis, "The dark energy spectrometer: a proposed multi-fiber instrument for the CTIO Blanco 4-meter Telescope", Proc SPIE 8446-198, 2012.

- Jeroen Heijmans, Martin Asplund, Sam Barden, Michael Birchall, Daniela Carollo, Joss Bland Hawthorn, Jurek Brzeski, Scott Case, Vladimir Churilov, Matthew Colless, Robert Dean, Gayandhi De Silva, Tony J. Farrell, Kristin Fiegert, Ken Freeman, Luke Gers, Michael Goodwin, Doug Gray, Ron Heald, Anthony Heng, Damien Jones, Chiaki Kobayashi, Urs Klauser, Yuriy Kondrat, Jon Lawrence, Steve Lee, Darren Mathews, Don Mayfield, Stan Miziarski, Guy Monnet, Rolf Muller, Naveen Pai, Robert Patterson, Ed Penny, David Orr, Andrew Sheinis, Keith Shortridge, Scott Smedley, Greg Smith, Darren Stafford, Nicholas Staszak, Minh Vuong, Lewis Waller, Denis Whittard, Elizabeth Wylie de Boer, Pascal Xavier, Jessica Zheng, Ross Zhelem, Daniel Zucker, "Integrating the HERMES spectrograph for the AAT", Proc SPIE 8446-213, 2012.

- Robert Content, Tom Shanks, Ray M. Sharples, David G. Bramall, Will Percival, "VXMS: the VISTA extreme multiplex spectrograph", Proc SPIE 8446-233, 2012.

- Peter Gillingham, "A 3 degree prime focus field for the AAT", Proc SPIE 8446-244, 2012.

- Michael Goodwin, Jurek Brzeski, Scott Case, Matthew Colless, Tony Farrell, Luke Gers, James Gilbert, Jeroen Heijmans, Andrew Hopkins, Jon Lawrence, Stan Miziarski, Guy Monnet, Rolf Muller, Will Saunders, Greg Smith, Julia Tims, Lew Waller, "MANIFEST instrument concept and related technologies", Proc SPIE 8446-289, 2012.

- Michael J. Ireland, "Aperture masking behind AO systems", Proc SPIE 8447-79, 2012.

- Theo A. ten Brummelaar, Stephen T. Ridgway, John D. Monnier, Michael J. Ireland, Harold A. McAlister, Laszlo Sturmann, Judit Sturmann, Nils H. Turner, Jean C. Shelton, Peter G. Tuthill, "Adaptive optics for the CHARA array", Proc SPIE 8447-127, 2012.

- J. Zheng, W. Saunders, J. S. Lawrence, F. Bastien, and F. Cantalloube, "Laboratory demonstration of a liquid atmospheric dispersion corrector", Proc SPIE 8450-14, 2012.

- Izabela Spaleniak, Nemanja Jovanovic, Simon Gross, Michael Ireland, Jon Lawrence, and Michael Withford, "Exploration of integrated photonic lanterns fabricated by femtosecond laser inscription", Proc SPIE 8450-40, 2012.

- Stan Miziarski, Jurek K. Brzeski, Joss Bland-Hawthorn, Michael Goodwin, Jeroen Heijmans, Anthony J. Horton, Jon S. Lawrence, Will Saunders, Greg A. Smith, Nicholas Staszak, James Gilbert, "Concepts for multi-IFU Robotic Positioning Systems", Proc SPIE 8450-42, 2012.

- James Gilbert, Jeroen Heijmans, Michael Goodwin, Rolf Muller, Stan Miziarski, Jurek K. Brzeski, Lewis G. Waller, Will Saunders, Alex Bennet, Julia Tims, "Starbugs: all-singing, all-dancing fibre positioning robots", Proc SPIE 8450-44, 2012.

- Simon C. Ellis, Joss Bland-Hawthorn, Jon S. Lawrence, Antoine Crouzier, "Potential applications of ring resonators for astronomical instrumentation", Proc SPIE 8450-53, 2012.

- Robert Content, Simon Blake, Colin Dunlop, Ray Sharples, Gordon Talbot, David Nandi, Tom Shanks, Danny Donoghue, Nikolaos Galiatsatos, Peter Luke, "CEOI microslice spectrograph", Proc SPIE 8450-59, 2012.

- Anthony J. Horton, Simon C. Ellis, Jon S. Lawrence, Australian Joss Bland-Hawthorn, "PRAXIS: a low background NIR spectrograph for fibre Bragg grating OH suppression", Proc SPIE 8450-65, 2012.

- Seong-sik Min, Christopher Trinh, Sergio Leon-Saval, Nemanja Jovanovic, Peter Gillingham, Joss Bland-Hawthorn, Jon Lawrence, Tim A. Birks, Martin M. Roth, Roger Haynes, Lisa Fogarty, "Multi-core fiber Bragg grating developments for OH suppression", Proc SPIE 8450-131, 2012.

- Marsha J. Wolf, Donald J. Thielman, Michael P. Smith, Kurt P. Jaehnig, Andrew I. Sheinis, Gregory Mosby, "Performance characterization of the near infrared detector system for RSS-NIR on SALT", Proc SPIE 8453-79, 2012.

# The AAO's Gemini High-resolution Optical SpecTrograph (GHOST) concept

Michael J. Ireland (AAO, Macquarie University), Matthew Colless, Anthony Horton, Jon Lawrence, Sarah Martell, David Orr, Stuart Ryder, Andrew Sheinis, Greg Smith, Nick Staszak, Julia Tims, Pascal Xavier, Jessica Zheng (AAO), J. Gordon Robertson (AAO, University of Sydney), Stuart Barnes (Stuart Barnes Optical Design), David Cochrane, Peter Connor, Steve Gibson, Tom Nicolle, Kathryn Nield (Industrial Research Ltd.), Peter McGregor, Peter Young (Research School of Astronomy & Astrophysics).

The Gemini consortium have prioritized the development of a new high-resolution optical spectrograph for the Gemini telescope(s). Gemini's design is problematic for high-resolution spectroscopy, because a large instrument has to either maintain adequate stability over a wide range of gravity vectors, or be placed a distance of tens of meters from the focus via a fiber cable. In late 2011, the Gemini Observatory awarded three competitive conceptual design studies for the Gemini High-Resolution Optical Spectrograph (GHOS) instrument. The AAO in partnership with the Australian National University (ANU) conducted one of these studies. Our concept, opting for the more easily pronounced "Gemini High-Resolution Optical SpecTrograph" (GHOST), is based on a fiber-fed design incorporating a Kiwispec spectrograph.

As a workhorse instrument, GHOST is expected to meet a wide range of science goals. Gemini provided 14 science cases derived from earlier White Papers, mostly relating to abundance determinations in various astrophysical contexts. As part of the bid process, the AAO added 3 additional science cases: Exoplanet detection via radial velocity, the study of time variation of fundamental constants through redshifted quasar absorption lines, and absorption spectroscopy of Gamma Ray Bursts (GRBs) with Rapid Target of Opportunity observations at Gemini. Approximately half the science cases would benefit from a a multi-object capability.

The science requirements resulting from these science cases are given in Table 1.

The GHOST baseline instrument concept comprises two integral field units (IFUs) with individual Atmospheric Dispersion Compensators (ADCs) that are positionable at the Cassegrain focal plane, feeding a fiber bundle that transports light to a gravity-invariant thermally-controlled spectrograph mounted on the rotating carousel. Key design features for GHOST are as follows:

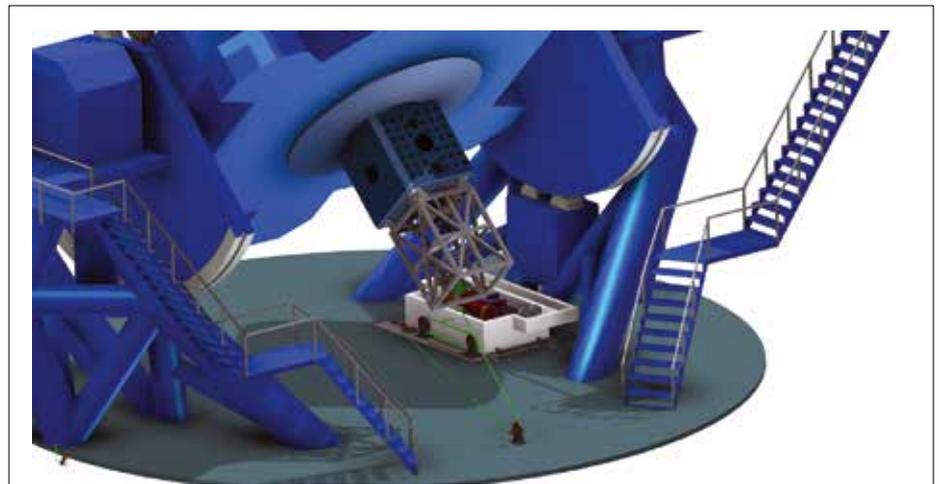

**Figure 1.** The AAO-GHOST spectrograph in its preferred location on the carousel, showing the fiber cable when the telescope is 45 degrees from zenith. The spectrograph is shown in its thermal enclosure but with its top surfaces removed.

| Title | Requirement |
|---|---|
| Wavelength range | GHOST shall provide simultaneous wavelength coverage from 363nm to 1000nm |
| Spectral resolution | GHOST shall have two selectable spectral resolution modes: standard resolution mode with R>50,000 and high resolution mode with R>75,000. |
| Sensitivity | GHOST shall obtain a sensitivity of m=18.0 in a 1 hour observation for 30 sigma per resolution element in standard resolution mode in dark time (50th sky brightness percentile) at a wavelength of 500 nm. |
| Targets and field size | GHOST shall have the capability to observe 2 targets simultaneously over a 7.5 arcmin diameter field of view. |
| Radial velocity precision | GHOST shall provide a radial velocity precision of 200 m/s over the full wavelength range in standard resolution mode and shall have the capability to provide a radial velocity precision of 2 m/s over the full wavelength range for the high spectral resolution mode. |
| Spatial Sampling | GHOST shall spatially sample each target object over a field size of 1.2 arcsec. |
| Spectro-polarimetry | GHOST should provide a spectropolarimetric capability that can distinguish all Stokes parameters. |

**Table 1.** The science requirements for GHOST.

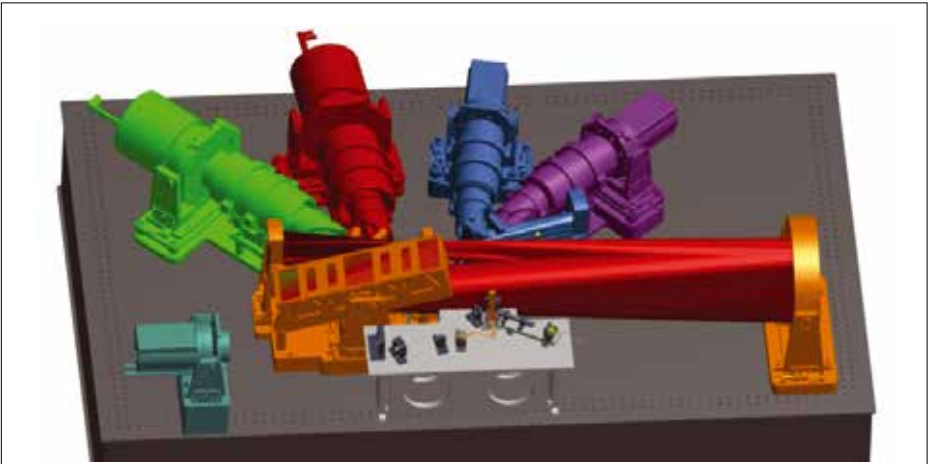

**Figure 2.** The opto-mechanical concept for the spectrograph, showing the common optical elements to all 4 arms (orange) as well as the Green, Red, Blue and UV (purple) arms. The slit assembly optics are on their own small optical table in the lower part of the figure. The pupil diameters are 100mm on the grating and 33mm on the VPH grisms.

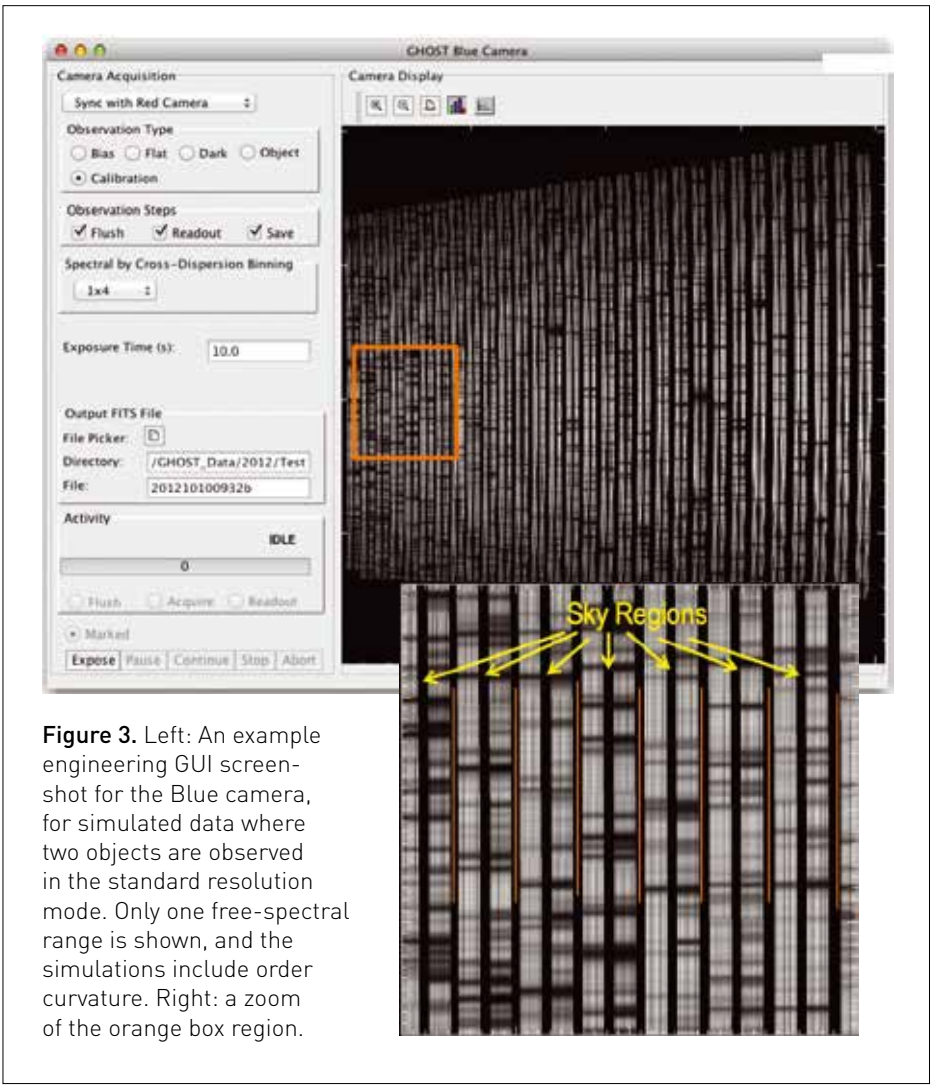

**Figure 3.** Left: An example engineering GUI screen-shot for the Blue camera, for simulated data where two objects are observed in the standard resolution mode. Only one free-spectral range is shown, and the simulations include order curvature. Right: a zoom of the orange box region.

- Broad Simultaneous Wavelength Coverage. A simultaneous wavelength coverage of 363 to 1000nm.

- Focal-plane image-slicing. GHOST finely samples the stellar PSF at the focal plane over a 1.2" diameter field of view with 7 elements (standard-resolution mode) or 19 elements (high-resolution mode) and reformats this into a spectrograph slit.

- Microlens-based IFUs. GHOST uses a pair of microlens arrays at the focal plane to inject light into the fibers. This approach gives added versatility in spectrograph and input optics design, and has high-efficiency (>90% of a fiber-only design).

- Miniature-ADCs. A dedicated Atmospheric Dispersion Compensator (ADC) is provided for each IFU.

- Dual slits. Two separate slits are fed from IFUs with a different scale of image sampling. This provides two separate spectral resolution modes: "standard" and "high".

- Multi-object positioning system. Two Commercial Off The Shelf (COTS) xy stages allow positioning of two IFUs within the Cassegrain field of view, providing a substantial gain in observing efficiency.

- Broadspectrum optical fibers. Broad bandwidth fibers with good blue throughput (>80% measured) and focal ratio degradation properties are used for the fiber bundle.

- Asymmetric white-pupil echelle spectrograph. The spectrograph is a variant on the Kiwispec 100mm pupil spectrograph with an R4 grating which is a relatively compact and tested design using a single catalog echelle grating.

- Four-arm spectrograph design. The spectrograph has 4 arms with a common collimator and main

dispersing element. This design maximizes throughput, enables thermo-electric cooling of relatively small (2K×2K) detectors in 2 of the 4 arms, and provides wide simultaneous wavelength coverage.

- Volume-phase holographic (VPH) grism cross-dispersers. The use of VPH grisms for the spectrograph cross-dispersion results in a relatively high throughput, especially given the relatively narrow wavelength range of each arm.

- Slit-viewing camera. A dedicated slit-viewing camera views a 1% reflection of the spectrograph input and a 100% output from a series of guide fibers. This obviates the necessity for an on-instrument wavefront sensor, and provides an exposure meter and spectral Point-Spread Function (PSF) monitor.

- Thermally controlled chamber. The spectrograph is enclosed inside a thermally-controlled chamber. This provides high spectral stability and maintenance/adjustment-free operation.

- High radial velocity precision mode. Simultaneous wavelength calibration, pressure calibration, and fiber agitation provide a very high precision radial velocity mode, with only minor impact to cost and complexity.

- Spectropolarimetry mode. A fixed module is proposed as an upscope option that provides polarimetric information.

The concept has many similarities to the VELOCE concept for AAT (presented by C. Tinney at the last future AAT instrumentation workshop), and builds on the microlens designs for KOALA and MANIFEST, plus the simultaneous Th/Xe reference source and fiber agitator for CYCLOPS2. Should this concept be selected by Gemini, commissioning will be anticipated in 2015. ✦AAO

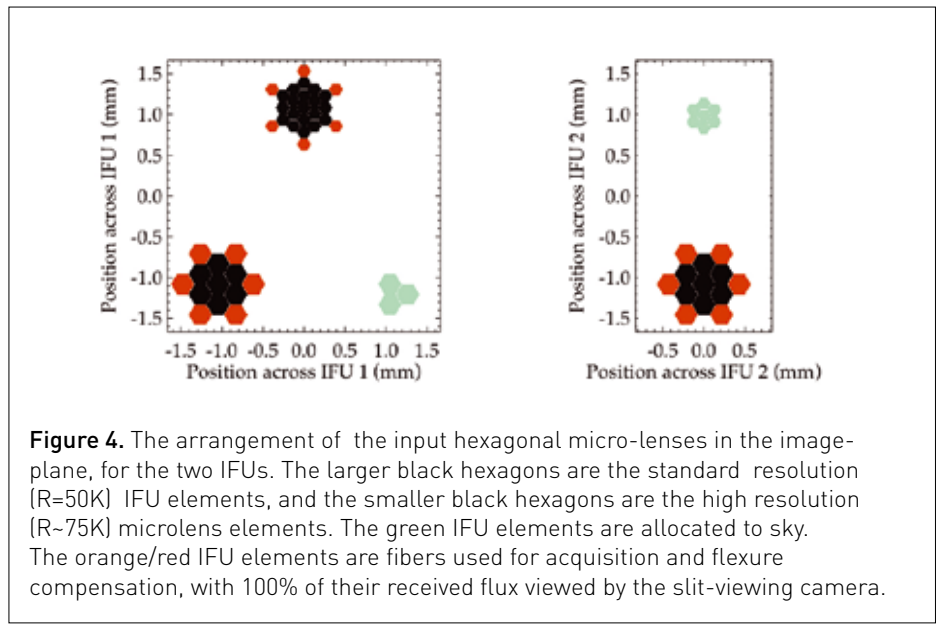

**Figure 4.** The arrangement of the input hexagonal micro-lenses in the image-plane, for the two IFUs. The larger black hexagons are the standard resolution (R=50K) IFU elements, and the smaller black hexagons are the high resolution (R~75K) microlens elements. The green IFU elements are allocated to sky. The orange/red IFU elements are fibers used for acquisition and flexure compensation, with 100% of their received flux viewed by the slit-viewing camera.

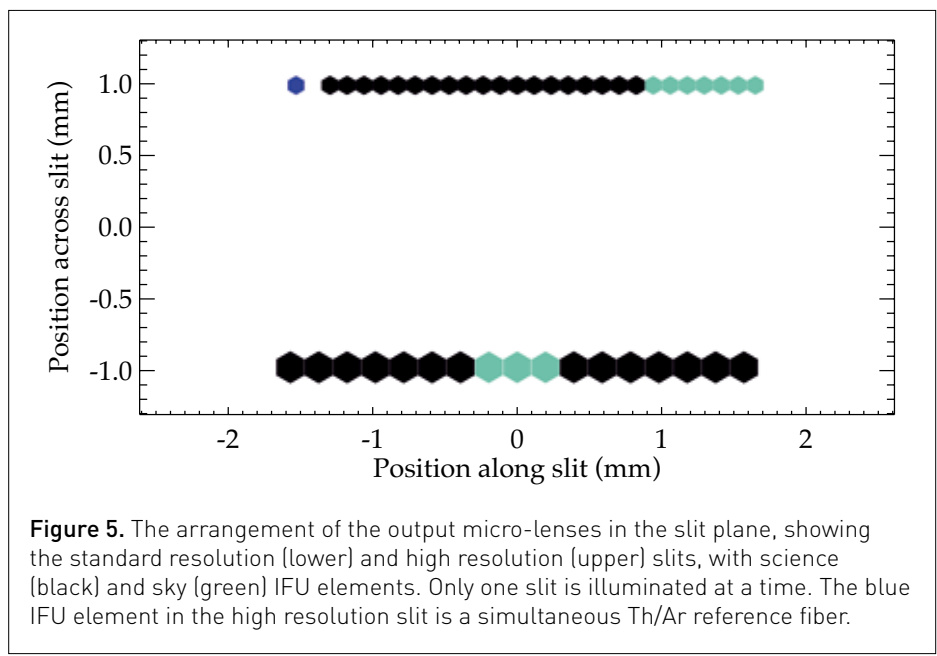

**Figure 5.** The arrangement of the output micro-lenses in the slit plane, showing the standard resolution (lower) and high resolution (upper) slits, with science (black) and sky (green) IFU elements. Only one slit is illuminated at a time. The blue IFU element in the high resolution slit is a simultaneous Th/Ar reference fiber.

# First Cosmological Constraints from the 6dFGS Peculiar Velocity Field

Christina Magoulas (AAO and University of Melbourne), Christopher Springob (AAO), Matthew Colless (AAO), Jeremy Mould (Swinburne) and the 6dFGS team

The 6-degree Field Galaxy Survey (6dFGS) is an all-southern sky spectroscopic survey that was carried out over six years, from 2001 to 2006, on the UK Schmidt Telescope [1]. It was devised to extend existing knowledge of the local galaxy population by combining measurements of both redshifts and peculiar velocities [2]. The primary redshift survey (6dFGSz) provides more than 125,000 galaxy redshifts in the final data release [3]. The wide sky coverage of 6dFGS galaxies maps the large scale galaxy structures in detail across the southern sky.

The 'cosmic web' of large-scale structure is formed through the gravitational collapse of density fluctuations in the early-universe. The over-densities in the distribution of matter induce peculiar motions that arise from the gravitational effects of neighbouring galaxies, motions that are in excess of the Hubble flow associated with the expansion of the universe. The redshift of a galaxy, cz, recovers its total motion, including the radial components from both the Hubble flow and its peculiar velocity. Disentangling the peculiar velocity of a galaxy from its recessional velocity requires the measurement of a distance in addition to the redshift.

Peculiar velocity studies allow us to determine these motions directly, providing an unbiased tracer of the underlying distribution of matter that is sensitive to mass fluctuations on the largest scales. They are also an independent probe of cosmology in the low redshift universe that can significantly improve constraints (and in some cases, remove degeneracies) on cosmological parameters that define models of large-scale structure formation.

The 6dFGS peculiar velocity survey (6dFGSv) comprises the brightest early-type galaxies in the primary redshift sample out to a redshift of cz < 16,500 km s$^{-1}$. Distances and velocities of the 6dFGSv sample are measured using the Fundamental Plane (FP) relation - a tight correlation between the properties of early-type galaxies [4,5]. The best-fit FP has been accurately determined for ~9000 6dFGS galaxies in near-infrared passbands [6]. The 6dFGSv sample therefore forms the basis of the largest and most far-reaching peculiar velocity sample to date. In this article we describe the derivation of the 6dFGSv peculiar velocities and the bulk galaxy motions we measure from the 6dFGS velocity field.

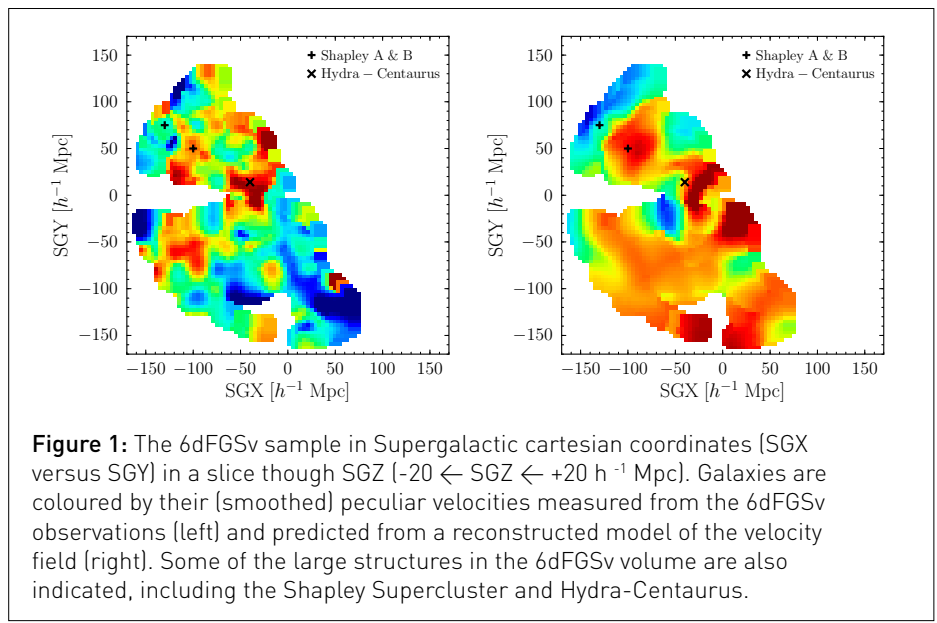

**Figure 1:** The 6dFGSv sample in Supergalactic cartesian coordinates (SGX versus SGY) in a slice though SGZ (-20 ← SGZ ← +20 h$^{-1}$ Mpc). Galaxies are coloured by their (smoothed) peculiar velocities measured from the 6dFGSv observations (left) and predicted from a reconstructed model of the velocity field (right). Some of the large structures in the 6dFGSv volume are also indicated, including the Shapley Supercluster and Hydra-Centaurus.

## Bayesian Peculiar Velocities

A robust Bayesian procedure was developed to model the 6dFGSv peculiar velocity field using a maximum likelihood approach and incorporating a 3D Gaussian model to fit the FP [7]. Not only does this Bayesian approach complement the 6dFGS FP analysis, but it is also an innovation with regards to previous work which simply treats velocity measurements as a single value with error bars. Instead we obtain the full Bayesian posterior probability distribution for each galaxy's peculiar velocity and apply a maximum likelihood method to determine the parameters defining the velocity field (such as the redshift space distortion parameter and the bulk flow, described below) [8]. This model was found to perform well under the demands of fitting to a variety of mock galaxy realisations designed to emulate the selection effects and distribution of the real 6dFGSv data.

In Fig. 1, we show the Supergalactic cartesian coordinates (SGX versus SGY) of the 6dFGSv galaxies in a slice though SGZ. The galaxies (at left) are coloured by their (smoothed) peculiar velocities measured from the 6dFGSv observations, which are in good agreement with those predicted from a reconstructed model of the velocity field (at right). A dipole motion is evident in the left panel of Fig. 1 as a prominence of positive peculiar velocities (in red) near the centre-left and more negative velocities (in blue) in the bottom-right corner.

## The redshift-space distortion parameter

Peculiar velocities distort the galaxy distribution in redshift-space relative to the distribution in real-space, affecting the determination of distance estimates. On large scales the clustering of matter is amplified by the coherent infall of galaxies towards over-dense regions. The effects of this type of distortion can be characterised by the linear redshift-space distortion parameter (β).

We can derive β from the comparison of the velocity field from the observed 6dFGS peculiar velocities and that reconstructed from the density field revealed by the redshift survey. The reconstructed velocity field we utilise is derived from the Two-Micron All-Sky Redshift Survey (2MRS) [9], which is largely based on the 6dFGSz. We measure a best-fit value for the redshift distortion parameter of $\beta$ = 0.29 ± 0.06. This is slightly lower than found by other recent studies, although consistent at the 1.5$\sigma$ level [10,11]. Discrepancies between $\beta$ values at this level may result from their sensitivity to the assumed value of linear bias between the mass and galaxy distributions, or from details such as the sample selection limits, or differences in fitting methods.

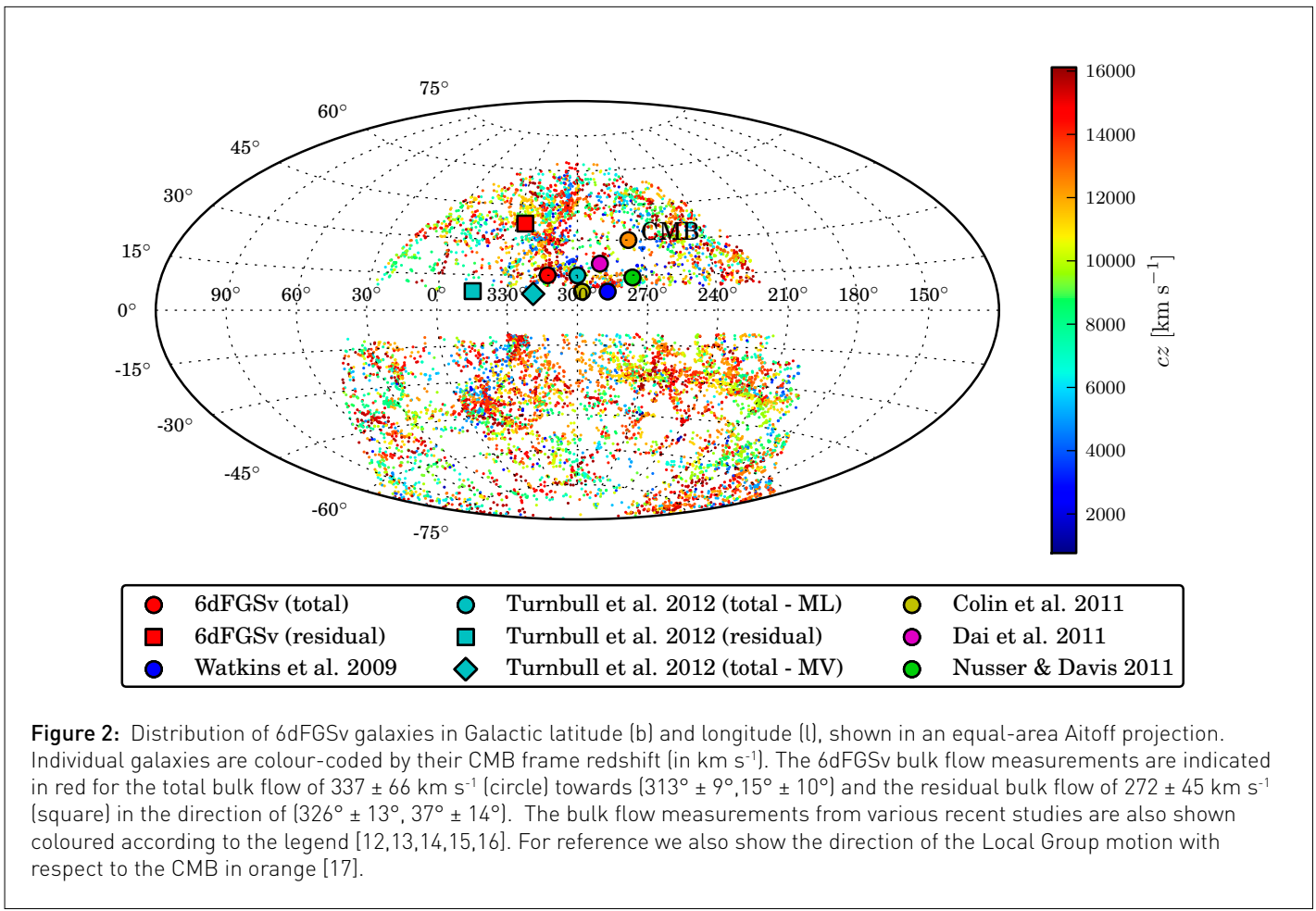

Figure 2: Distribution of 6dFGSv galaxies in Galactic latitude (b) and longitude (l), shown in an equal-area Aitoff projection. Individual galaxies are colour-coded by their CMB frame redshift (in km s$^{-1}$). The 6dFGSv bulk flow measurements are indicated in red for the total bulk flow of 337 ± 66 km s$^{-1}$ (circle) towards (313° ± 9°,15° ± 10°) and the residual bulk flow of 272 ± 45 km s$^{-1}$ (square) in the direction of (326° ± 13°, 37° ± 14°). The bulk flow measurements from various recent studies are also shown coloured according to the legend [12,13,14,15,16]. For reference we also show the direction of the Local Group motion with respect to the CMB in orange [17].

### The 6dFGS bulk flow motion

Peculiar velocities are also used to study the coherent peculiar motion in a volume, or bulk flow, with respect to the CMB rest frame. From the peculiar velocity field derived from the 6dFGSv, we have determined both the total bulk flow motion and the residual bulk flow after subtracting the 2MRS predicted velocity field.

Fig. 2 shows an all-sky projection of the 6dFGSv sample, colour-coded by the CMB frame redshift of each galaxy. The large coloured circles or squares represent the direction of the bulk flow estimates for 6dFGSv (in red) and other recent studies as indicated by the legend. The direction of the bulk flow derived in these studies is, in general, located in a narrow 10°-wide strip parallel to the Zone of Avoidance. The bulk motion from deeper surveys, such as 6dFGSv, tend to point towards the most massive structures probed by the survey volume, in our case in the direction l = 313° ± 9° and b = 15° ± 10° towards the Shapley supercluster.

We compare the amplitude of the 6dFGSv total bulk flow (337 ± 66 km s$^{-1}$) to other measurements at different scales in Fig. 3. The bulk flow measurements from previous studies are shown at the effective scale of the samples from which they were derived (approximately the limiting radius of the sample). We also plot the theoretical prediction of the rms bulk flow calculated in a flat $\Lambda$CDM model defined by the WMAP7 results [12]. In general, the measured bulk motions of most studies shown in Fig. 3 are within the 90% range of theoretical expectations and are consistent with the expected trend of decreasing bulk flow amplitude at large radii.

If we account for the fact that our survey volume is limited to the southern hemisphere we find the 6dFGSv total bulk flow is not consistent with the predictions of $\Lambda$CDM at greater than 2$\sigma$. The residual bulk flow of 6dFGSv after subtracting the 2MRS prediction is 273 ± 45 km s$^{-1}$ accounting for most of this motion and is in the direction of l = 326° ± 13° and b = 37° ± 14°, very close to Shapley. The large amplitude of the residual dipole suggests that the bulk flow motion of 6dFGSv is dominated by mass distributions unaccounted for by the 2MRS volume or that the contribution of the Shapley supercluster is underrepresented in the 2MRS reconstruction.

### Cosmological Implications of a Large Bulk Flow

Recently, there have been conflicting measurements of the bulk flow in the local volume, some of which are in excess of the predictions of $\Lambda$CDM [12,14,16]. If the bulk flow of the local volume differs significantly from the theoretical expectations of $\Lambda$CDM cosmology and the latest WMAP7 constraints, this could indicate a departure from standard cosmological models.

An alternative explanation for the divergence in the bulk flow is that the interpretation of the origin of the CMB dipole motion may be incorrect. In a 'tilted universe' model, some fraction of the CMB dipole is due to fluctuations from the pre-inflationary universe, rather than the Local Group's motion with respect to the background. This results in an intrinsic bulk flow that does not converge and continues beyond the local volume [19]. A more extreme proposal suggests that the CMB dipole is partially induced by differential expansion histories in different directions on the sky, undermining the assumption of a CMB rest frame [20]. Further observation and theoretical investigation is required to establish the validity of both these models.

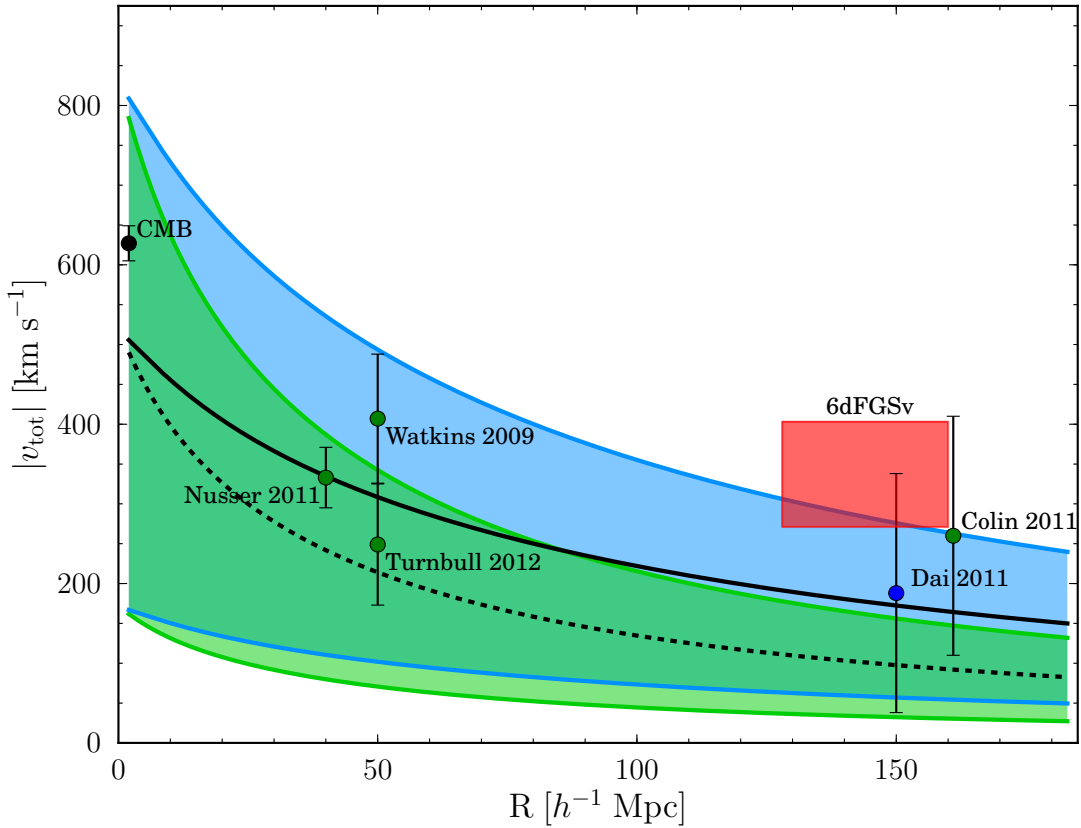

**Figure 3:** The bulk flow amplitude as a function of scale. The 6dFGSv bulk flow measurement in indicated by a red rectangle whose height is the ±1σ uncertainty in the fitted motion and whose width runs from the outer limit of the survey at 161 $h^{-1}$ Mpc to 129 $h^{-1}$ Mpc, which is the radius of a sphere having the same volume as the hemispherical 6dFGSv survey. The predicted rms bulk flow in a flat ΛCDM model ($\Omega_m$ = 0.274, $\Omega$ = 0.704 and $\sigma_8$ = 0.811) is shown as the solid black line for a top-hat window function (90% scatter from cosmic variance shaded in light blue) and as the dashed black line for a Gaussian window function (90% scatter from cosmic variance shaded in light green). Bulk flow measurements from recent studies are coloured according to the most appropriate window function (blue for top-hat, green for Gaussian).

## Summary and Future Work

The 6dFGSv provides the largest homogeneous peculiar velocity sample to date, leading to improved measurements of the motions in the local universe due to its wide coverage.

Using 6dFGSv, we map the velocity field in the local universe and compare to the density field derived from redshift surveys. This leads to new constraints on the redshift distortion parameter β combining mass density and galaxy bias as well as the cosmological bulk flow in the 6dFGSv volume. The amplitude of the 6dFGSv bulk flow (337 ± 66 km s⁻¹) is high compared to other measurements and to the theoretical prediction of a ΛCDM model, leaving the origin of this large bulk motion a open question.

Substantial improvement in the precision of bulk flow measurements and cosmographic description of the nearby universe are gained from such densely distributed and deep surveys such as 6dFGSv. We present here only the preliminary analysis of the 6dFGSv velocity field, with further detailed investigations still to come. The 6dFGSv has a wide range of cosmological applications and we expect further analysis of the survey will reveal new insights in the near future.

# Galaxy And Mass Assembly (GAMA): GAMA Announces Second Data Release

Andrew Hopkins (AAO) for the GAMA collaboration

## Introduction

The Galaxy And Mass Assembly (GAMA) survey is a multiwavelength photometric (ultraviolet, optical, infrared and radio) and (optical) spectroscopic survey aimed at producing the most comprehensive database of galaxies and their properties over the past third of the age of the Universe (Driver et al., 2009, 2011). To date GAMA has measured over 220000 spectra, which in combination with existing spectra from SDSS, 2dFGRS, the MGC and other surveys totals almost 300000 individual spectra in GAMA's three equatorial and two southern fields. Of these, about 230000 are unique galaxy targets, and of those 215000 (92%) have high quality redshift measurements. The goal of the survey is to measure about 300000 unique galaxy targets over the 280 square degrees comprised by the five survey fields. GAMA aspires to address fundamental questions in cosmology related to the nature of gravity and dark energy, along with providing the best measurements of all aspects of galaxy evolution, giving us the most complete picture yet possible of how our Universe works.

The second public data release from the GAMA survey (GAMA DR2) will be announced in October this year. Access to the public GAMA data is available from the GAMA website http://www.gama-survey.org/. This article provides an update on the GAMA survey, with highlights from recent scientific results.

## GAMA Survey and Data Releases

The GAMA team began observations with the AAT in 2008, with 68 nights of AAT time awarded from 2008-2010. This initial phase of the survey, usually referred to as "GAMA I," allowed the acquisition of over 112000 new galaxy spectra and redshifts, for a total of over 130000 redshifts in the original GAMA survey area. Subsequently the survey has been extended, with the award of 110 nights of AAT time over 2010–2012 (referred to as "GAMA II"), to allow the inclusion of two southern fields, broaden the three equatorial fields, and achieve a uniform depth of $r_{pet}$ < 19.8 mag over the full survey region.

The first public data release of 59000 GAMA spectra occurred on 25 June 2010, announced and demonstrated at the AAO

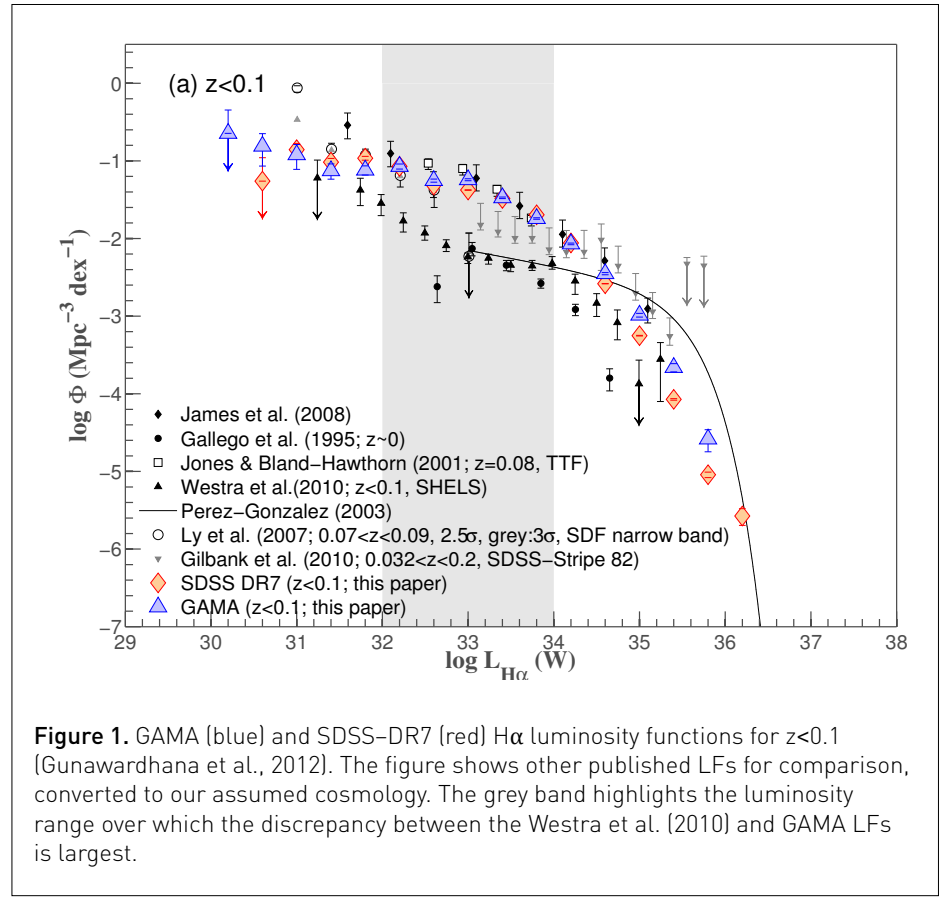

**Figure 1.** GAMA (blue) and SDSS–DR7 (red) H$\alpha$ luminosity functions for z<0.1 (Gunawardhana et al., 2012). The figure shows other published LFs for comparison, converted to our assumed cosmology. The grey band highlights the luminosity range over which the discrepancy between the Westra et al. (2010) and GAMA LFs is largest.

Symposium in Coonabarabran (see AAO Observer issue 118, Aug. 2010). Since that time, the GAMA dataset has more than doubled in size, and the value-added measurements of the spectra, the galaxy multiwavelength photometry and quantifying the structure of the galaxy distribution have all progressed substantially (some early results were highlighted in AAO Observer issue 119, Feb. 2011). The GAMA second data release will include not only the flux-calibrated spectra and redshifts associated with the targets released in GAMA DR1 (the originally released spectra were not flux-calibrated), but will include those for fainter targets in the GAMA G15 field (to rpet < 19.4 mag) as well. For all these targets DR2 will also include the reprocessed GAMA photometry (Hill et al., 2011), Sersic index fits (Kelvin et al., 2012), stellar mass estimates (Taylor et al., 2011), redshifts corrected for local bulk flow effects (Baldry et al., 2012), GAMA Group Catalogue associations (Robotham et al., 2011), emission line measurements (Hopkins et al., 2012), derived properties such as obscuration and aperture corrected H$\alpha$ luminosities and star formation rates (Gunawardhana et al., 2012), and more.

## GAMA Science Highlights

Since the last GAMA update in the AAO Observer, there have been many exciting results from the survey. Here we take the opportunity to showcase a selection of these. We note that all the work described here has been done using the GAMA I dataset only. The full list of GAMA publications is available from the GAMA survey web site: http://www.gama-survey.org/pubs/.

The H$\alpha$ luminosity function (LF) and its evolution over 0<z<0.35 have been explored by Gunawardhana et al., (2012). The depth and area of the GAMA observations allow us to probe the low redshift LF incredibly broadly. At z<0.1 the H$\alpha$ LF spans six orders of magnitude in H$\alpha$ luminosity, and encompasses both the faintest and the brightest H$\alpha$ emitters (Figure 1). Two key results come out of this analysis. First is that the shape of the Hα LF at the bright end is not well-matched to the traditional Schechter function, but drops off more slowly, as a Gaussian rather than an exponential. We find that the Saunders functional form used to fit far-infrared and radio luminosity functions for star forming galaxies provides a much better model, and for the first time show that a consistent functional form can be used to explain the shape of the star-formation-sensitive luminosities at each of these wavelengths. Second, we highlight the shortcomings of a broad-band magnitude selection for a spectroscopic sample. Not being able to target galaxies fainter than the magnitude limit leads to a sample that unavoidably misses some fraction of nonetheless very bright line emitters. The impact of this selection effect is outlined by Gunawardhana et al., (2012), and will be quantified in more detail in a forthcoming analysis of the bivariate luminosity function (Gunawardhana et al., in prep.).

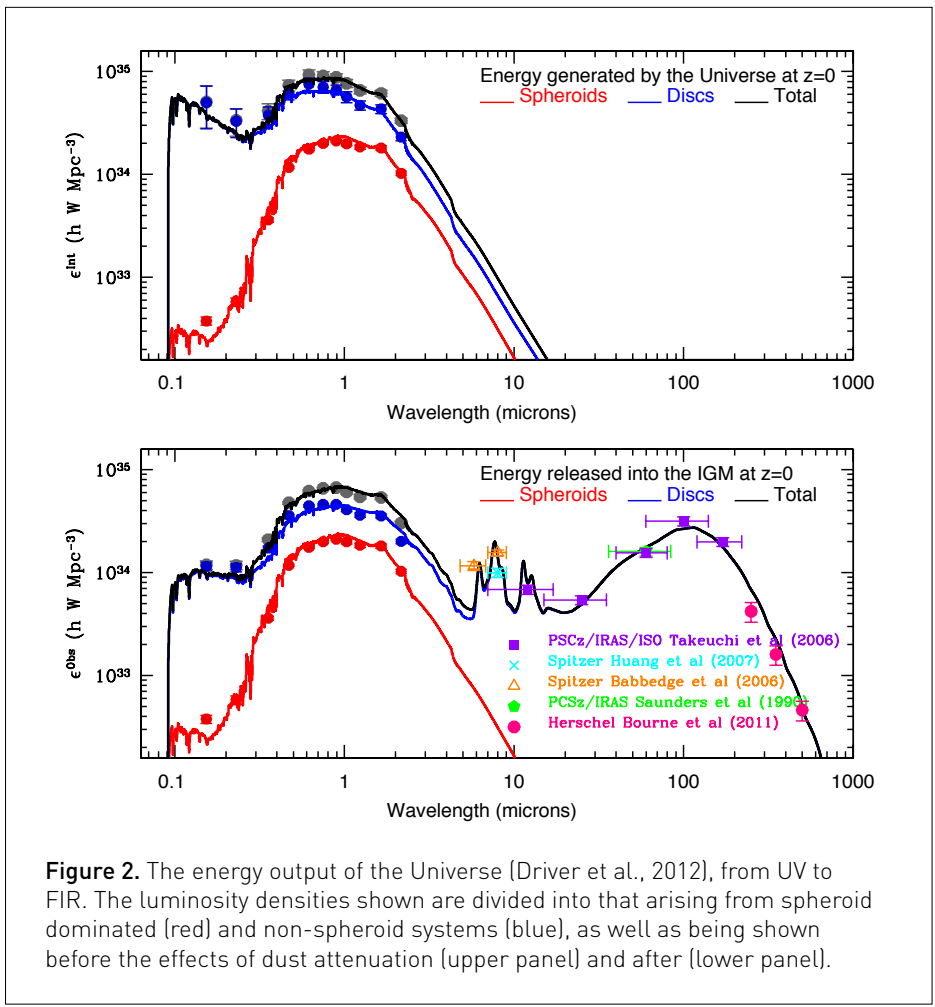

**Figure 2.** The energy output of the Universe (Driver et al., 2012), from UV to FIR. The luminosity densities shown are divided into that arising from spheroid dominated (red) and non-spheroid systems (blue), as well as being shown before the effects of dust attenuation (upper panel) and after (lower panel).

The cosmic spectral energy distribution has been analysed by Driver et al., (2012), who show the self-consistency of the energy balance between dust absorbing emission at ultraviolet wavelengths and re-radiating it in the mid- to far-infrared (Figure 2). The attenuation model of Driver et al., (2008) allows the prediction of the mid- to far-infrared luminosity densities based on the observed UV to optical luminosity densities. This agrees remarkably well with the measurements. After accounting for dust attenuation, the total luminosity density of the universe at the current epoch is $1.8\pm0.3 \times 10^{35}$ W h Mpc$^{-3}$, generated almost exclusively by starlight. About two thirds of this energy is directly released into the intergalactic medium, and one third is absorbed by dust and re-radiated in the mid- to far-infrared.

The star formation histories of galaxies have been explored by Bauer et al., (2012), using the evolution of galaxy specific star formation rates (star formation rate per unit stellar mass). We find clear evidence that low-mass galaxies ($M^* < 10^9$ M$_\odot$) cannot have formed with smooth exponential star formation histories, as higher mass galaxies may do, but rather require more intermittent bursty episodes of star formation (Figure 3). This Figure shows the location of the Milky Way (MW) galaxy as it is now (in the lowest-redshift panel), and where it would have been seen at the highest redshift able to be probed by GAMA, given a smooth, exponentially declining star formation history. The location of a hypothetical dwarf galaxy is shown as the filled magenta circles, with a similar exponential star formation history, and a different end point (magenta triangle) when the dwarf system is only forming stars 25% of the time under the same exponentially declining model. The existence of low-mass galaxies with even higher SSFRs are evidence that such extreme mass-formation events cannot be maintained, otherwise such galaxies would necessarily appear at much higher mass. The apparently smooth star formation histories of massive galaxies may in fact simply be a consequence of the integration over many such small, intermittent and stochastic star formation events, that only become apparent in very low mass galaxies due to the small number of individual events they can host.

Using the GAMA groups catalogue (Robotham et al., 2011), an investigation into how typical our own Local Group might be has identified a number of so-called "Milky Way Magellanic Cloud Analogues" or MMAs (Robotham et al., 2012). The best such example is shown in Figure 4. This investigation looks in detail at the statistics of galaxy pairs and groups, where a galaxy of similar mass to the Milky Way has low mass companions of similar mass to the Magellanic Clouds. We find that the particular configuration of the Local Group is rare, with only 0.4% of Milky Way mass galaxies having two late-type companions at least as massive as the Small Magellanic Cloud.

GAMA is producing many other exciting results as well, including measurement of galaxy luminosity functions (Loveday et al., 2012) and mass functions (Baldry et al., 2012), that demonstrate the sensitivity of GAMA to the faint galaxy population at low redshift, and the evolution of both red and blue galaxy populations. We have explored the impact of environment on galaxy star formation rates (Wijesinghe et al., 2012) finding, consistent with other recent work, that for the actively star-forming galaxy population the distribution of star formation rates has no dependence on local environment. We have looked at systematics associated with the measurement of gas-phase metallicities in galaxies (Foster et al., 2012), demonstrating that the mass-metallicity relationship is robust to a variety of measurement limitations, and that there is tentative evidence for evolution in the mass-metallicity relation over 0<z<0.3 (explored further by Lara-Lopez et al., in prep.). We have used the Sersic index fits to galaxy profiles, combined with optical colours, to identify a new and very clean galaxy-bimodality separator, in $(u-r)_{rest}$-n parameter space (Kelvin et al., 2012). We have explored an independent estimator of halo masses using a caustic analysis (Alpaslan et al., 2012), to complement the GAMA groups halo-mass estimates. We have also begun detailed exploration of the WISE mid-infrared survey data to provide independent and complementary analyses of star formation processes in GAMA galaxies (Cluver et al., in prep.).

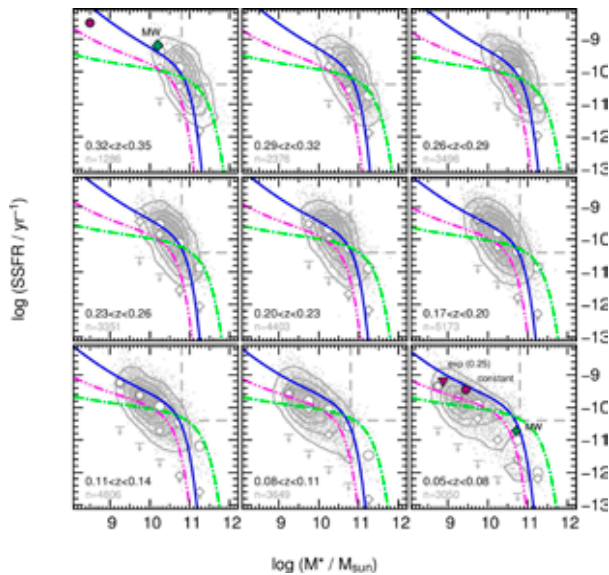

Figure 3. SSFR vs stellar mass as a function of redshift (Bauer et al., 2012). The median SSFR in bins of stellar mass for the star forming and full sample are shown as circles and diamonds, respectively. Two parametrizations of exponentially declining star formation histories (SFH) derived from the GAMA sample are shown as the solid blue and dash-dot-dot magenta curves. Gilbank et al. (2011) SFHs derived from a $z \sim 1$ sample are shown as green dash-dot curves.

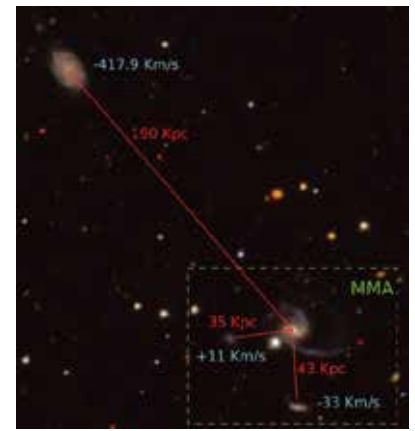

Figure 4. SDSS image of the best Local Group analogue found in GAMA. The spiral galaxy in the bottom right with the two indicated brighter companions is the MMA (GAMA-MMA1), where all three galaxies are late-type and star-forming, making this system similar to the MW Magellanic Cloud system in the most fundamental ways. An associated nearby spiral galaxy is also shown.

## GAMA in 3D

In addition to these science highlights, the GAMA team continues to pursue a variety of collaborative projects with many other survey teams, including VISTA VIKING, VST KIDS, CFHTLenS, Herschel ATLAS, GALEX-GAMA, WISE, and ASKAP-DINGO. This coordination of collaborations continues with the development of new projects now building on the GAMA survey itself. Some of the most exciting current developments are coming from a selection of recent or upcoming SPIRAL observing projects with the AAT, supplemented in some cases by WIFES on the ANU 2.3m (PI: Croom; PI: Brough; PI: Bland-Hawthorn; PI: Robotham). These include:

- An investigation into the spatial distribution of star formation within galaxies as a function of local environment. Preliminary results imply that only the densest environments seem to have an impact, and that this primarily acts to suppress star formation in the outskirts of galaxies, (Brough et al., in prep.).

- Exploration of some of the most extreme star formation rate galaxies in GAMA and SDSS to identify the mechanisms associated with the production of star formation rates that, in some cases, exceed 100 M$_\odot$ yr$^{-1}$. Preliminary results suggest that merger events are not dominant, and clumpy star formation may not be prominent in all cases. Whether galactic winds are in evidence is an ongoing question, waiting on more detailed analysis of the data (Fogarty et al., in prep.; Gunawardhana et al., in prep.).

- Further exploration of the properties of Milky Way Magellanic Cloud analogues, to explore their properties in more detail (observations scheduled for January 2013).

In summary, the GAMA survey continues to be highly successful, with numerous exciting scientific results in hand and ongoing. GAMA continues to develop new and valuable collaborations and to generate new projects growing from the existing survey. We are looking forward to many exciting new results to come.

SCIENCE HIGHLIGHTSSCIENCE HIGHLIGHTS

# Evidence for Significant Growth in the Stellar Mass of the Most Massive Galaxies in the Universe

Chris Lidman, (AAO), Janette Suherli (AAO/Bosscha Observatory, Indonesia), Sarah Brough (AAO), Gillian Wilson (University of California Riverside, USA), Adam Muzzin (Leiden Observatory, The Netherlands) and Ricardo Demarco (Universidad de Concepcion, Chile)

Located in the cores of rich galaxy clusters are the brightest, most luminous and most massive galaxies in the universe – the so-called Brightest Cluster Galaxies, or BCGs for short. Since they are so large and bright and since they lie in the centre of clusters, BCGs are easy to spot. They can also be spotted out to large distances, which means that we can observe them at a time when they were much younger. This makes them an attractive target for testing our understanding of the processes that drive galaxy evolution in extreme environments.

A few years ago, a serious discrepancy between the predictions of numerical simulations and observations was uncovered. Numerical simulations predict that the stellar masses of BCGs should increase dramatically with time. For example, between redshift 1.0 (corresponding to a time when the universe was around half as old as it is now) to today, numerical simulations predicted that BCGs should increase their stellar mass by a factor of a few [1]. On the other hand, observations suggested no growth at all [2,3].

During her time as an AAO student fellow , Janette Suherli examined this discrepancy afresh using a new sample of BCGs. In contrast to earlier results, she found that stellar mass does increase with time and that the increase is at a rate that is only slightly slower than that predicted by the simulations.

## The sample, estimating stellar masses and comparisons with theory

We use galaxy clusters from SpARCS, CNOC and the literature [1,2] to assemble a sample of over 100 BCGs. From the nearest BCG in the sample (at z=0.03) to the most distant (at z=1.63), almost 10 billion years of the 13.7 billion year history of our universe is covered. Some examples are shown in Fig. 1.

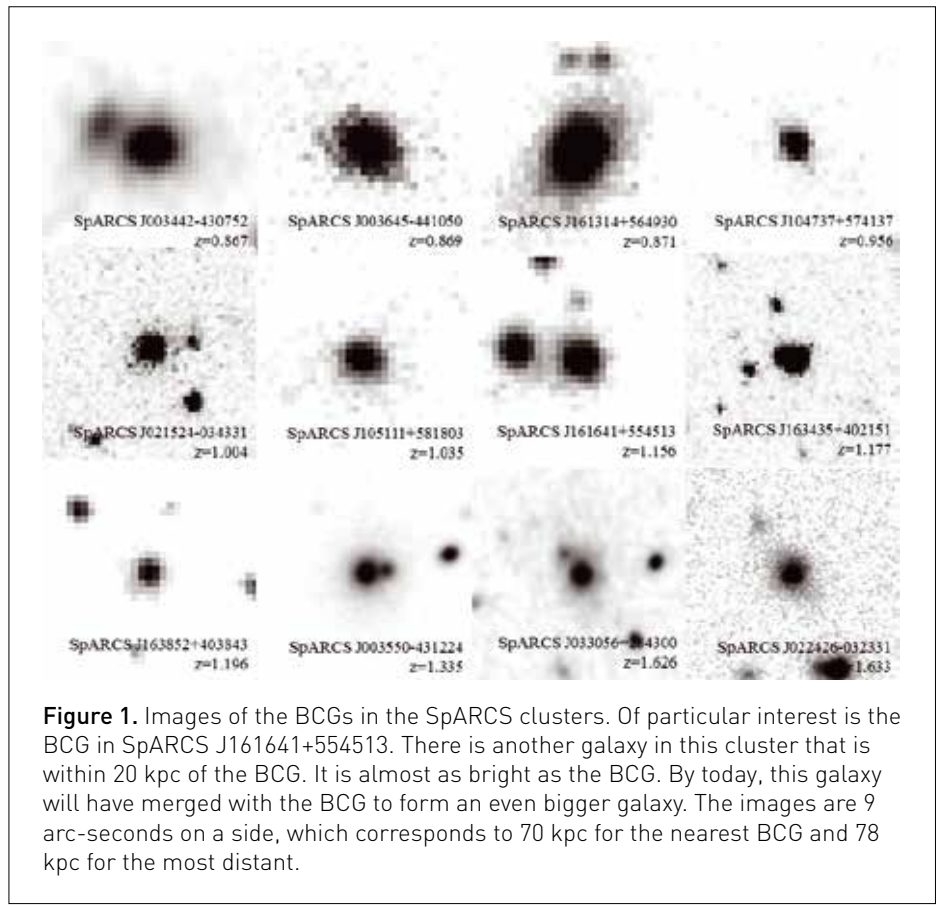

**Figure 1.** Images of the BCGs in the SpARCS clusters. Of particular interest is the BCG in SpARCS J161641+554513. There is another galaxy in this cluster that is within 20 kpc of the BCG. It is almost as bright as the BCG. By today, this galaxy will have merged with the BCG to form an even bigger galaxy. The images are 9 arc-seconds on a side, which corresponds to 70 kpc for the nearest BCG and 78 kpc for the most distant.

Stellar masses are estimated by comparing the observer-frame K-band flux with the predictions of a synthetic stellar population model that reproduces the J-Ks colour of BCGs over the entire redshift range covered by the BCGs. We then compare the median mass of BCGs in three broad redshift intervals: a low-redshift interval up to z=0.3, an intermediate-redshift interval between z=0.3 and z=0.8 and a high-redshift interval beyond z=0.8.

Two comparisons are made. The first comparison is shown as the black squares in Fig. 2. While it is clear that the median mass of BCGs in the high-redshift subsample is lower than the median mass of BCGs in the low-redshift subsample, there is little difference between the intermediate-redshift subsample and the low-redshift subsample.

However, the first comparison neglects the correlation between the stellar mass of BCGs and the mass of clusters. Bigger clusters tend to host bigger BCGs. If the average masses of the clusters in the three subsamples differ, then the black squares in Fig. 2 will be biased low or high.

Since clusters in the intermediate and high-redshift sub-samples will be by today more massive than the clusters in low-redshift subsample (the bias comes from the way clusters were selected), the black squares in Fig. 2 are biased high. When the correlation is taken into account, the black points move down to the green triangles, as shown in Fig. 2.

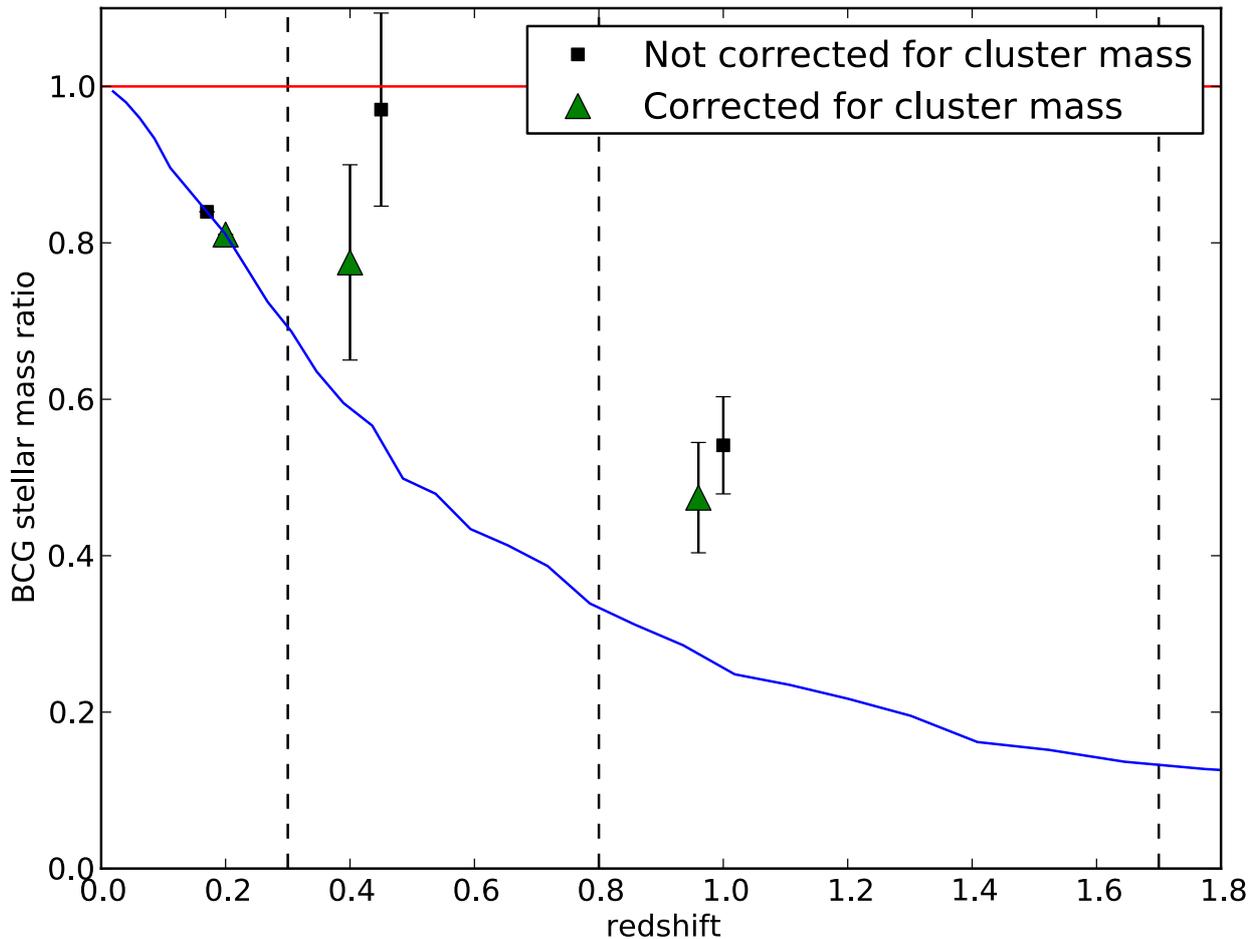

**Figure 2.** The evolution in the stellar mass of BCGs as a function of redshift. The black squares take no account of the correlation between the masses of BCGs and the masses of the clusters. The green triangles, on the other hand do take this correlation into account. The blue line represents how the stellar mass grows in semi-analytic models [1]. The red line represents no evolution. Note how the green points in the intermediate and high-redshift bins lie below the black points and how these points are a better match to the semi-analytic models. All points are normalised so that their low-redshift points land on this model. The vertical dashed lines mark the boundaries of the low, intermediate and high-redshift subsamples. The points are plotted at the median redshifts of the subsamples. They differ slightly between the green and black points because a more restricted range of clusters is selected when matching cluster masses.

Overall, between redshift 0.9 and 0.2, we find that the stellar mass of BCGs increase by a factor of almost 2. Compared to previous results [3], our result is in much better agreement with the predictions of numerical simulations. However, a small discrepancy with simulations, which predict a three-fold increase in the stellar mass between these redshifts, persists. Some of the discrepancy may come from stars that are lost to the intra-cluster medium during mergers. These stars may lie outside the aperture that was used to compute the K-band flux and estimate the mass.

**Future Work**

It is likely that most of the growth in the stellar mass of BCGs between redshift one and today is due to galaxy merging rather than in-situ star formation, since the star formation rates that are inferred from the [OII] emission line at 3727 Angstroms that can be seen in a small fraction of BCGs are too low. However, it is not yet clear if most of the stellar mass is built up through a small number of major mergers or a large number of minor mergers. Major mergers do take place, as can be seen in Figure 1. We are now in the process of computing the likely number of major mergers in clusters using the extensive data that we have collected on GCLASS clusters. Stay tuned for the results of this work.

# The 5th Southern Cross Conference: A Joint CASS/AAO Conference

Michelle Cluver and Amanda Bauer (AAO)

The theme of this year's Southern Cross Conference, "Multiwavelength Surveys: A Vintage Decade", was befitting of its location in the exquisite Hunter Valley, and also timely in the era of massive data sets and large international collaborations. Seeking to bring together observation and theory across the breadth of the electromagnetic spectrum, tied to both large-area and more focused surveys, it produced a well-stocked cellar of experts, as well as more fruity younger wines.

Matthew Colless kicked off proceedings with an inspiring look at the importance of surveys and survey astronomy. And amongst the vignobles, a vig-Nobel! I refer of course to Prof. Brian Schmidt, recent recipient of a Nobel Prize in Physics, who presented an update on the SkyMapper survey. Other invited talk highlights were from Dr Naomi McClure-Griffiths on the HI Galactic All-Sky Survey, and Dr James Jackson presenting results from the MALT90 millimetre astronomy survey. These studies are making it possible to probe Galactic processes and the ISM life-cycle in our own galaxy.

At the other end of the wavelength spectrum, the Galaxy and Mass Assembly (GAMA) Survey presented an impressive slew of talks, indicative of the richness of science made possible with multiwavelength, spectroscopic surveys. And, another high point, Dr. Loretta Dunne presented an overview of the latest results from the Herschel Extragalactic surveys, illustrating how the far-infrared wavelengths are revolutionizing our view of nearby and distant galaxies.

The smooth flow of the conference was testimony to the adept organization of Jill Rathborne (CSIRO), and a quality programme of invited and contributed sessions fortified with variety and depth. The calibre of talks and pioneering results certainly created an atmosphere of a "Golden Age" of surveys, some well-established and some in their infancy, but commonly spearheading the

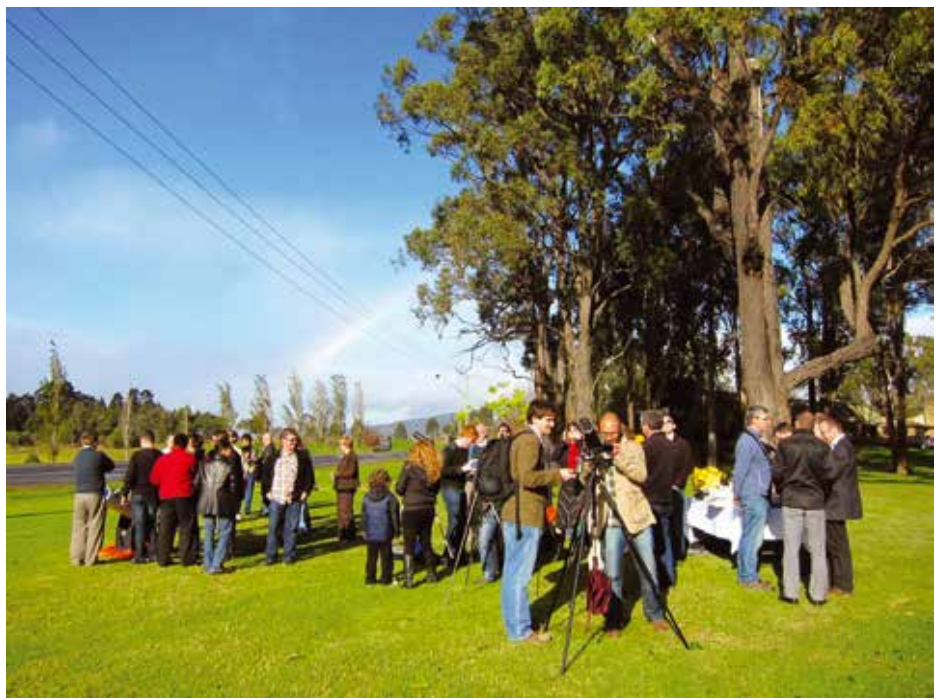

Southern cross conference attendees enjoy a chance to view the transit of Venus during a tea break.

studies of the future. From SkyMapper to HOPS, LOFAR to HERMES, GAMA to GASKAP, no part of the sky or wavelength is safe from astronomers.

On Wednesday, June 6, the backdrop of the Hunter Valley vineyards was a suitably memorable setting for a most rare event -- the Transit of Venus. 105 years is a long time to wait for another one, but fortunately the weather (mostly) cooperated and afforded astronomers the opportunity of witnessing this historic occurrence, with Dr Robert Hollow kindly providing telescopes for the event.

Not as rare, but an absolute treat nonetheless, was David Malin's lecture on "Photography and the Discovery of the Universe". Exceedingly informative and equally entertaining, it was definitely a conference highlight.

The philosophy of the Southern Cross Conference Series is to bring Australian and international researchers from different specializations together, to remind us of the common ground that exists, and to be informed and inspired by each other's work. From galactic scales to simulations of the universe, and through each wavelength window, our ultimate goal is to understand the physics of the universe - to work in isolation only slows progress. And therein lies the success of a conference such as this, where tea times, lunches and dinners become fertile territory for forging new connections and reinvigorating discussion and collaboration. I suspect the next decade will yield some spectacular harvests because of it!

Many thanks to the organizers and participants for the smooth running of the meeting, and for making it an exciting overview of the current state of multiwavelength surveys internationally. And thanks also to the all-important sponsors, CSIRO and AAO.

The conference website is available at: http://www.atnf.csiro.au/research/conferences/2012/SCCSV/index.html

## Next Year's Southern Cross Conference

Late in June 2013, the AAO will be holding an international astronomy conference which will be hosted at a coastal resort near Brisbane, Australia. The meeting, entitled "Feeding, Feedback, and Fireworks: Celebrating Our Cosmic Landscape," will be the 6th of the Southern Cross Conference Series, jointly supported by the AAO and the Australia Telescope National Facility (ATNF).

The Southern Cross Astrophysics Conferences are held annually around Australia with the aim of attracting international experts with wide ranging skills to discuss a particular astrophysical topic. This conference will focus on galaxy evolution and how various feedback and feeding processes transfer energy into and out of galaxies. We intend to bring together observations, from radio to X–rays, and the best available theoretical models, to create a more complete picture of our cosmic landscape.

# GALAH: Preparing for Flight

Sarah Martell (AAO)

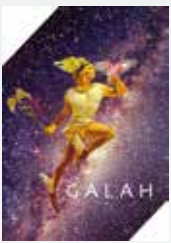

On April 30, the GALAH survey team held a meeting at AAO. GALAH is a large survey planned for the HERMES spectrograph, aimed at unravelling the star-formation history of the Galactic disk by identifying unique groups in stellar abundance, and topics of discussion at the meeting ranged from the HERMES construction schedule to spectrum-analysis software and target selection strategy.

### Instrument updates

Jamie Gilbert from AAO's instrumentation group showed a time-lapse movie of the assembly of the HERMES frame, and Jeroen Heijmans, also from instrumentation, reported that 3 of 4 CCDs have been delivered, and testing had begun on optical components for the blue camera. AAO's software group are also contributing major efforts to HERMES, with three specialized programs that allow astronomers to simulate HERMES data before the instrument is built, to reduce the raw data frames and extract spectra, and to pre-schedule series of science and calibration observations. This last program is based on a strong recommendation from Chris Tinney from UNSW, who has written similar software for the Anglo-Australian Planet Search, a long-running AAT program.

### Science software development

There are two major software-development projects in GALAH: the stellar parameters and abundance analysis pipeline, known as GA3P, and the database back-end that will contain target information, observing logs, reduced data and science products. This database, coordinated by Stefan Keller at ANU and hosted by the National Computational Inftrastructure , will be the structure for both internal and public releases of GALAH data, which are currently expected to occur annually. The spectrum-analysis pipeline, led by Liz Wylie de Boer at ANU, works in two phases, first choosing stellar parameters based on full-spectrum information and then determining abundances of individual elements from synthesis of small regions of spectrum.

### Looking ahead: commissioning and observing plans

The instrumentation group's current schedule has HERMES being delivered to AAT in March 2013, with assembly and commissioning continuing through Semester 13A. Science Verification, a short phase for testing all aspects of HERMES science observations, is expected to follow early in Semester 13B. The head of GALAH's working group on commissioning and science verification, Dan Zucker from Macquarie University, will be working with AAO to issue a call for SV proposals to the Australian community. The GALAH team anticipates proposing a pilot survey for late in Semester 13B, after Science Verification, to demonstrate that the target selection, observing strategy, and data analysis are all ready for full survey operations. Membership in GALAH is open to all members of the Australian astronomical community, assuming a certain level of participation in survey operations. Further information on survey goals and the various working groups are available at the survey web site.

# Fire in the Sky

Fred Watson (AAO)

If you want to see the effects of subatomic particles interacting with a magnetic field, you can do no better than head towards a pole. Preferably, one of the Earth's, where vivid auroral displays can make the polar regions, well … a magnet for tourists.

This was what took me and two small groups of Australians to far-northern Scandinavia at the beginning of 2012, for our 'Fire in the Sky' study tours in celebration of the Sun's peak of auroral activity. Over the past few years, astronomy tourism has become a significant adjunct to my day-job, and there's nothing quite like the prospect of taking interested people to places where scientific ideas originate. These folk come from all walks of life, but their common passion is a hunger for learning.

From our icy vantage point at Lyngenfjord in arctic Norway, a succession of clear nights allowed us to discern a distinct pattern in the appearance of the Aurora Borealis. The display would start early in the night, soon after the long twilight had ended, with a glow on the northern horizon. Then, around eight pm or so, we'd detect a thin greenish band, snaking from east to west through the far northern constellations and, if we were lucky, the show would begin ramping up to its evening crescendo. Swirling curtains of light, twisted into impossible shapes, would sweep in waves across the sky, taking only seconds to form, brighten into prominence, and then fade again.

At their height, these displays could bring a rare climax, with the formation—almost overhead—of an auroral corona. Like crystals precipitating out of a super-saturated chemical solution, green, finger-like rays would burst from the zenith before our astonished eyes. This extraordinary effect is a trick of perspective, caused by parallel columns of light forming along the near-vertical lines of the Earth's magnetic field. Except in the most extreme conditions of solar activity, it is exclusively the province of those who venture to the planet's polar regions.

## Particles from the sun

In earlier times, the Sami people who inhabited this frostbitten country regarded the Aurora Borealis with fearful awe. Legends about the origin of the eerie lights were intertwined with traditions concerning the souls of the departed, their fabled camp fires flickering mysteriously beyond the northern horizon.

It was only at the turn of the twentieth century that the first glimmer of understanding began to emerge as to the true cause of the aurora—and it had its origins not far from the small Norwegian town of Alta, where our study tours had begun.

Just to the west of Alta, in the spine of mountains that dominates Norway's arctic coastline, a high plateau called Haldde became the site of the world's first auroral observatory. It was set up by a visionary Oslo University scientist called Kristian Birkeland, who had a wild notion that the Aurora Borealis is caused by electric currents in space, and is somehow linked with the Earth's magnetic field.

Birkeland postulated that the Sun was a source of the subatomic particles then known as cathode rays (electrons, today), and that their interaction with the Earth's magnetic field near the poles caused them to excite the atoms of the upper atmosphere into a frenzy of luminescence. He later suggested that positive ions, too—protons—might also be emitted from the Sun, and play their part in the celestial light-show.

Rivalling the first-quarter Moon in intensity, an auroral corona bursts over the skies of Lyngenfjord in the Norwegian Arctic (Anne Spencer).

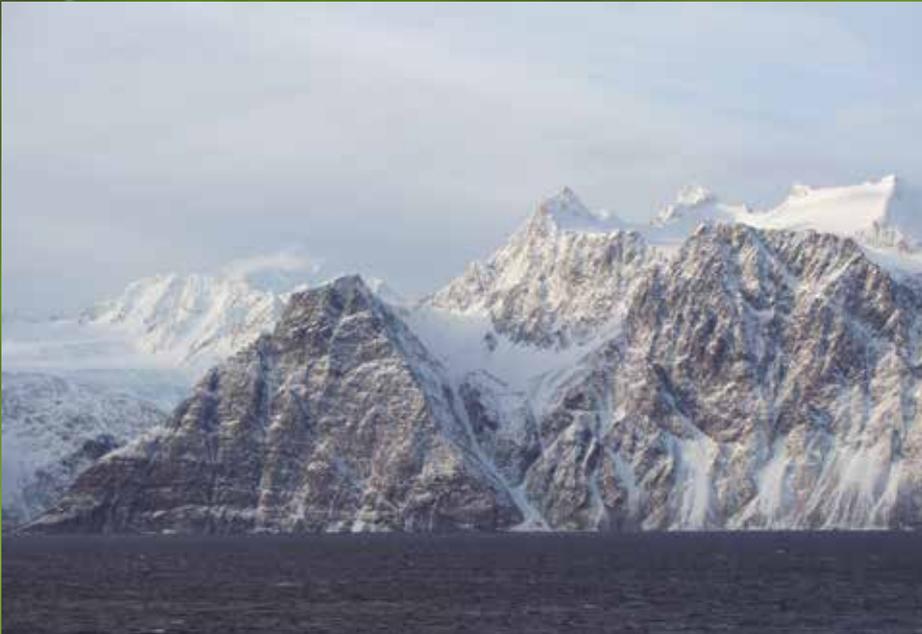

In daylight, Nature's magnificence is revealed in the peaks and glaciers of the Lyngen Alps across the fjord.

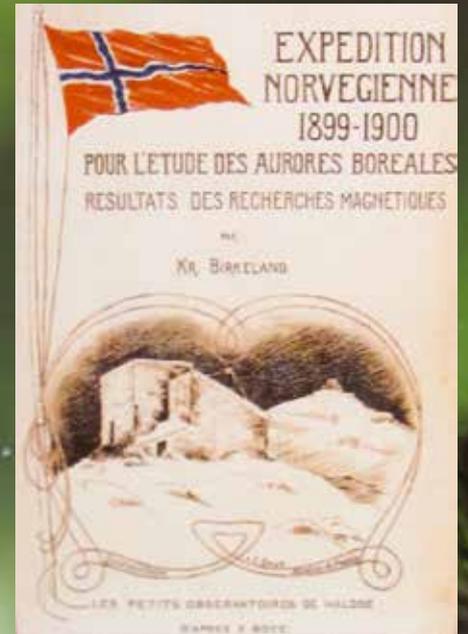

Kristian Birkeland's book on his first expedition to Haldde made a political statement as well as reporting on his scientific observations of the Aurora Borealis. Controversially, the Norwegian flag was shown without its Swedish counterpart in the corner, which was mandatory in the days before Norway gained its independence.

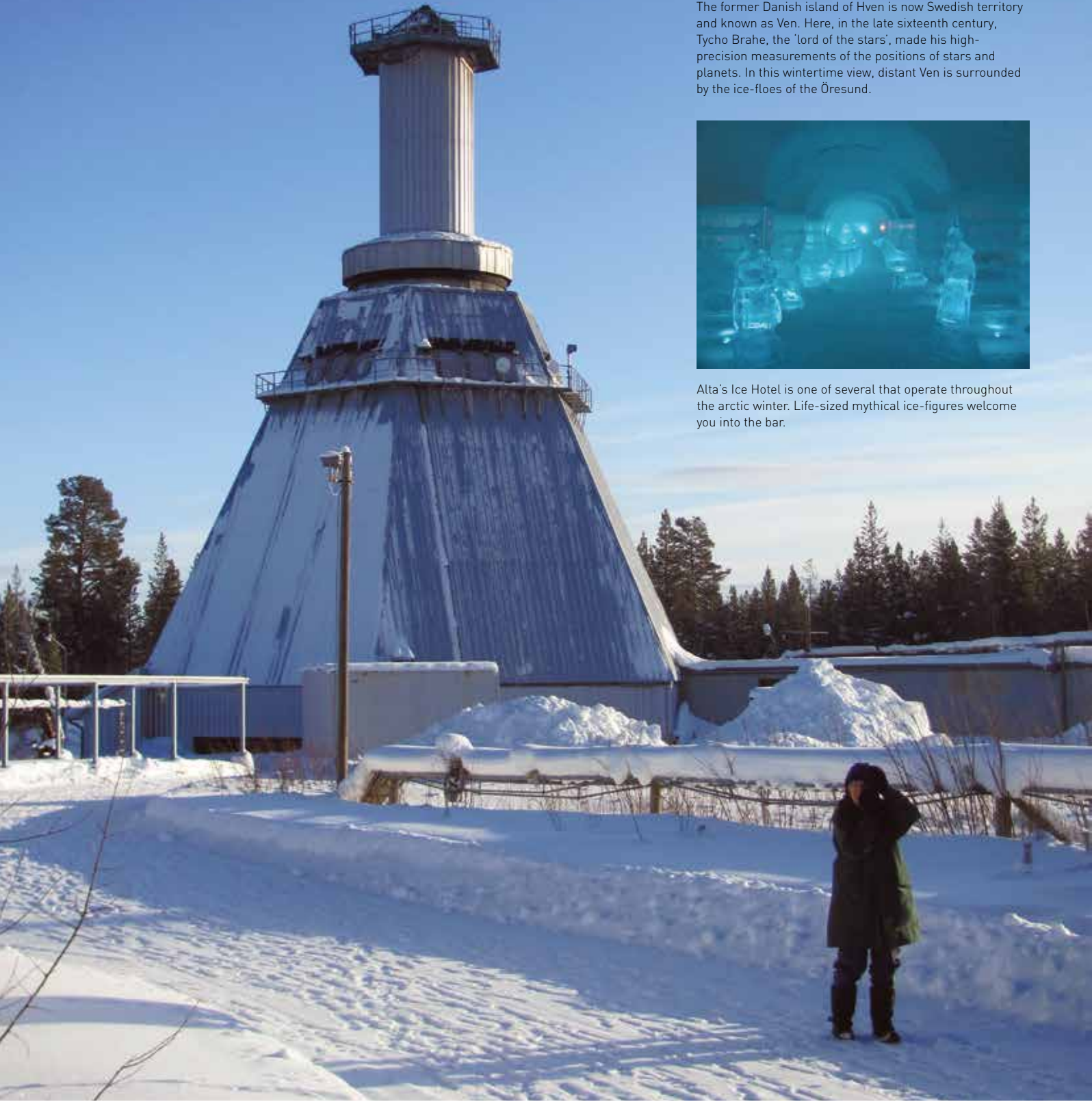

Launch tower at the Esrange Space Centre at Kiruna in northern Sweden. Until the mid-2000s, the tower was used to launch Skylark sounding rockets for upper atmosphere research.

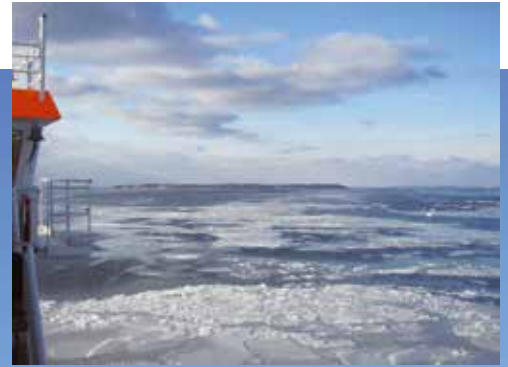

The former Danish island of Hven is now Swedish territory and known as Ven. Here, in the late sixteenth century, Tycho Brahe, the 'lord of the stars', made his high-precision measurements of the positions of stars and planets. In this wintertime view, distant Ven is surrounded by the ice-floes of the Öresund.

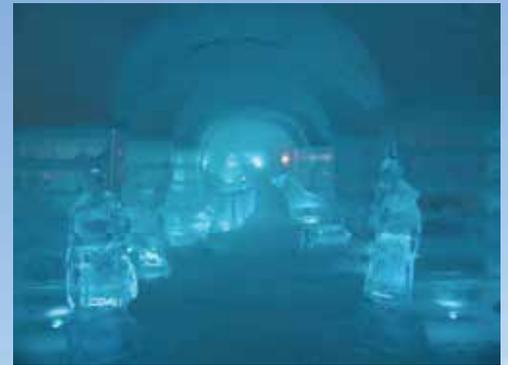

Alta's Ice Hotel is one of several that operate throughout the arctic winter. Life-sized mythical ice-figures welcome you into the bar.

However, the idea that anything other than heat, light, and gravity could come from the Sun incensed Birkeland's critics, particularly in Britain's Royal Society, which regarded itself as the de facto owner of these phenomena.

With the outbreak of the First World War, and amidst growing paranoia about persecution by the British, Birkeland began to sink into a mire of mental instability from which he never recovered. Eventually, in 1917, he died in Tokyo at the age of only 49. It was another half-century before magnetic measurements made from Earth-orbit revealed what ground-based observations had failed to detect. Space is, indeed, full of subatomic particles—negatively and positively charged. On both counts, Birkeland had been right.

Today, we know that aurorae are the result of a highly complex interaction between the solar wind and the Earth's magnetic field. Our modern understanding includes such subtleties as the aurora's occurrence in a circular zone around each magnetic pole rather than a concentration of light at the pole itself—something that was not explained by Birkeland's theory. And our knowledge of the energies carried by the solar particles lets us understand why we see aurorae of different colours at different heights.

### Adventures in the north

The secret of successful touring is to have an itinerary that also caters for the sometimes not-so-interested partners of the study tourists. Thus it was that after our rendezvous with the northern lights, we embarked on a tour of the other attractions of the region, both scientific and otherwise.

Alta and Lyngenfjord are within striking distance of Narvik, which boasts a spectacular railway line linking the famous port with Kiruna in northern Sweden. And Kiruna has a rocket range—the Esrange, operated by the Swedish Space Agency, whose restricted airspace will eventually be used by Richard Branson in connection with his Spaceport Europe. From Kiruna airport, Virgin Galactic will fly well-heeled passengers through the Aurora Borealis, rather than simply observing it from below.

Southern Scandinavia, too, has much of interest. Not far from Stockholm is Kvistaberg, where the historic University of Uppsala has custody of one of the largest Schmidt telescopes in the northern hemisphere. With an aperture

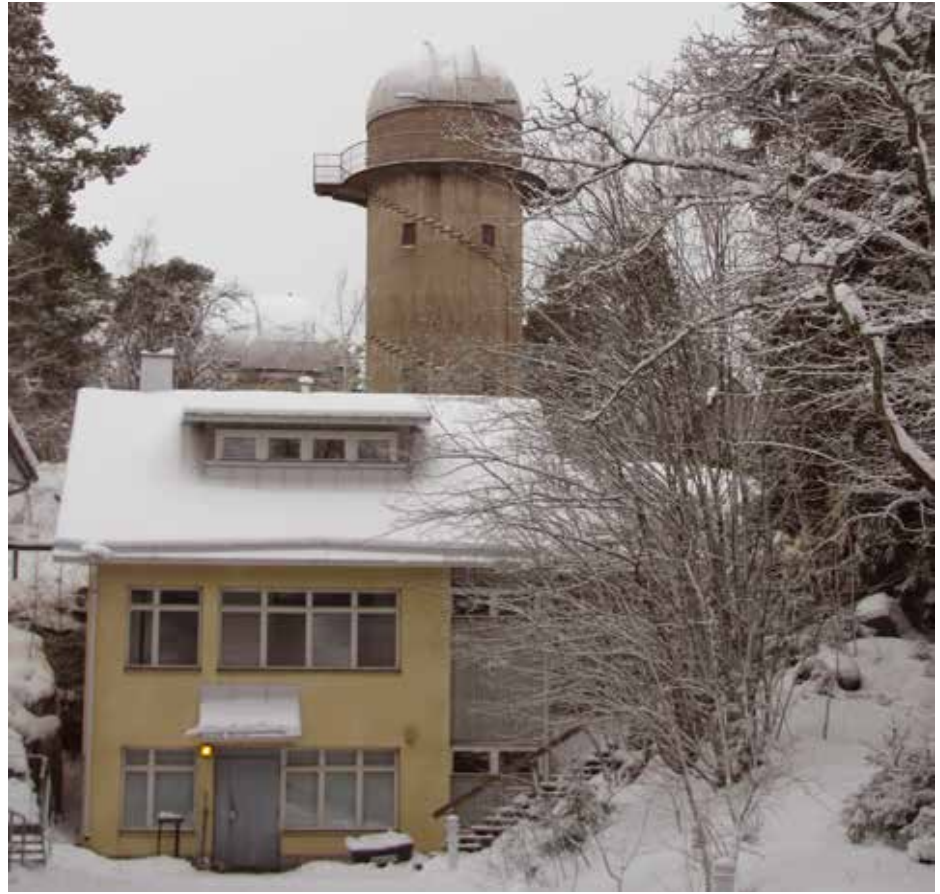

The one-metre telescope of the Tuorla Observatory in Finland is mounted on a high tower to raise it above the level of the surrounding forest. Tuorla is operated by the University of Turku, and is the largest astronomical research institute in the country.

of one metre, the Kvistaberg Schmidt bears a distinct family resemblance to the AAO's 1.2-m UKST at Siding Spring. The most striking difference, apart from the smaller size, is the steep polar angle of the telescope's equatorial mounting—equal to its latitude—of some 60 degrees, compared with the laid-back 31 degrees of the UKST. Although equipped with a CCD camera, the Kvistaberg Schmidt is little used today due to indifferent weather conditions compared with overseas sites.

The astronomical jewel of southern Scandinavia, however, is the little island of Ven in the Öresund—the strait separating Sweden and Denmark. Here, in the late sixteenth century, the greatest astronomer of the pre-telescopic era plied his trade. The noble Tycho Brahe built two observatories, whose few broken remains make Ven a mecca for astronomy enthusiasts. In summertime, the island is enchanting, but a winter mantle of snow, and Öresund ice-floes sporting the occasional basking seal, made our February visits even more spectacular.

And so on. There is little space here to mention our visit to Tartu in Estonia to see Joseph Fraunhofer's masterpiece: his 24-cm 'Great Dorpat Refractor' of 1824, which set the pattern for refracting telescopes throughout the nineteenth century. Nor our visit to Tuorla Observatory in Finland, where the tall tower housing its one-metre telescope of 1959 sprouts incongruously from a snow-laden pine forest. Nor a magical night-time excursion to Copenhagen's curious Round Tower of 1642, whose 209 metre long internal spiral ramp allowed horse-drawn carriages to access an astronomical observing platform high above the streets of the city. Nor even our ventures into astronomical music, which featured the great Finnish composer, Jean Sibelius, and his contemporary Estonian counterpart, Urmas Sisask, who gave us an extraordinary recital in his private planetarium near Tallinn.

Oh, and then there was Iceland, with its volcanoes, hot springs and glaciers. But that is a story for another time… and, indeed, for another trip. The next 'Fire in the Sky' tour through Scandinavia will take place in January 2013. There are still a few spaces left…

# HERMES: Assembly in Pictures

Gayandhi De Silva, Anthony Heng, Tony Farrell, Keith Shortridge (AAO)

## Introduction

Over the last year, the Mechanical, Electronics, Software, Optics, and Fibre Optics sections have all been heavily involved in the HERMES instrument. It was a challenging time for the team as many major components of the spectrograph arrived. HERMES science meetings were also conducted. The mechanical team has integrated and tested the spectrograph frame (Figure 1).

Most of the mounts for the optics have been manufactured and tested. The slit systems (Figure 2) were received and tested.

Leak testing for all the cryostat casings has been completed. The integration of the cryostat for the blue channel of the spectrograph has been completed and its ability to maintain vacuum and temperature is now under test (Figure 3).

We have also tested the Collimator mirror (Figure 4).

## Electronics

The electronics team has fitted out six of the seven electronics cabinets (One cabinet for each of the blue, green, red and infra-red cameras, one for the main control panel, one for the slit systems and one for the Dewar systems). The Dewar system cabinet is still to be wired up. The Hartmann and Bonn Shutters (Figure 5) have been received and fully tested.

## Software

The software team have resolved a long-standing problem with CCD readout speed calculations whilst soak testing the HERMES controllers. The spectrograph software can now control the focus hardware, having previously been tested with the slit hardware. The control task modifications to support four CCDs are now complete.  Most HERMES-related software interlocks have been enabled. The software for the CCD controllers is now capable of reading out through all four amplifiers at non-astro speed in 10 seconds - this is the most extreme test possible with only one controller.

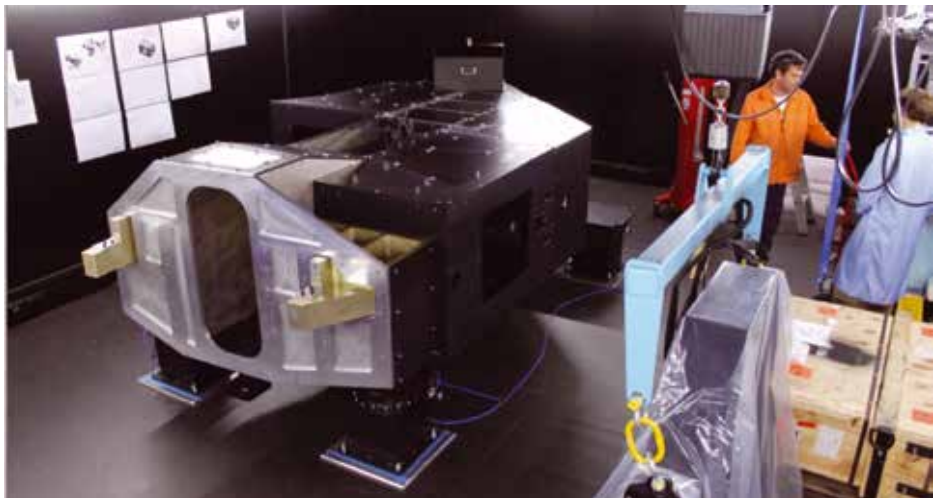

Figure 1. HERMES spectrograph frame.

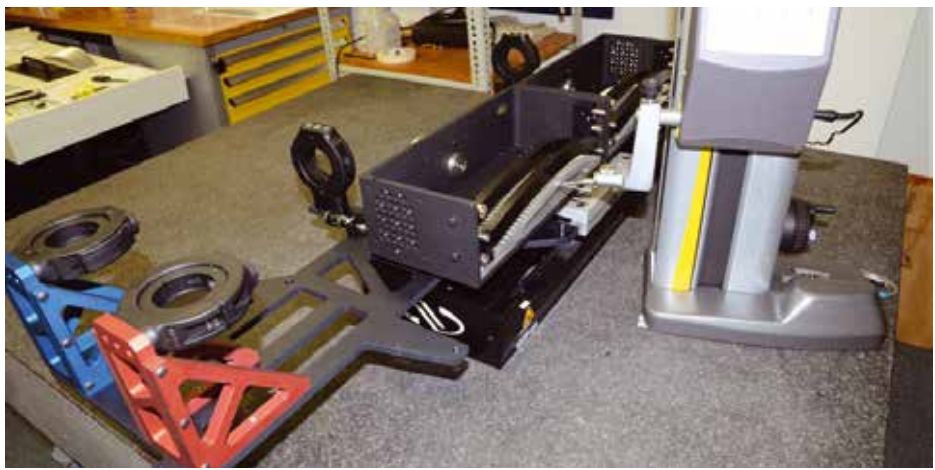

Figure 2. HERMES slit systems.

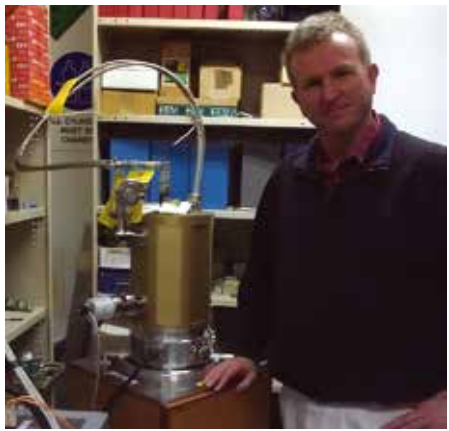

Figure 3. Lew Waller with the blue channel cryostat.

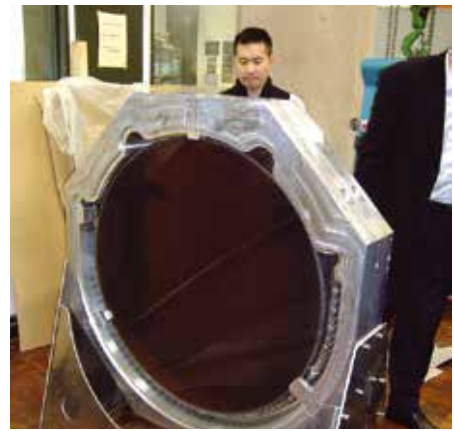

Figure 4. Anthony Heng inspecting the Collimator mirror.

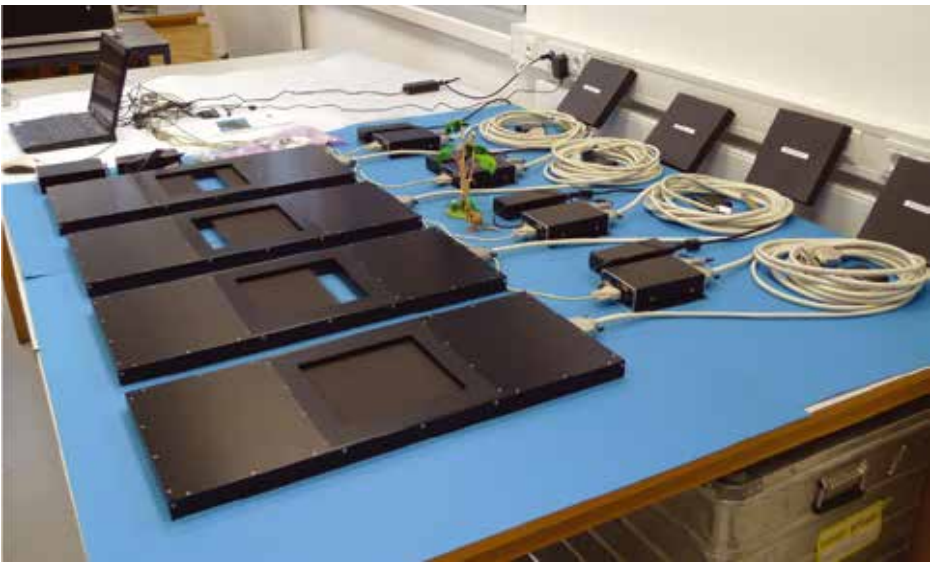

Figure 5. Four Bonn shutters received and fully tested.

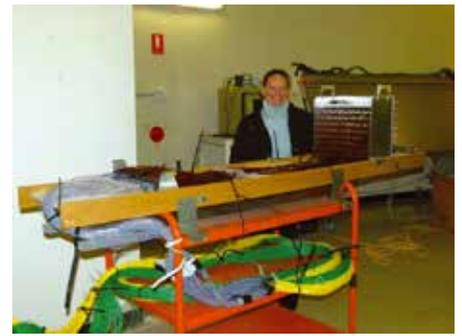

Figure 6. Kristin Fiegert (fibre optician) with the HERMES and AAOmega fibre optic cables, ready to be inserted into the conduit.

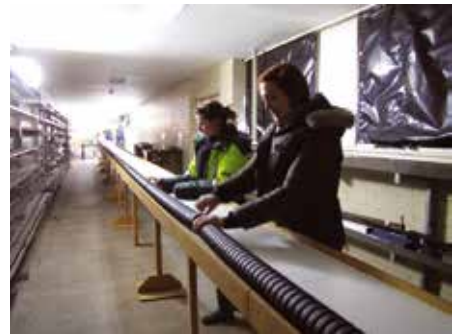

Figure 7. Carolin Barth and Sue Wilson (fibre optic assistants) inserting the cable into the Triflex conduit.

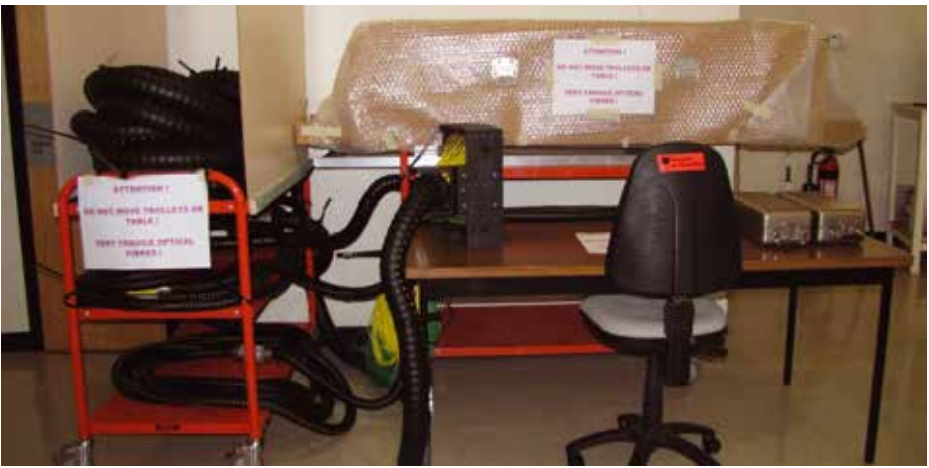

Figure 8. The completed fibre optic cables for HERMES and AAOmega waiting to be installed on 2dF

## Optical

The optical team has also been busy testing optic components received from the vendors. The blue grating and beam splitter have been received and are being mounted in the frame. Fold mirrors have been tested and one of them has been mounted. The blue camera has passed acceptance tests and has been received at Epping. The other three camera lenses (green, red and infrared) have been manufactured and polished and are waiting to be coated. The collimator mirror has been received by AAO and tested. The collimator field lens has also been received.

## Fibre Optic cable

The fibre optic team have completed the preparation of the 800-fibre, 49.95-metre HERMES cable and the new 800-fibre, 39.15-metre AAOmega cable. Each of the optical fibres has been inserted into polyimide tubing, integrated with its ferule and prism, and finally fed into the overall conduit (Figures 6 and 7). The cable is now ready for installation on 2dF, which will be done between August and October (Figure 8).

## Issues and challenges

At this stage, the most challenging part of this project is dealing with late deliveries from vendors. In many cases, this is caused by resource constraints within the vendor. Such delays have a direct impact on the project schedule. To mitigate the problem, the HERMES team is constantly communicating with and visiting the vendors to check and discuss progress, and to help them resolve issues. The optics still to be delivered to AAO included the collimator corrector lenses, the green, red and infrared cameras, and the green and red beamsplitters. All these will be delivered in October 2012. Because of this delay we have changed our testing strategy and will be testing the complete blue channel first, followed later by the green, red and infrared channels. Our aim is to observe a solar spectrum with the blue channel and process it through the data reduction software and GALAH abundance pipeline, allowing most of the major tests to be completed and most of the outstanding risks to be retired as early as possible.

## HERMES' Science

The GALAH survey team has been making steady progress with the survey preparations. Interfacing of the data reduction pipeline with the abundance analysis pipeline has started, and simulated HERMES data are being used to verify and test both pipeline specifications. Various selection and tiling options for the input catalogue are being investigated and co-ordinated with practical operational aspects. The second 'all-hands' team meeting was held on the 30th April 2012 at the ATNF lecture theatre. The working groups shared progress and plans, and the policies and operational procedures of the survey were discussed.

# AAT REFURBISHMENT: BRAKING THE AAT SHUTTER

Doug Gray (AAO)

Built in the 1970s, the AAO telescopes are subject to regular upgrades and ongoing maintenance to provide astronomers with state of the art capabilities, as well as to preserve and extend the life of this major Australian scientific asset. As part of a NCRIS $4.2M refurbishment program the emergency brakes on the Anglo Australian Telescope main shutter were identified as needing replacing.

The shutter, weighing close to 17 tonnes is driven by a single electric motor and gearbox located on top of the dome. Two large chains located on both sides of the shutter and anchored at each end are used to pull the shutter over the drive system to expose the telescope to the night sky.

The shutter was originally designed to have two independent mechanically operated emergency brakes, one either side, to prevent the shutter falling to the ground in the event of a drive system failure. The emergency brakes proved unreliable over the years with one of them being removed shortly after commissioning in the late 1970s.

The concept of mounting the brake callipers on the top of the dome and braking onto steel ribs welded to the main shutter was originally developed in-house by Brendan Jones and Allen Lankshear.

P.M.Design, an engineering company based in Geelong were engaged to develop the concept and design and build the new emergency brake system.

The four tonne brake arrived early April and was lifted onto the top of the AAT dome with a 200 Tonne crane from G.B.P. Cranes of Gunnedah, which used its entire 90m boom to reach the centre of the AAT dome, some 50m from ground level. The whole operation to secure the brake unit on top of the dome lasted a day and a half.

The rest of April was taken up welding nine ribs weighing 150kg each to either side of the shutter. In total the project lasted six weeks and did not hinder observing.

The four brake callipers located at either side of the shutter use large springs to provide the braking force that is applied to the disc plates on the shutter.

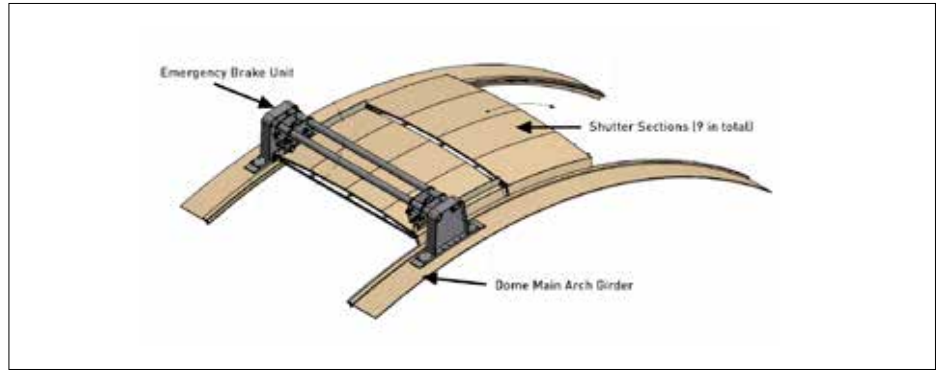

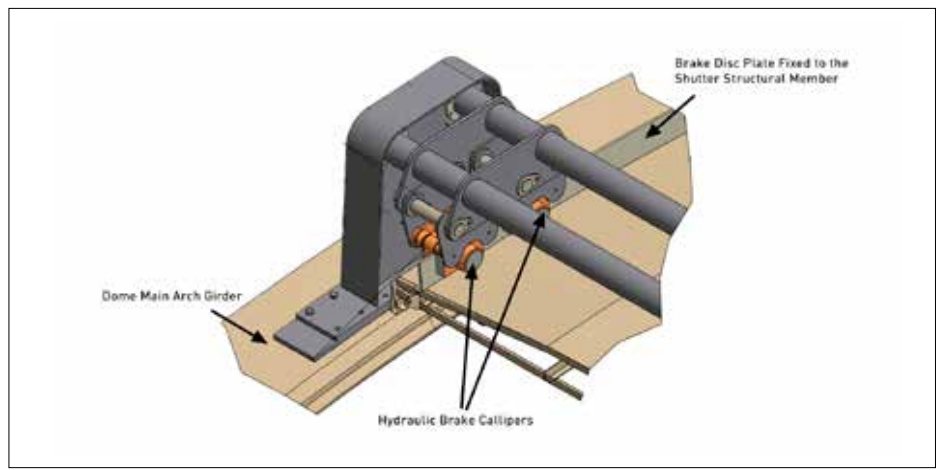

When the shutter receives a signal to open, hydraulic pressure builds up in the brake system to release the brakes. When all four brakes are released a signal is then sent to the drive system to drive the shutter to the desired position. Once at the shutter is at the desired position, the drive system drops out, the hydraulic pressure is released and the emergency brake are applied. Should a failure of the drive system occur, two encoders, one either side of the shutter monitor the shutter speed, if either one detects over speed the drive is disengaged and the emergency brakes are applied.

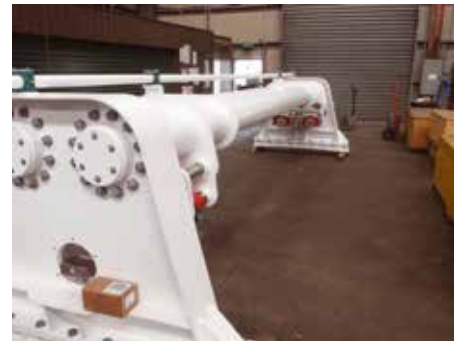

The dome shutter emergency brake before hoisting onto the top of the AAT dome. The brake itself weighs four tonnes.

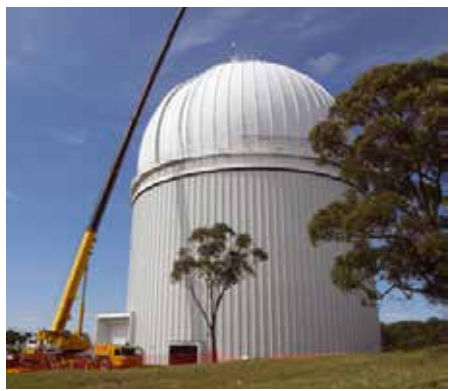

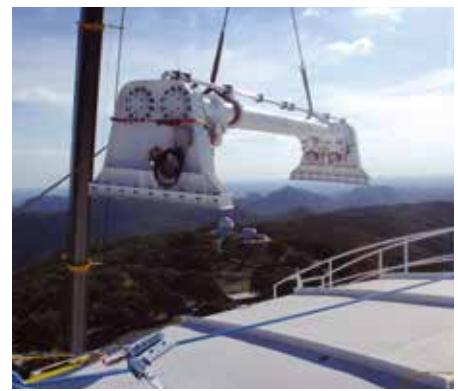

# AusGO CORNER

Stuart Ryder (Australian Gemini Office, AAO)

**AusGO Strategic Plan 2012–2015**

As part of the AAO's Forward Look process, AusGO has been tasked with carrying out a strategic planning exercise on the future staffing, resourcing, and mission of AusGO. This was motivated by a number of factors including the recent transition of AusGO funding from the National Collaborative Research Infrastructure Strategy (NCRIS) to AAO's operations funding; the imminent completion of the terms of the incumbent Deputy Gemini Scientists and Magellan Fellows; and the need to plan for a potential change in the portfolio of large offshore telescopes supported by the AAO, involving one or more of Gemini, Magellan, ESO, and ultimately GMT.

Throughout April and May 2012 AusGO staff participated in a series of strategic planning sessions examining our core mission; stakeholder needs; a Strengths, Weaknesses, Opportunities, and Threats (SWOT) analysis; and finally translating all these into a set of goals, objectives, strategies, and tactics. We identified six key goals:

1. Enhance scientific productivity and impact of Australian use of Gemini and Magellan;
2. Add extra value to the Australian Gemini and Magellan users' experience;
3. Expand scope of AusGO service provision to meet future needs;
4. Enhance AusGO knowledge base and staff development;
5. Raise profile of AusGO within the Australian community; and
6. To be recognised as the best National Gemini Office within Gemini.

To achieve these goals AusGO will:

- Organise a workshop on Australian Gemini and Magellan science highlights, followed by a new large program strategy session.
- Streamline the proposal process and encourage ATAC to set aside a fraction of time on each telescope for large, multi-telescope programs.
- Fund student travel to Gemini sites in conjunction with allocated queue or scheduled classical programs.
- Participate in trials of remote eavesdropping with Gemini and Magellan, leading to more frequent use in future.
- Develop data reduction cookbooks and tutorials based on actual Gemini and Magellan data.
- Initiate discussions with ESO and organisations like FINCA (Finnish Centre for Astronomy with ESO).
- Look to redefine AAT, Gemini, and Magellan support positions from instrument-specific (UCLES/UHRF, GMOS, etc.), to mode-specific ("high-resolution spectroscopy"; "adaptive optics", etc.)
- AusGO staff will be cross-trained on Magellan when visiting Gemini South.
- Trade services with, or pay other NGOs for their expertise in critical areas.
- Implement an RSS feed for news items.
- Solicit a host institution for one Deputy Gemini Scientist external to AAO.
- Identify a more appropriate name for AusGO, and job titles for AusGO staff.
- Collect feedback from Gemini queue users.
- Organise an Australia-wide roadshow of new observing capabilities in early-2013.
- Maintain and develop Joint Proposals Database, new Phase 2 checklists, data reduction cookbooks, and AAO-developed software.

More details on these and other initiatives will be released in due course but if you have any comments or suggestions on these then please contact us at ausgo@aao.gov.au.

## Proposal Statistics

For Semester 2012B ATAC received a total of 34 Gemini proposals, of which 16 were for time on Gemini North, 3 for exchange time on Subaru, 9 were for time on Gemini South, and 6 were for time on both Gemini North and Gemini South. The oversubscription for Gemini North went from 3.2 in 2012A to 2.5, while demand for Gemini South was up from 1.3 in 2012A to 1.6. Although total demand was up 15% on 2012A, the amount of science time available also increased by 35%. Magellan time in 2012B was oversubscribed by a factor 4.6, with 15 proposals (2 more than in 2012A). MIKE is still the most-requested instrument, followed by IMACS and FourStar.

In 2011B, all but one of the ten Band 1 programs were completed or had insufficient Target of Opportunity (ToO) triggers; all 8 Band 2 programs were completed/triggered; and just 2 of 5 Band 3 programs were completed/triggered. The fraction of Australian time which went to programs that were completed dropped to 74%, due in part to the large amount of unused ToO time in 2011B (over 45 hours). From Semester 2012B each partner will be charged for half of any unused ToO time so as to cut down on the amount of unused time effectively "banked" to future semesters.

## Funding for Magellan and Gemini activities

Since the conclusion of the Access to Major Research Facilities Program in mid-2011, the AAO has been reimbursing observers allocated time on Magellan by ATAC for the cost of their travel. AAL has recently indicated that they would no longer sponsor the costs of the Australian Gemini Undergraduate Summer Studentship (AGUSS) program from their Overseas Optical Reserve. AusGO is pleased to announce that the AAO and AAL have secured funding from the Department of Industry, Innovation, Science, Research and Tertiary Education sufficient to cover the costs of all ATAC-allocated Magellan/Gemini classical observer travel, as well as AGUSS in FY 2012/13. An announcement on extending Australian access to Magellan beyond mid-2013 is expected shortly from AAL. The application deadline for the 2012/13 AGUSS program is Friday 31 August 2012 – please direct any interested Australian undergraduate students to http://ausgo.aao.gov.au/aguss.html.

## Gemini Science and User Meeting 2012

Gemini held its 4th Science and User Meeting in San Francisco from 17–20 July 2012. Australia was well-represented at this meeting with 10 attendees: Stuart

Ryder, Chris Lidman, Alan Alves-Brito, Karl Glazebrook, Lee Spitler, Sarah Brough, Warrick Couch, Amanda Bauer, Sebastian Haan, and Andy Sheinis. The first five listed all gave talks; Karl and Sarah also represented Australia's interests at meetings of the Science and Technology Advisory Committee (STAC), and Users Committee for Gemini (UCG – see below) respectively. An important aspect of this particular meeting was the 3 sessions dedicated to initiating a long-range development plan for Gemini.

## Users Committee for Gemini

With the transition of the former Gemini Science Committee into the STAC, Gemini has also convened its own Users Committee for Gemini (UCG). The role of the UCG is "To provide feedback to the Gemini Observatory on all areas of its operations that affect current users of the facility, based on the experience of the committee members as well as input collected from the larger community of Gemini users. The Observatory will use this information to improve the service it provides to users." The UCG's charter and membership is available at http://www.gemini.edu/science/#ucg. At the recommendation of AAL's Optical Telescopes Advisory Committee, Sarah Brough from the AAO has been appointed as Australia's representative on the UCG for a term of 3 years. While the roles of the UCG and the AAO Users Committee will inevitably overlap to some degree, it is AusGO's expectation that user issues regarding and under the control of AusGO should continue to be directed to the AAOUC, while anything specific to the Gemini Observatory and its performance are best directed to the UCG. Anyone with comments or issues they would like to have discussed by the UCG should contact Sarah at Sarah.Brough@aao.gov.au.

## Gemini Forum at the ASA ASM

During the recent ASA Annual Scientific Meeting held at UNSW, AusGO arranged for a Gemini Forum during one of the lunch breaks attended by 30–40 people. Gemini's Associate Director for Operations Dr Andy Adamson attended the ASA meeting, and during the forum presented an update on Gemini facilities and the transition to the UK withdrawal at the end of 2012 (see photo above). Stuart Ryder gave an overview of the AusGO Strategic Plan. Karl Glazebrook (STAC rep) and Sarah Brough (UCG rep) also attended, and

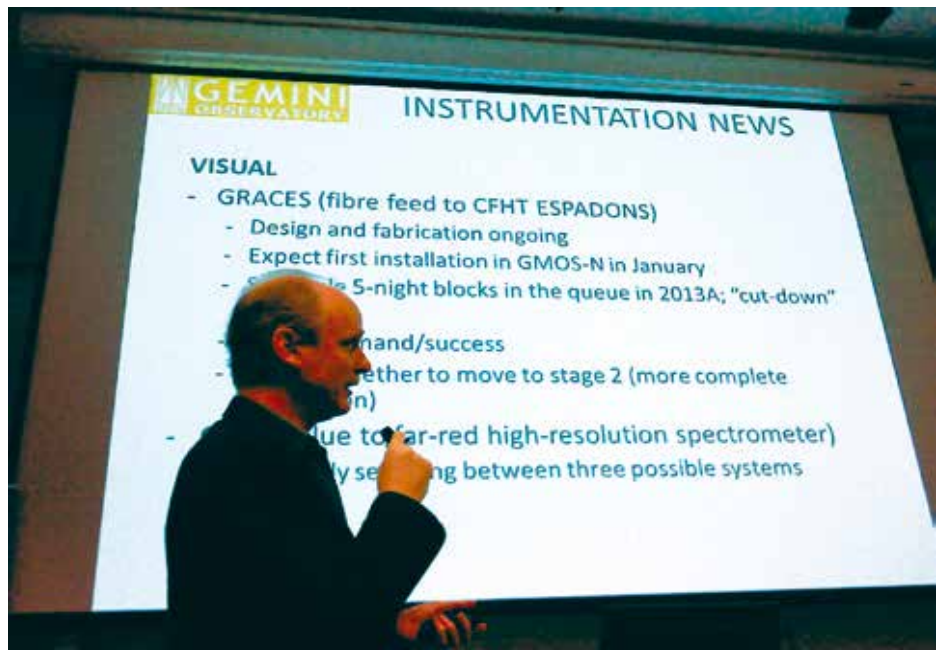

Dr Andy Adamson from Gemini Observatory gives an update on upcoming instrumentation at the July 2012 ASA meeting. Image credit: Stuart Ryder solicited input from the community for the Gemini Science and User Meeting.

## Contest and Imaging Tutorial

For the past 4 years AusGO has run the Australian Gemini School Astronomy Contest. The winning entry for 2012 came from Ryan Soares at Trinity College in Perth to observe the interacting galaxy pair NGC 7232/7233. Observations of this target are now in the Gemini South queue. The runners-up were The Heights School in Adelaide, and the Astronomy Club from St. Margaret's Anglican Girls School in Brisbane.

Prof. Travis Rector (University of Alaska Anchorage) is an expert in turning multi-filter Gemini images into "true colour" renditions, and has assisted AusGO in producing all the final contest images. He has produced an excellent tutorial exercise demonstrating how the 2009 image of the "Glowing Eye" planetary nebula was created, and giving anyone the opportunity to try making their own from the reduced individual images using the "FITS Liberator" software and Photoshop package. Contest coordinator Dr Chris Onken has produced a similar exercise for use with the free GIMP package. Users are encouraged to upload their own results to a Flickr gallery. The tutorial is available at http://ausgo.aao.gov.au/ImageTutorial/.

## Instrumentation Update

- FLAMINGOS-2: Following the catastrophic failure of the main collimator lens in January just as System Verification was about to begin, Gemini has procured both a replacement and a spare lens. All lens mountings have been redesigned to reduce the stresses during warm-up and cool-down. An improved thermal mount for the detector has also been installed. Recommissioning is planned for late-2012B.

- GeMS/GSAOI: GSAOI and the Canopus optical bench are in shutdown mode for the winter and while various upgrades are carried out, in particular improvements to the Natural Guide Star Wavefront Sensor to improve its sensitivity. Although inclement weather in the April commissioning block prevented observations of any Community Commissioning Targets, a GSAOI cooling issue in March provided an opportunity to test the potential of GeMS in the optical using GMOS. Excellent performance was obtained, with FWHM as good as 0.08 arcsec in the i' band.

- GMOS CCDs: A noise issue with the new Hamamatsu CCDs has been resolved. Replacement of the interim E2V deep-depletion CCDs in GMOS North with Hamamatsu CCDs should occur early in 2013, with installation in GMOS South to follow as soon as FLAMINGOS-2 is ready to play a major role in queue observations.

- GPI: Assembly and integration of the Gemini Planet Imager is well-advanced at UC Santa Cruz.

The failure of an actuator in the deformable mirror will delay delivery to Gemini South until mid-2013.

- GHOS: Following the Concept Design Review for the Gemini High-resolution Optical Spectrograph at the end of June, an announcement on which of the 3 designs (one of them led by the AAO and RSAA) should proceed to the Preliminary Design phase is expected shortly.

- GRACES: Testing of a fibre-feed system from Gemini North to the ESPaDOnS spectrograph at CFHT is underway. A limited block of observing time may be offered in 2013A once commissioning results are available.

- Michelle/T-ReCS: As flagged in recent calls for proposals, both Michelle and T-ReCS will no longer be offered for community use after 2012B. This is driven by a combination of low demand, and the need to move to a maximum of 4 instruments + AO at each telescope in the post-UK withdrawal environment.

- GIROS: Given the modest response to the call for White Papers on science cases for the Gemini Infra-Red and Optical Spectrometer and a number of other uncertainties, the STAC is not prepared to recommend moving forward with GIROS at this time.

### Remote eavesdropping

The Gemini Observatory has accepted a recommendation from the STAC that they offer remote eavesdropping by PIs of highly-ranked queue programs, to engage users more closely with observing and provide real-time input, such as assisting with difficult acquisitions or deciding on whether sufficient signal-to-noise has been achieved. Each NGO (including AusGO) has identified a subset of Band 1 programs with targets accessible late in 2012B for trials of remote eavesdropping. For Australian PIs this should be a particularly appealing option, as the observing night at Gemini South is pretty well aligned with the standard working day in Australia, while a Gemini North observer would not need to stay up beyond ~2am. It is envisaged that during this trial AAO/Sydney-based PIs and their AusGO contact scientist would be on stand-by with a few hours' notice that there was a high likelihood of their program being executed (conditions permitting), then would connect to the Gemini control room via the AAO's Polycom videoconferencing system while the observations are underway. Magellan Fellow Dr Francesco Di Mille will be conducting similar remote eavesdropping trials from Sydney to Las Campanas later this year, allowing Australia-based collaborators to participate in Magellan observations while one of their team members is at the telescope.

# DECam available to Australian astronomers
Andrew Hopkins (AAO)

The AAO and NOAO/CTIO have entered into an ongoing agreement to trade observing time between the AAT and the Blanco telescope at CTIO.

Following the initial allocation of 5 nights in semester 12B traded between the AAT and CTIO, the AAO and NOAO/CTIO are pleased to announce that we are continuing to offer time in coming semesters, to allow our respective communities to maximise the scientific facilities and opportunities to which we have access.

In Semester 13A there will be 15 nights available to the Australian community to apply for time on the 4m Blanco telescope using the DECam imager. The time-swap agreement allows for 20 nights per year to be traded between the Blanco and the AAT.

For Australian users of the Blanco, this will be distributed as 15 nights in the A semester, and 5 in the B semester. This balance is due to the demand on DECam by the Dark Energy Survey in B semester time. For NOAO applicants proposing to use the AAT, the time will be distributed as 10 nights each semester.

The Blanco nights available to the Australian community will be announced in the regular AAT Call for Proposals. Users should submit their observing proposals through the regular AAT online proposal form, selecting CTIO/DECam in the "instrument" section. Proposals will be assessed by ATAC, and the graded proposals provided to CTIO for scheduling. The deadline to apply for observing time in 13A is 5pm AEST Friday 14 September 2012.

From Semester 13A onwards, DECam time for Australian users will be allocated in "classical" observing mode, with appropriate telescope support being provided by CTIO.

More information about DECam can be found here:

www.ctio.noao.edu/noao/content/dark-energy-camera-decam

Details about obtaining AAT observing time can be found here:

www.aao.gov.au/astro/obs.html

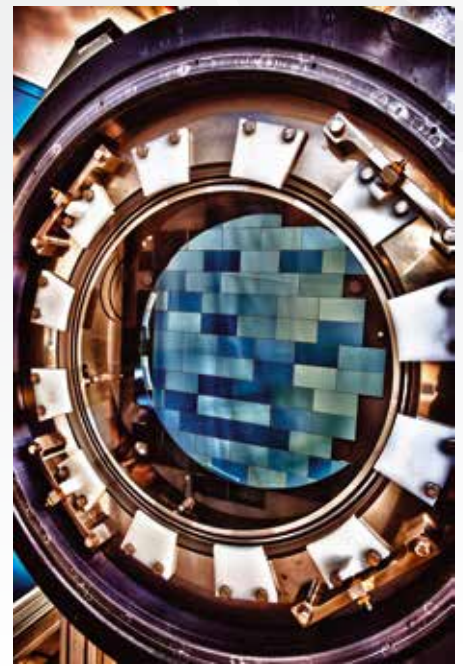

The Dark Energy Camera, complete with 74 science-grade CCDs. Ready to observe 300 million galaxies and discover thousands of supernovae. Credit: Reidar Hahn/Fermilab.

**OBSERVATORY** NEWS

# Transit of Venus photos

James Gilbert (AAO)

*In a year with several major astronomical events, the transit of Venus was a highlight for many AAO employees. The transit lasted most of the day on Wednesday, 6 June in Australia, and there was plenty of interest in the last chance to see this spectacle for 105 years. Below, James Gilbert tells us the story of his transit experience, along with both his own photos, and those from other AAO members observing the transit in different parts of Australia.*

I have to admit I completely forgot about the transit as I was booking flights to Cairns for a June holiday. As it happened, I would land in north Queensland the day before the event, and so with less than two weeks to go I frantically ordered some solar filter film online and checked the weather forecast. Cairns was given a fifty per cent chance of cloud cover, Sydney significantly less. It was pretty disappointing – having not seen the 2004 transit, I'd told myself months earlier that I would make a special effort to not only view but photograph the entire event and make a time lapse movie of it. Still, despite the short notice and iffy weather I thought I'd try my luck at hiring the gear I needed to do so. Incredibly I managed to find a suitable telephoto lens that wasn't booked out, so I snapped it up and hoped for the best.

Port Douglas, a beautiful place, had just been on the receiving end of torrential downpours when my better half and I arrived on the evening of June 5. Perhaps it was our British blood that told us to ignore the rain, go to the beach, and – in the dark – find a suitable spot to set up early next morning in spite of it all. Well, thank heavens we did: Wednesday June 6 dawned without a cloud to be seen. From our little camp among (but not underneath) the coconut palms, I rigged up my camera and settled in for the next seven hours.

My setup was a Canon 50D digital SLR and a Canon 400 mm f/5.6 L lens with a 2x extender. This amounted to an 800 mm f/11 lens, which with the 50D's sensor gave a field of view about three suns wide. The optical quality of this combination wasn't brilliant, but it was cheap, small, and most importantly

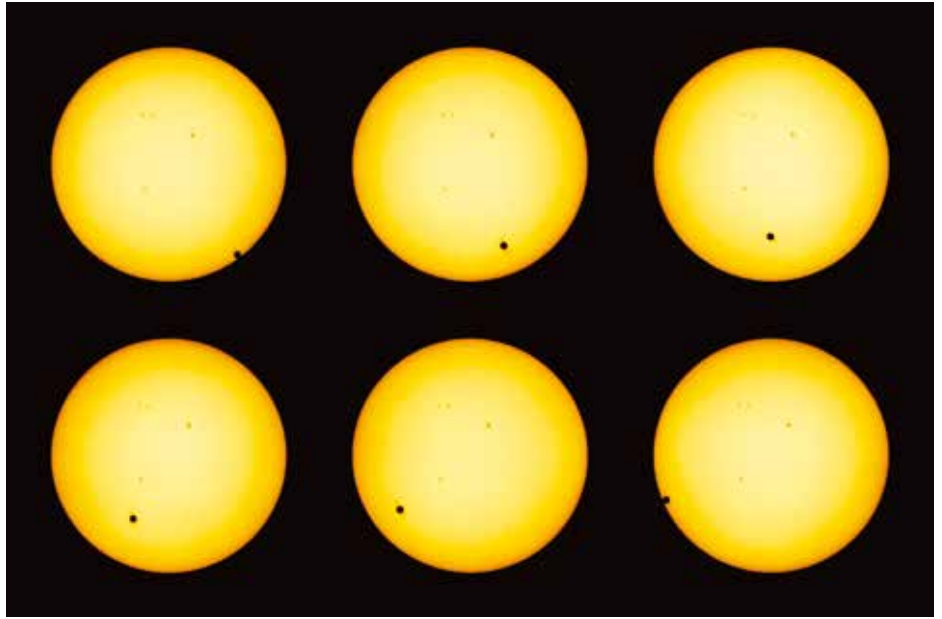

The transit of Venus 2012, as seen by James Gilbert. Six frames at equal intervals throughout the transit, starting here with second contact (top left) at 8.34am, then along each row until third contact (bottom right) at 2.27pm. To give it an Australian flavour, the sun is shown in 'southern hemisphere mode', meaning that its south pole is at the top. The photos were white light images which were given a sunny false colour for added pop.

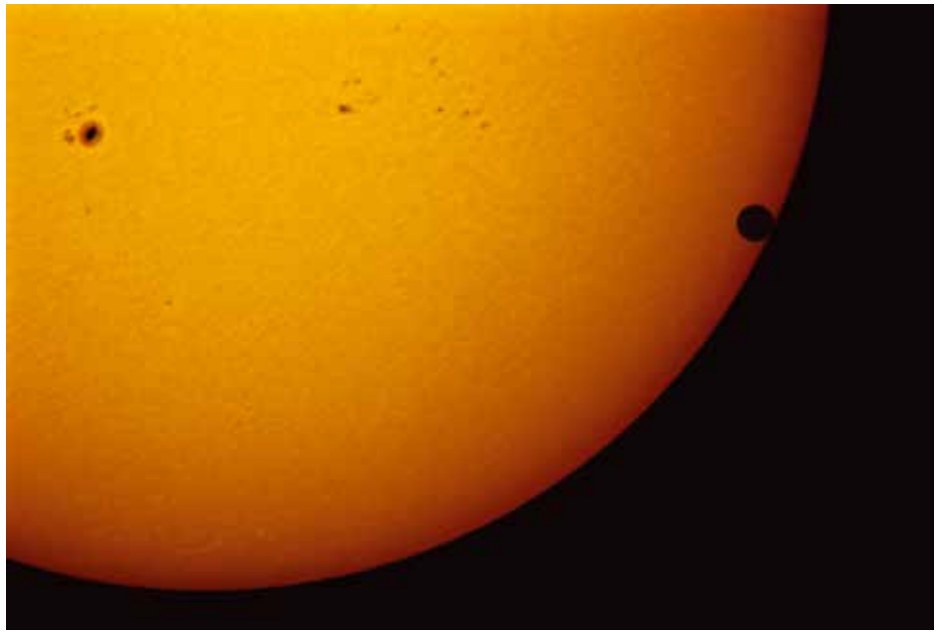

This stacked image shows the second contact of the last Transit of Venus. Ángel López-Sánchez used a Skywatcher Black Diamond Telescope, with an aperture of D = 80 mm and focal f = 600 mm, with a 20 mm eyepiece projection feeding a CANON EOS 600D. This image actually stacks 25 frames, the stacking was performed using the freeware Lykeos software for Mac. Details of the sunspots, the granulation of the Sun and the limb darkening, can be appreciated in this image.

Credit of the image: Ángel R. López-Sánchez (Australian Astronomical Observatory / Macquarie University) and Enric Pallé & Roberto López López (Instituto de Astrofísica de Canarias).

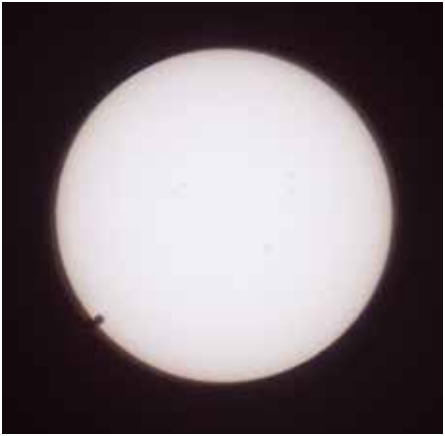

An unprocessed photo, as seen by James Gilbert's camera (Canon 50D, 800 mm at f/16, 1/200 s, ISO 250).

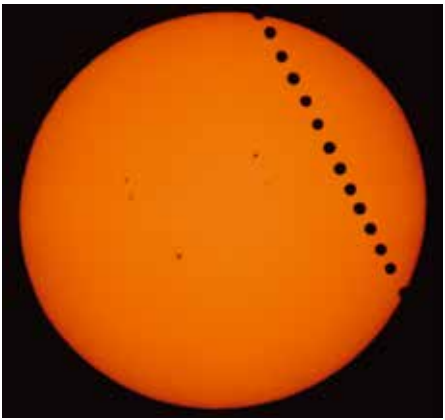

Ángel R. López-Sánchez took this sequence of the transit from the Warramunga Seismic Station near Tennant Creek (Northern Territory, Australia). This image shows Venus in 14 different positions over the solar disk, each separated by about 30 minutes. Ángel took over 3000 images of the transit. The site was the destination of the joint IAC (Instituto de Astrofísica de Canarias) / AAO Scientific Expedition to study the Transit of Venus. Credit of the image: Ángel R. López-Sánchez (Australian Astronomical Observatory / Macquarie University) and Enric Pallé & Roberto López López (Instituto de Astrofísica de Canarias).

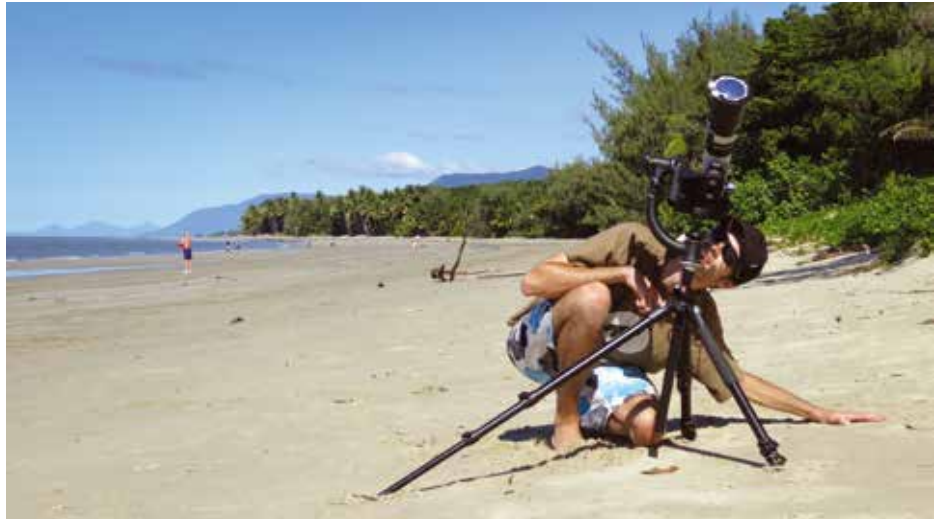

Not a bad observatory: James Gilbert photographing the transit from Four Mile Beach in Port Douglas, Queensland.

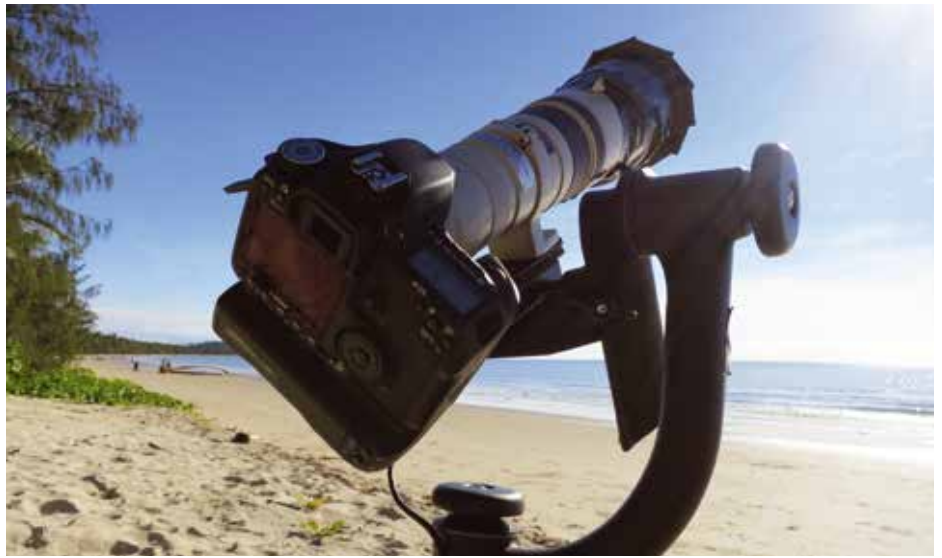

James Gilbert's setup was a Canon 50D digital SLR, a Canon 400 mm f/5.6 L lens with a 2x extender, and a homemade Baader film solar filter gaffa-taped to the front of the lens hood.

it was available at the time. I'd made a solar filter for the front of the lens using Baader solar film and I also hired a gimbal tripod head to make tracking the sun (by hand!) a little easier. I used a timer remote to take a photo every 60 seconds, giving me over 400 frames to stitch together for a time lapse movie. Not a single cloud got in the way that day – I was extremely lucky.

Something I really enjoyed, and that I didn't expect, was talking to passers-by about what I was doing. A surprising number of people didn't know about the transit at all, and almost all of the ones that did had left home that morning thinking they wouldn't get to see it for themselves. It was terrific to give them that opportunity. Needless to say I left the beach that afternoon with a smile on my face.

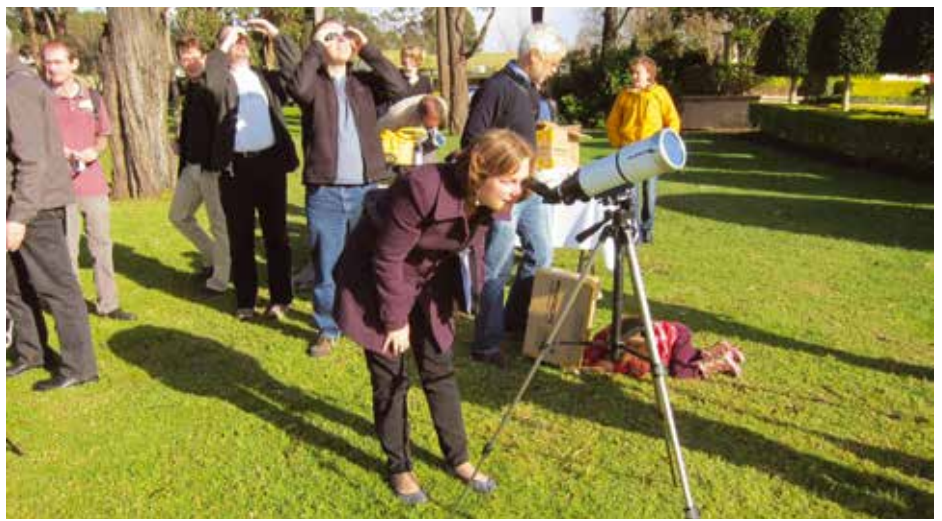

During the morning tea break, Amanda Bauer has a good look at the transit of Venus already in progress from the lawn of the Grand Mercure, Hunter Valley. She was attending the Southern Cross conference (see page 18 of this issue). In the background, Scott Croom also has a look with specially designed transit viewing sunglasses. Rob Hollow brought along a collection of telescopes and other equipment to entertain the conference goers between clouds for most of the day.

# Farewell to a distinguished observer

Fred Watson (AAO)

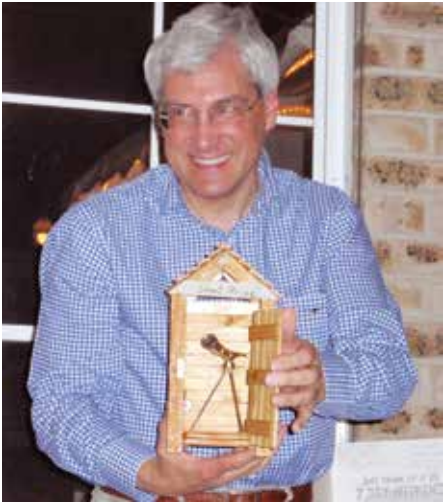

Paul Cass holds up the ultimate pun on the name of the Schmidt Telescope, crafted by Mick Kanonczuk for the occasion of Paul's retirement. (Helen Goodyear.)

The end of July saw a significant milestone in the history of the UK Schmidt Telescope – the retirement of Paul Cass. Paul's earlier career was in science administration, a role that originally brought him to Siding Spring in 1981 as the UKST's Administrative Officer. Following spells back in the UK and at the Isaac Newton Group of Telescopes in La Palma, Paul returned to Coonabarabran in 1991 as a Schmidt photographic specialist, but quickly graduated to become a fully-fledged observer. He has ably fulfilled that role ever since, and he made the transition from photographic to multi-fibre observing in the early 2000s with flair and panache. Having the twin strings of astronomy and administration to his bow, Paul made a value-added contribution to the work of the AAO, including 6dF data archiving and distribution, and playing a key role in the Observatory's Dark Skies Committee. He will be missed by all at Siding Spring.

Paul's retirement party at the Acacia Motor Lodge on 6 July brought another of his talents to the fore. Not many of today's AAO staff members would have known that both Paul's parents were on the stage, and in the past, his theatrical pedigree was often evident in his performances with the Coonabarabran Amateur Dramatic Society. That legacy surfaced once more in Paul's performance of his 'Ode to the 6dF Robot'. Reprinted here, it is a heartfelt plea that will be familiar to anyone who has spent the night with one of the AAO's delinquent fibre positioning robots...

### The Schmidt Observer's Ode to the 6dF Robot (with apologies to William Blake)

By Paul Cass

Robot, Robot, do it right
lest I curse you through the night.
Place those buttons, one by one,
accurately, stamping none.

What's that hissing ? What's that clunk?
Don't pretend you like steam-punk!
What the theta ? what the r?
How many microns moved too far?

Please don't stop or hesitate
whilst I unload the previous plate.
The telescope is working fine;
the sky is clear - is this a sign?

Thank you Robot : work well done.
Do the next field now : oh what fun!
Take it easy : no vibration –
let's do a good configuration.

When from RAVE stars photons vector
through the fibres to detector,
did he smile his work to see?
Did he who made OzPoz, make thee?

Robot, Robot, do it right,
lest I curse you through the night.
Park those buttons, one by one.
No looping fibres - no, not one.

# A 2dF night at the Anglo-Australian Telescope

Ángel R. López-Sánchez (AAO/Macquarie University)

One of the most complex astronomical instruments nowadays available is the Two-Degree Field (2dF) system on the AAT. The main part of 2dF is a robotic gantry, which can position up to 400 optical fibers on astronomical objects anywhere within a two-degree field of the sky. These fibres feed into the AAOmega spectrograph, providing a broad range of spectral resolutions and wavelength ranges. 2dF was designed at the AAO in the 1990s and has since been used by many astronomers around the world. Indeed, this sophisticated instrument has conducted observations for hundreds of astronomical projects, including galaxy surveys such as the 2dF Galaxy Redshift Survey, the WiggleZ Dark Energy Survey and the Galaxy And Mass Assembly survey (which is still on going). In a clear night, 2dF can obtain high-quality optical spectroscopic data of more than 2,800 objects.

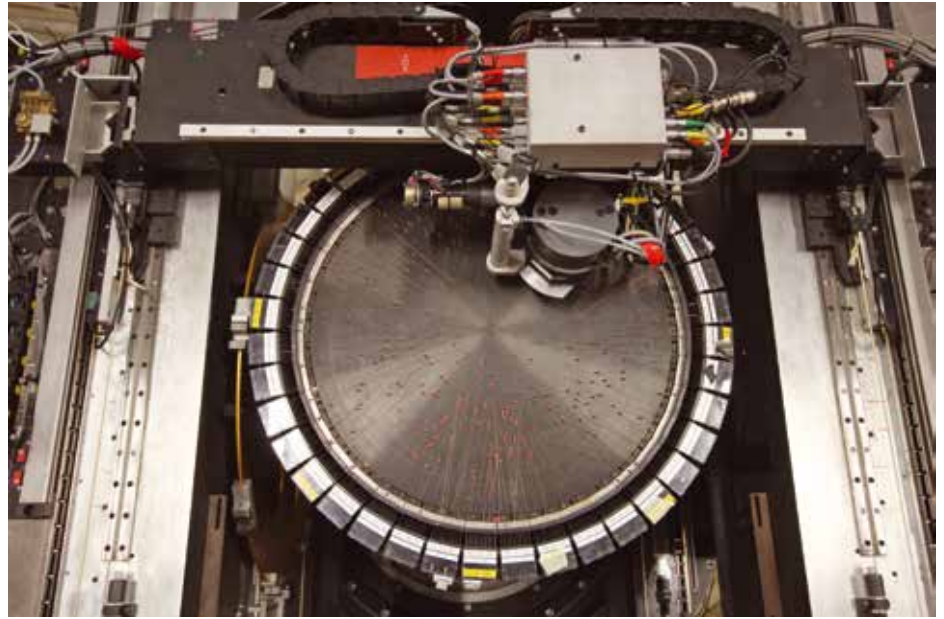

The 2dF robotic fibre positioner as featured in the film. The tips of the fibres are lit red to the robot can see them in the dark as it moves them around the circular magnetic surface of the telescope's 2-degree focal plane. (Credit: Ángel R. López-Sánchez)

How does 2dF move and position the optical fibers? A very nice way of explain it is using the time-lapse technique, that is, taking many images and then adding all to get a movie of the robot while moving and positioning the fibers. That is why I decided to create the video, A 2dF night at the AAT, which assembles 14 new time-lapse sequences taken at the AAT. This new time-lapse video shows not only how 2dF works but also how the AAT and the dome move and the beauty of the Southern Sky in spring and summer. The film, which combines more than 4000 still frames, is about 3 minutes. The individual frames were captured with a CANON EOS 600D fitted with a 10-20mm wide-angle lens. The sequences were taken during September and November 2011 while I was working as support astronomer for the 2dF instrument.

The video consists in three kinds of sequences created at 24 frames per second (fps). The first 3 sequences show how the 2dF robot gantry moves the optical fibers over a plate located at the primary focus of the telescope. Although in real life 2dF needs around 40-45 minutes to configure a full field with 400 fibers, the process is compressed by time-lapse. The first 2 sequences have been assembled taking 1 exposure per second, therefore 1 second of the video corresponds to 24 seconds in real life. The third sequence considers an exposure each 3 seconds, and hence it shows the robot moving very quickly. The next four sequences show the movement of the telescope and the dome. All of them were obtained taking 2 images per second (a second in the movie corresponds to 12 seconds in real life). The long black tube located at the primary focus of the telescope is 2dF. The remaining sequences, all obtained during the night, were created taking exposures of 30 seconds, and hence each second in the video corresponds to 12 minutes in real life.

As in my previous time-lapse video, The Sky over the AAT (see issue 121, page 17), I have included time lapses of the night sky. This time I paid particular attention to the colours of the stars, a detail often lost in this kind of processing. Some sequences required more than 12 hours of computer time to process, including 3 or 4 iterations per sequence, to get a good combination of low noise and fine detail of the sky. Aldebaran and Betelgeuse appear clearly red, while the stars in the Pleiades and Rigel have a blue color. Other details to watch out for are satellites and airplanes crossing the sky, the nebular emission of the Orion and Carina nebulae, the moonlight entering in the AAT dome, and kangaroos jumping around the telescope. Finally, I chose an energetic soundtrack which moves with both 2dF and the sky. It is the theme "Blue Raider" of the group "Epic Soul Factory", by the composer Cesc Villà. Actually, all sequences were created to fit the changes in the music, something that also gave me some headaches. But I think the result is worth all the effort.

# The AAO Headquarters move to North Ryde

Andy Green (AAO)

As part of the AAO's incorporation within the Department of Innovation, Industry, Science, Research and Tertiary Education, we were given space for our headquarters in a new Commonwealth building, shared with the National Measurement Institute. The state of the art building features large spaces for instrument assembly, integration and testing, a large electronics workshop, as well as optics and metrology labs, dedicated IT assembly labs and server rooms. There are new offices and open plan seating for astronomers and support staff, and the building features modern facilities and nearby public transit via train and bus. This has been an excellent demonstration of Department's strong support for the AAO.

The primary move from our old premises in Marsfield/Epping was undertaken over the weekend of 10-13 August. First to move was the library in the week before, with the staff offices and IT equipment moved over the weekend. Staff arrived Monday morning to receive new identification badges and a small gift from the Director. Naturally, the move has involved a lot of resettling, but the dedicated efforts of both AAO and Department staff much of the transition has been smooth.

Remaining in the old premises are the HERMES Instrument and the team working to complete that for delivery to the AAT early in 2013. This allows them to continue largely uninterrupted in preparing that instrument, and those remaining staff will be moved to the new building once the instrument has been delivered.

Because of our co-location with the National Measurement Institute, there is increased security at our new building.

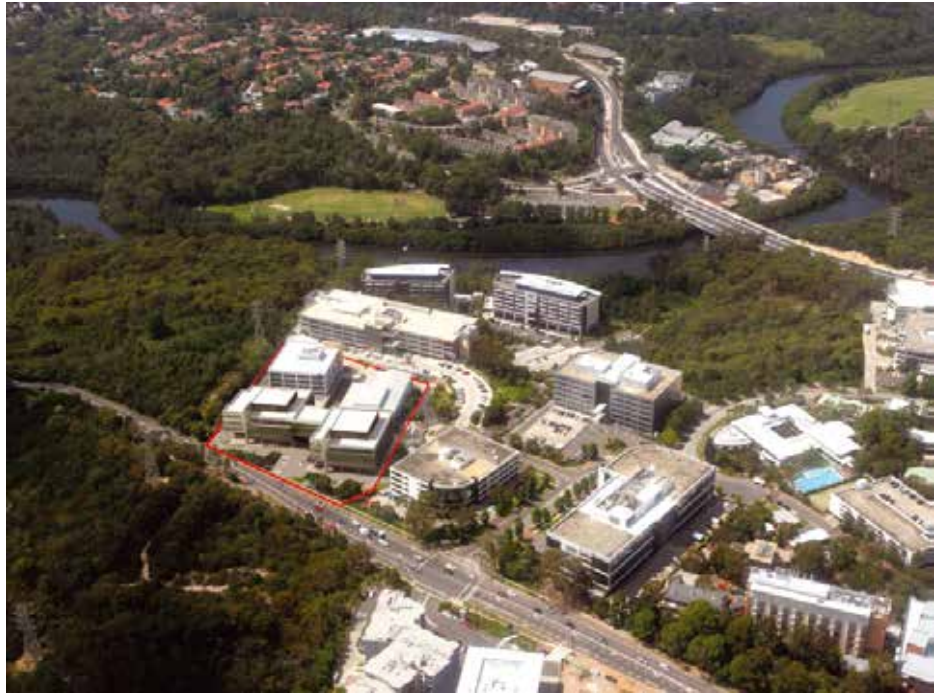

An arial photo of the Dehli road complex. The AAO is located in the building at the bottom left, which fronts onto Delhi Road. Epping Road can be seen at the top right, crossing the Lane Cove River. Much of the campus is surrounded by the Lane Cove River National Park.

Visitors should, if at all possible, contact the person they are visiting before they arrive. Both following the move, and while some staff remain in Marsfield, some patience from the community will be greatly appreciated while we settle into our new home.

We will retain many fond memories of our old building, which was the AAO's first home.

New Offices Address:
105 Delhi Road
North Ryde NSW 2113

New Postal Address
PO Box 915
North Ryde NSW 1670
Australia

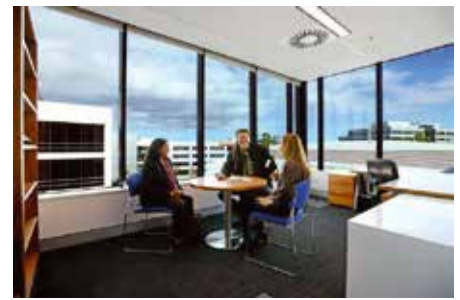

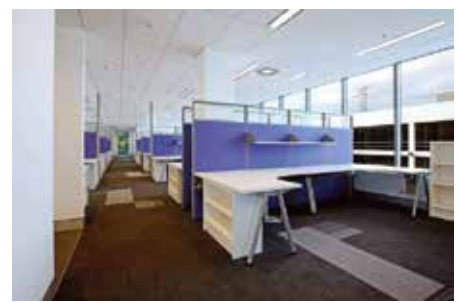

Views of a corner office and open plan workstations in the AAO's new Dehli Road headquarters.

# Letter from North Ryde

Andy Green (AAO)

I write to you from the AAO's new Sydney offices located in North Ryde. As this newsletter goes to press, the move is complete and everyone is settling into the new building.

In 2011, Gayandhi De Silva was awarded an **Innovation Scholarship** award granted by DIISRTE. Normally these grants are for overseas conference travel, but fundamentally they are for professional development. Given family responsibilities, Gayandhi is unable to travel regularly, so instead she applied for funding to host a meeting in Sydney to bring experts in her research field of Galactic Archaeology to her. In late July 2012, with supplementary contributions from Macquarie University, Gayandhi led the organisation of the Workshop on Galactic Archaeology Surveys: Past, Present and Future[1]. The general theme was along the lines of What we have learned, What we are now learning, and What we will learn, both in terms of characterizing the Milky Way, as well as technicalities of carrying out successful large scale surveys. This workshop was particularly timely with the imminent start of several large Galactic Archaeology surveys, including the Australian GALAH survey using the HERMES instrument.

## Staff

The astronomy group welcomes its newest member, Iraklis Konstantopoulos to the John Stocker Research Fellowship at the AAO. Iraklis will be involved with the SAMI survey. Iraklis carried out his doctoral studies at University College London and the European Southern Observatory (Santiago, Chile), with a short placement at Gemini Observatory (La Serena, Chile). He joins us after working for three years at The Pennsylvania State University as a Research Associate. His work involves mostly anything to do with star formation, be it in star clusters or larger scales; the interpretation of the star formation history of galaxies; and the dynamical interplay of galaxies in small groupings. Outside the office Iraklis likes to escape to nature for hikes, bike rides, mountain treks, and the odd road trip, with his wife Millie—except for when he goes solo, suits up, and pretends to be a duelist on the fencing strip.

Having been awarded a position at the University of Western Australia, **Chris Springob** will be leaving the AAO at the

---

[1] http://physics.mq.edu.au/astronomy/ga-surveys/

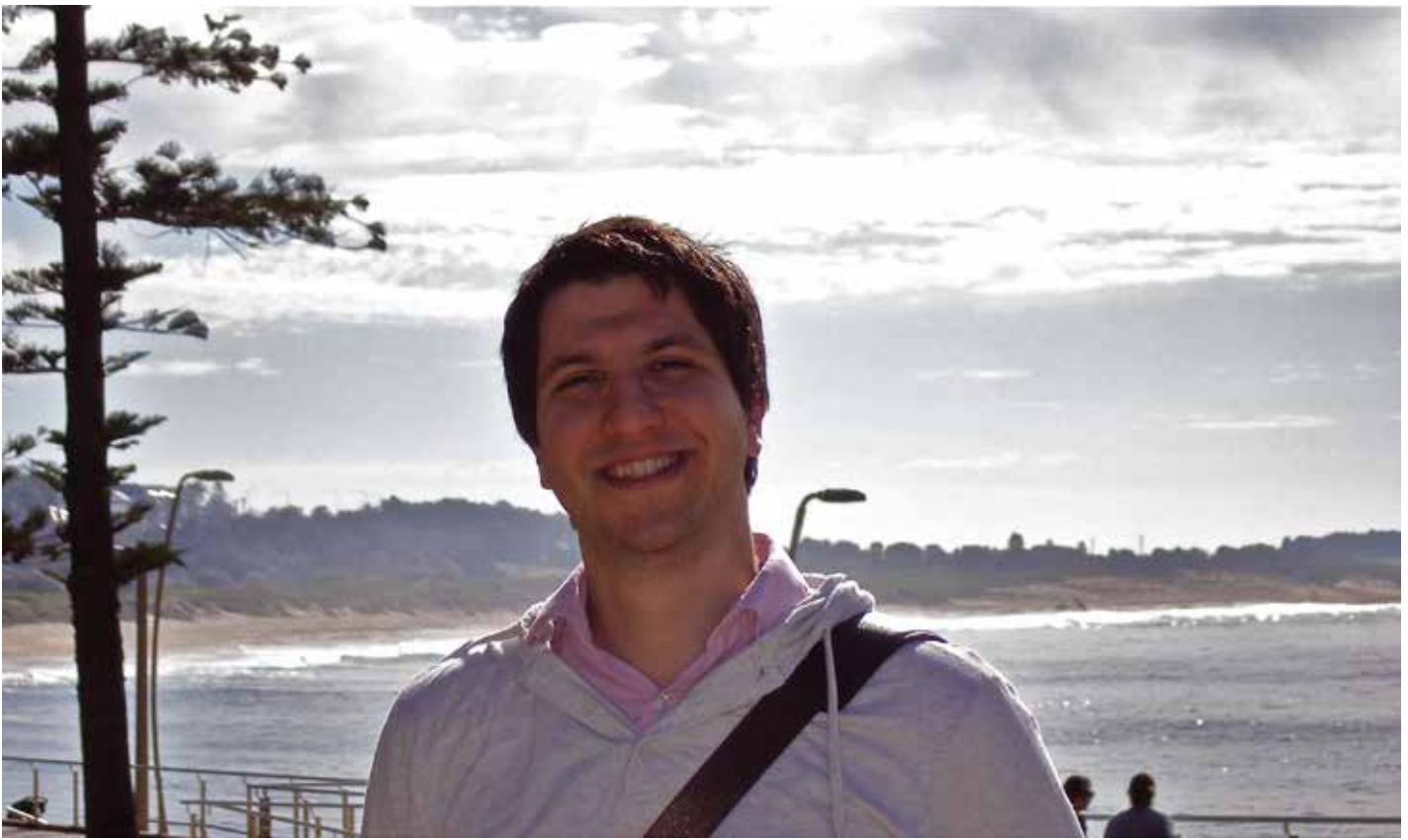

Iraklis Konstantopoulos has recently joined the Astronomy Group at the AAO.

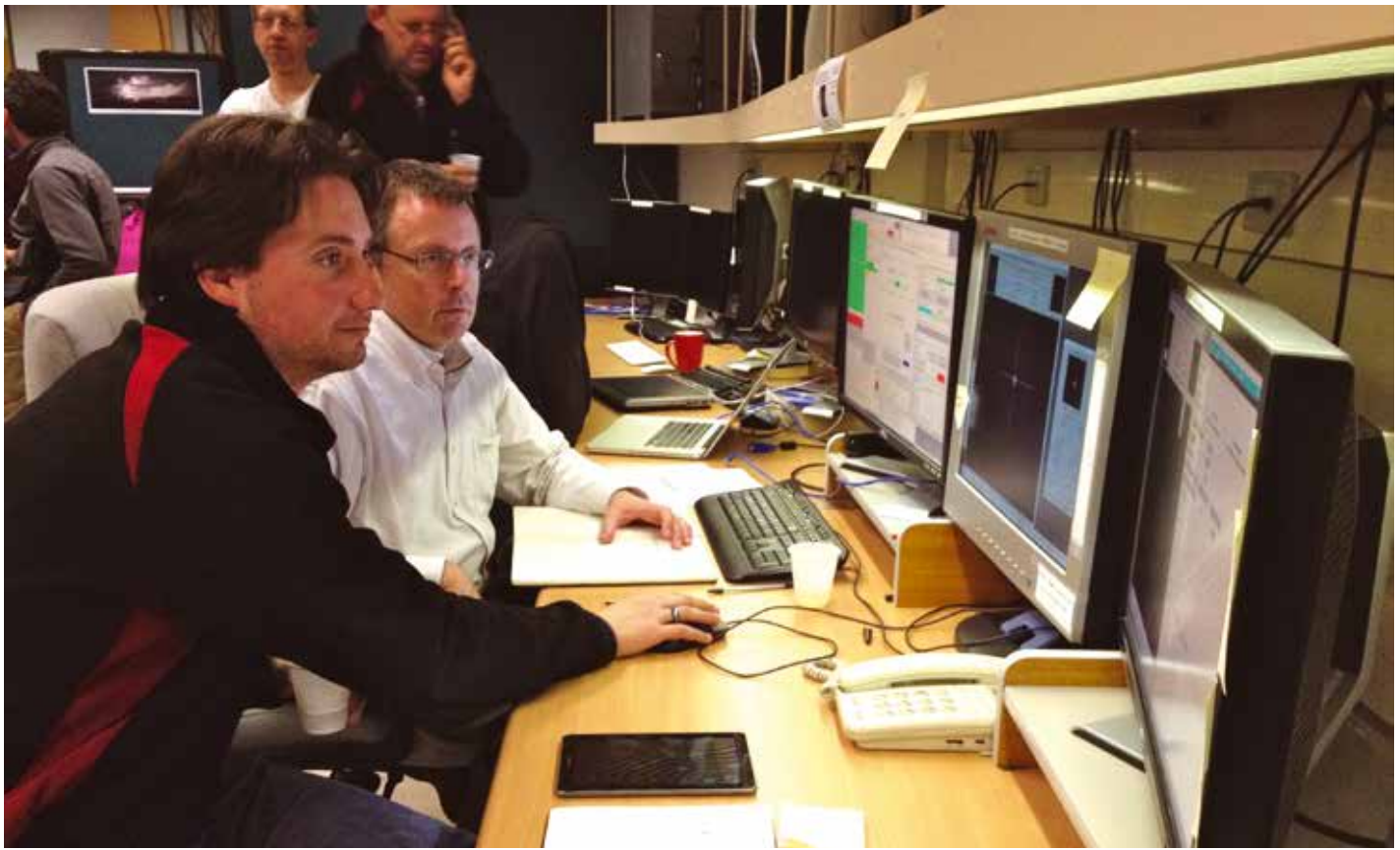

In the AAT control room, Anthony Horton and Chris Tinney review first light images from CYCLOPS2. The instrument is an integral field unit and fibre feed for UCLES, and is mounted to CURE, the new Cassegrain acquisition and guiding system, which was simultaneously commissioned.

start of September. His new position, jointly funded by CAASTRO[2] and the AAO, will involve work on the Tully-Fisher relation in the 2-micron All-Sky Survey (2MASS), as well as continuing work he began at the AAO on the 6dF Galaxy Survey fundamental plane. We wish Chris the best of luck in his new position.

**Katrina Tapia-Sealey** has rejoined us to look after the update of our extensive web presence to improve the overall usability, look and feel, and also to meet department standards. Joining the Instrument Science group is **Robert Content**.

## Instrument Updates

After a considerable effort, both the **AAOmega** web pages and the 2dF+AAOmega Observing procedure have been updated. The latter acts as both a reference for the support astronomer and as a detailed guide for anyone who will be using this instrument.

Since the last edition of the Observer, the 2dF robot unfortunately suffered some major problems with its position encoders. These have all now been replaced by Robert Patterson and Darren Stafford, along with many other older parts and related software. These upgrades are now resulting in an even faster, more reliable robot than reported in the last issue, excellent for both AAOmega and in preparation for HERMES.

In this semester we look forward to the installation of a new fibre run for AAOmega and HERMES, which will result in significantly lower fringing and higher throughput. The AAOmega team will also be releasing a significantly updated version of the 2dFdr data reduction software to the community.

**SAMI** is awaiting a new fibre run and new connectors to address blue throughput issues and make the instrument more user friendly. The team expect to commission these early in 2013, which should allow SAMI to become a facility instrument open to the community.

With some luck, both **CURE** and **CYCLOPS2** had first light on the first night of their commissioning. CURE provides a new instrument mount point, guiding and acquisition capabilities which will be used by both CYCLOPS2 and the upcoming KOALA integral field unit. CURE mounts on the bottom of the existing Acquisition and Guiding Unit at the Cassegrain focus of the AAT.

---

2  Centre for All Sky Astrophysics

# Letters from Coona

During March and through to June of this year we had a few new faces on-site- Caro Barth and Sue Wilson, who were helping Kristin with assembling the fibres for HERMES. It was sad to see Caro and Sue leave and they both will be missed. Kristin got a great shot of Caro and Sue working very hard.

As I'm sure you are all aware Paul Cass retired at the end of June, but before he left we squeezed in a scrumptious morning tea, which was held at the UKST. There were scones, cupcakes and jelly cakes galore, all were really yummy and many of us had to walk back to the AAT to try and drop some of the extra calories that were gained.

Katrina Harley

On Friday 22nd June Siding Spring Observatory hosted a mountain top finish for Stage 3 of the Inaugural Santos North Western Cycling Tour. Visitors including some SSO staff were thrilled to watch 120 male and 40 female cyclists emerge from thick fog and rain as they rode up the final mountain approach to cross the finish line near the Visitor Centre car park. Food was in plentiful supply with the Rotary club putting on a terrific BBQ while hot soup'n'bread and plenty of other goodies were available at the Visitor Centre. Bob Dean manned the local radio station's outside broadcast caravan and managed to scoop an exclusive interview with SBS TV Cycling Central's Mike Tomalaris. The men's race (127km) winner was Mark O'Brien (Budget Forklifts) who took 3 hrs and 15m with Ruth Corset (Pensar-Hawk-Racing) winning the 76km womens race in 2 hrs 31 mins. Stage winners were later awarded jersey's on the mobile stage truck by local Major Peter Shinton and other VIP's. Stage 4 kicked off on Saturday morning with the guys leaving from SSO while the girls departed from Coona cycling to Gunnedah.

John Goodyear

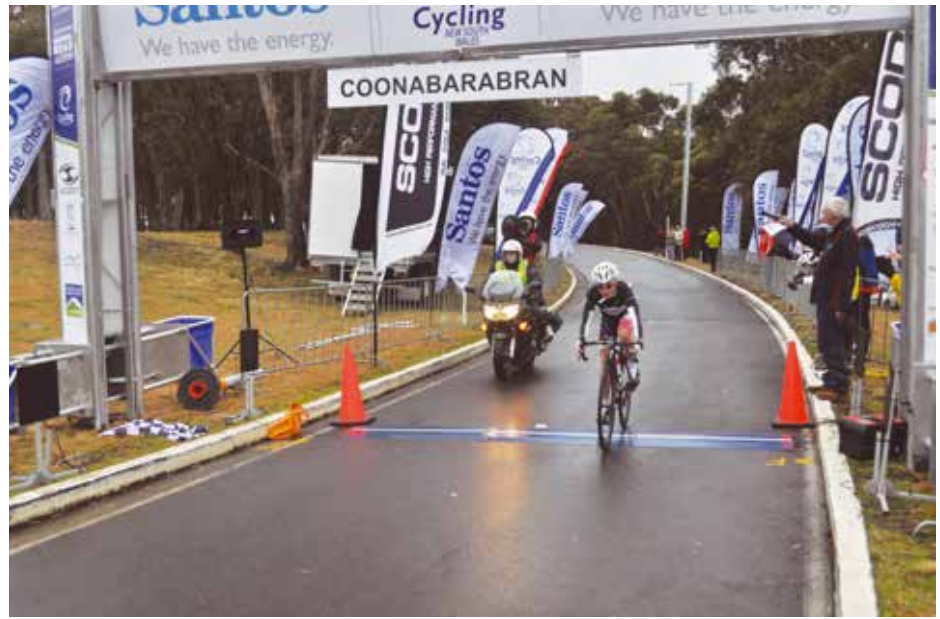

First across the line for the women's race is Ruth Corset, with Bob Dean looking on in the unfortunately wet weather on the summit of the mountain. Credit: Milton Judd

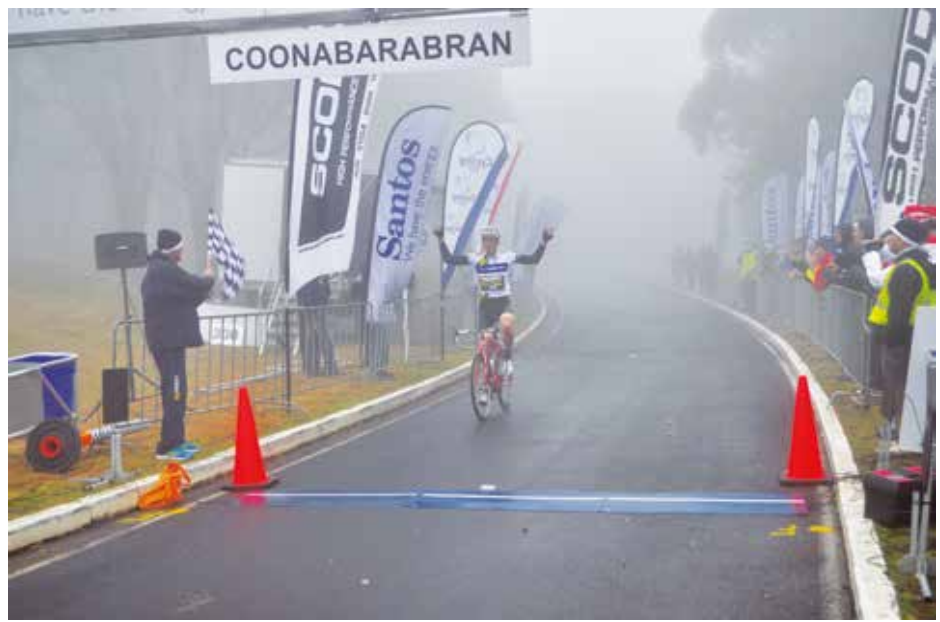

Mark O'Brien rides to the stage victory outside the Exploratory at Siding Spring Observatory. Credit: Milton Judd

With the emergency brakes installed there was no time to waste in getting the dome painted before the end of the financial year and the onset of winter.

The dome was last painted in 1991 and was starting to oxidise and in certain areas lichen was growing. Programmed Property Services of Tamworth were successful in winning the contract and quickly set about abseiling down the dome washing the entire surface prior to painting it with a Dulux product called InfraCool™ that claims to have a total solar reflectance of 90% and a thermal emittance of 0.88. The dome is now looking as good as new for hopefully another 20 years.

The UK Schmidt dome has also been repainted.

Doug Gray

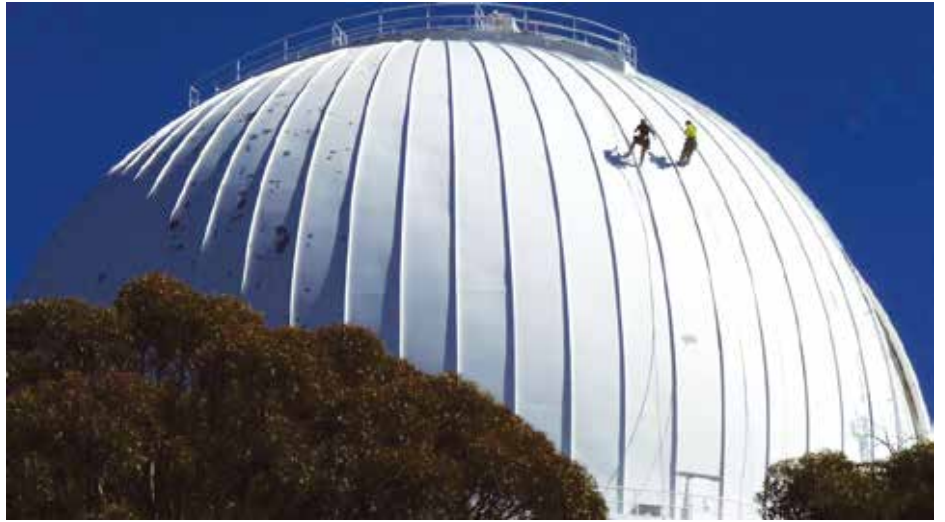

The guys from Programmed Property Services abseil down the outside of the dome, spraying paint while they go. Conveniently, the dome rotates so they can always be working in the warmth of the sun. Credit: Guy Albertson

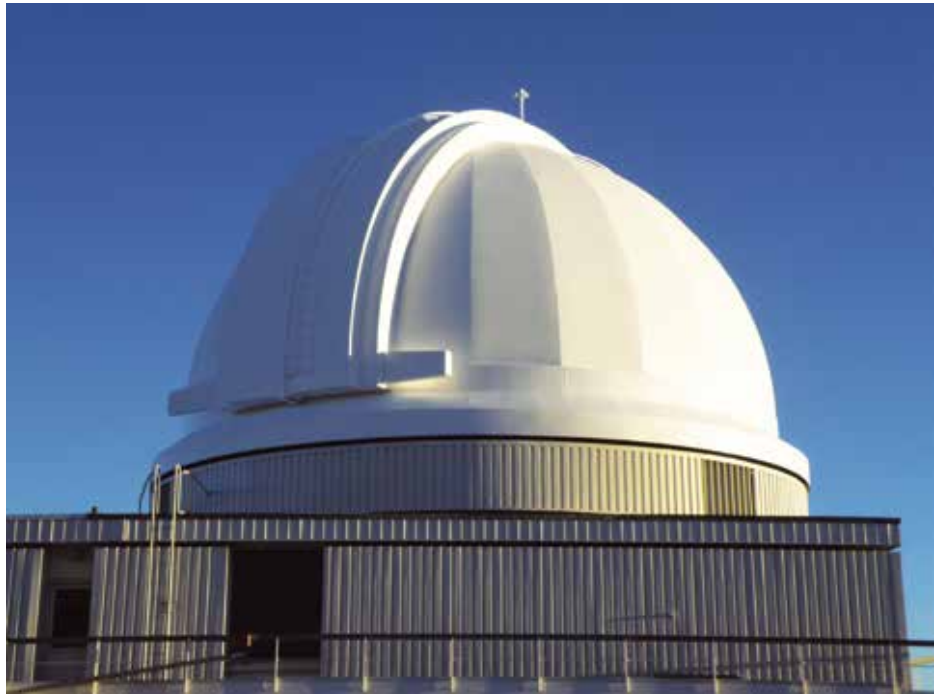

The newly painted UK Schmidt Dome.

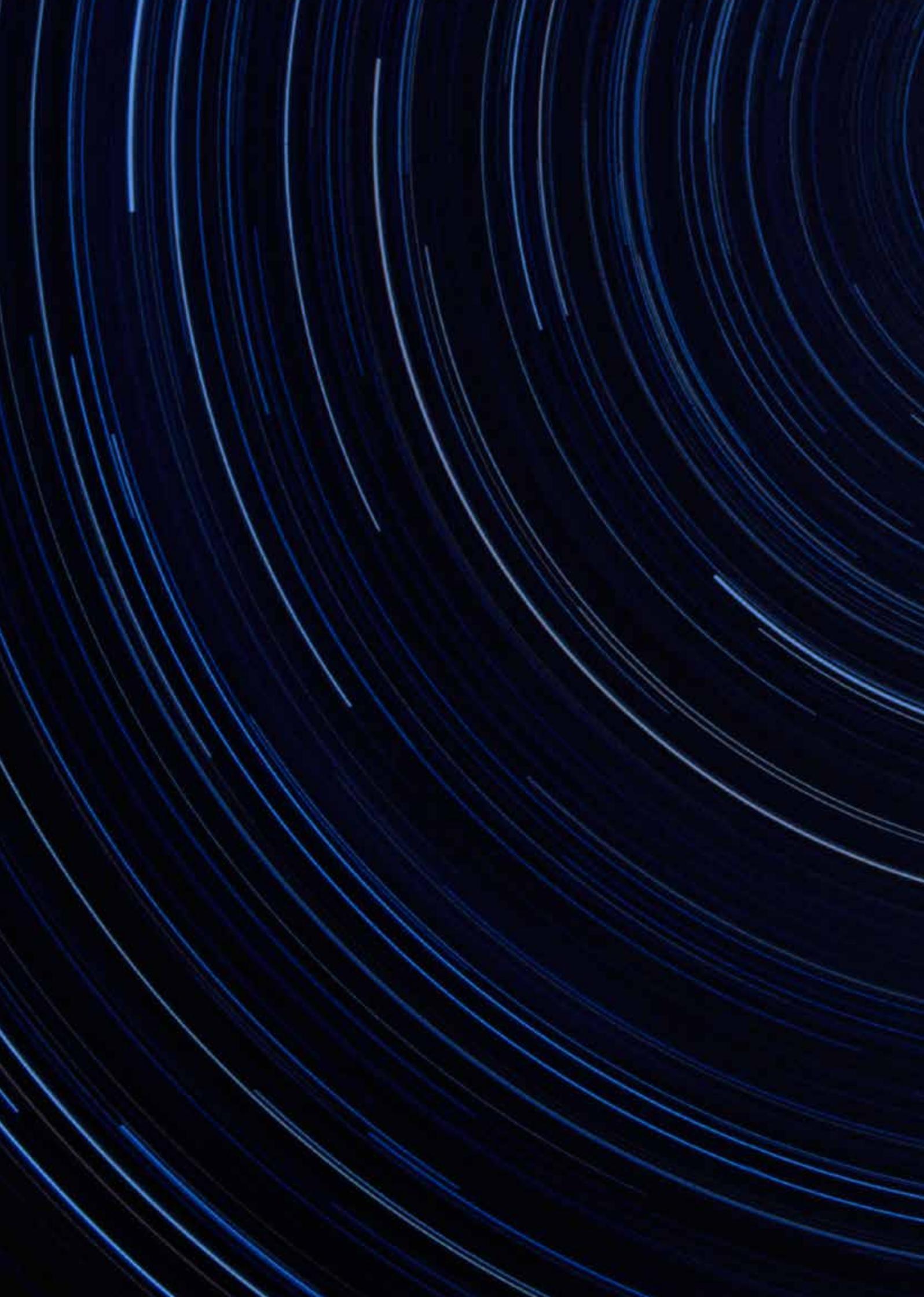

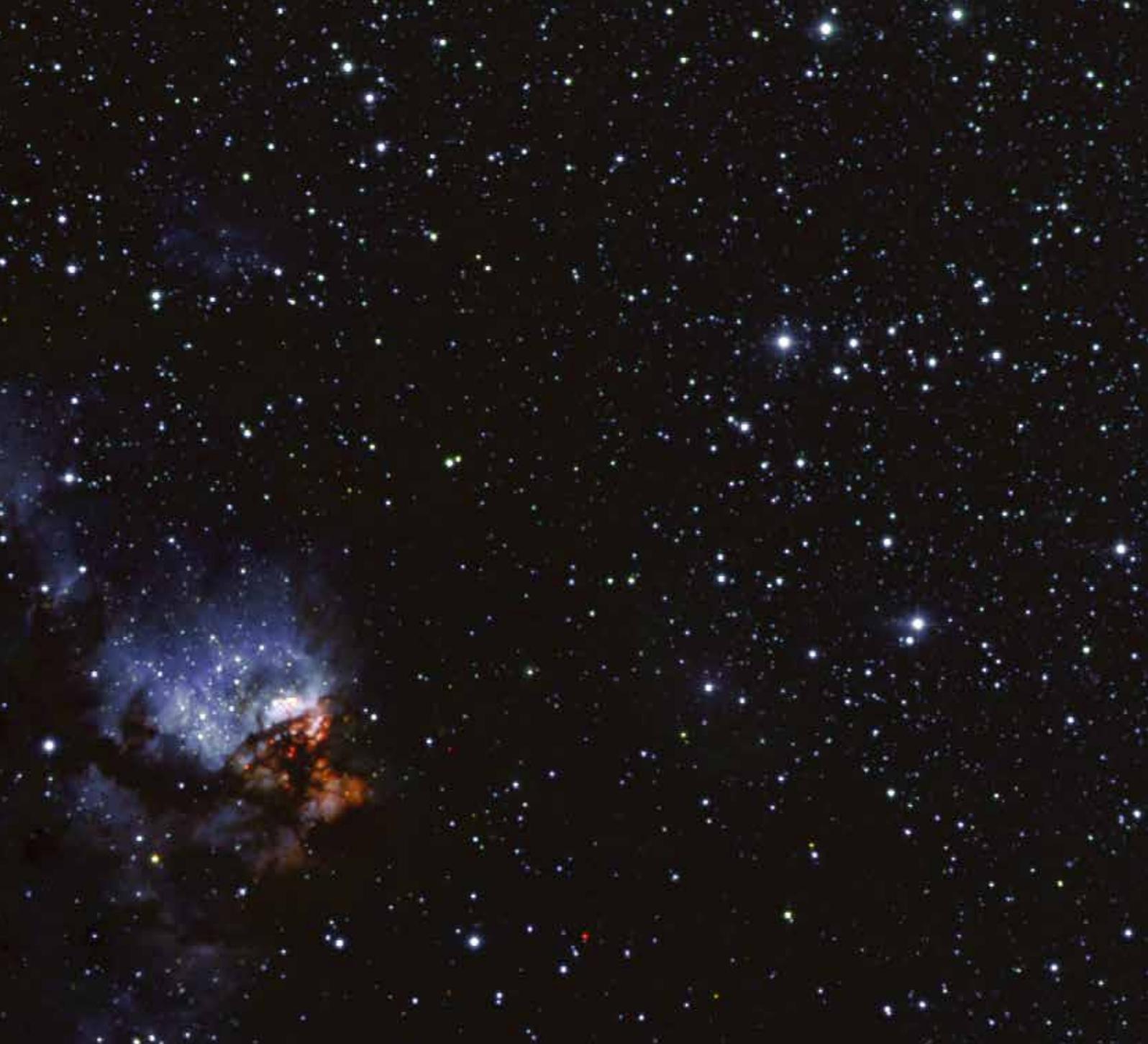

Near-infrared JHK colour composite image of the NGC 3581 star-formingregion in Carina. The images were obtained with IRIS2 in April 2012 by Stuart Ryder and Peter Barnes as part of the IR-CHaMP project (InfraRed Census of High- and Medium-mass Protostars; Barnes et al. 2010, MNRAS, 402, 73). Two adjacent IRIS2 pointings were reduced by the ORAC-DR pipeline then stitched together automatically with the Starlink CCDPACK software.

Colour processing courtesy of Angel Lopez-Sanchez, AAO.

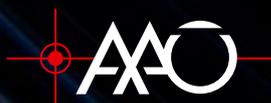